\documentclass{emulateapj}
\usepackage{epsfig}

\newcommand{\beq}{\begin{equation}}
\newcommand{\eeq}{\end{equation}}

\def\be{\begin{equation}}
\def\ee{\end{equation}}
\def\bea{\begin{eqnarray}}
\def\eea{\end{eqnarray}}



\def \logTd6 {\hbox{log$( T/6 \kev)$} }

\def\myputfigure#1#2#3#4#5%
{\vskip#5pt\makebox[0pt]{\hskip#2in
\includegraphics[width=#3\textwidth]{#1}}\vskip#4pt\hfill}

\def \arcmin     { ^{\prime} }
\def \arcsec    {^{\prime\prime}}

\def \kms            {~{\rm km~s}^{-1}}

\def \etal      {et al.}

\def \kev       {{\rm\ keV}}

\def \hMpc      {h^{-1}{\rm\ Mpc}}

\input epsf
\bibpunct{(}{)}{;}{a}{}{,}

\begin{document}

\lefthead{Giant Arcs}
\righthead{Hennawi \etal}

\title{A New Survey for Giant Arcs}

\author{Joseph F. Hennawi\altaffilmark{1,2}, 
  Michael D. Gladders\altaffilmark{1,3,4}, 
  Masamune Oguri\altaffilmark{5,6}, 
  Neal Dalal\altaffilmark{7},
  Benjamin Koester\altaffilmark{8},
  Priyamvada Natarajan\altaffilmark{9,10},
  Michael A. Strauss,\altaffilmark{6}
  Naohisa Inada,\altaffilmark{11}
  Issha Kayo,\altaffilmark{12}
  Huan Lin\altaffilmark{13},
  Hubert Lampeitl\altaffilmark{13},
  James Annis\altaffilmark{13}
  Neta A. Bahcall,\altaffilmark{6}
  Donald P. Schneider\altaffilmark{14} 
} 

\altaffiltext{1}{Hubble Fellow}
\altaffiltext{2}{Department of Astronomy, University of California
  Berkeley, Berkeley, CA 94720}
\altaffiltext{3}{Carnegie Observatories, Pasadena, CA 91101}
\altaffiltext{4}{Department of Astronomy \& Astrophysics, University of Chicago, Chicago, IL 60637}
\altaffiltext{5}{Kavli Institute for Particle Astrophysics and
Cosmology, Stanford University, 2575 Sand Hill Road, Menlo Park, CA
94025.}
\altaffiltext{6}{Princeton University Observatory, Princeton, NJ 08544}
\altaffiltext{7}{Canadian Institute for Theoretical Astrophysics, University of Toronto, 60 St. George St., Toronto, Ontario, Canada M5S 3H8}
\altaffiltext{8}{Physics Department, University of Michigan, Ann Arbor, MI 48109}
\altaffiltext{9}{Department of Astronomy, Yale University, P. O. Box 208101, New Haven, CT 06511-8101}
\altaffiltext{10}{Department of Physics, Yale University, P. O. Box 208120, New Haven, CT 06520-8120}
\altaffiltext{11}{Institute of Astronomy, Faculty of Science,
University of Tokyo, 2-21-1 Osawa, Mitaka, Tokyo 181-0015, Japan.}
\altaffiltext{12}{Department of Physics and Astrophysics, Nagoya
University, Chikusa-ku, Nagoya 464-8062, Japan.}
\altaffiltext{13}{Fermi National Accelerator Laboratory, P.O. Box 500, 
 Batavia, IL 60510}
\altaffiltext{14}{Department of Astronomy and Astrophysics, Pennsylvania State
University, 525 Davey Laboratory, University Park, PA 16802.}

\begin{abstract}
  We report on the first results of an imaging survey to detect strong
  gravitational lensing targeting the richest clusters selected from
  the photometric data of the Sloan Digital Sky Survey (SDSS) with
  follow-up deep imaging observations from the Wisconsin Indiana Yale
  NOAO (WIYN) 3.5m telescope and the University of Hawaii 88-inch
  telescope (UH88). The clusters are selected from an area of
  8000~${\rm deg^2}$ using the Red Cluster Sequence technique and span
  the redshift range $0.1 \lesssim z \lesssim 0.6$, corresponding to a
  comoving cosmological volume of $\sim 2 {\rm Gpc}^3$. Our imaging
  survey thus targets a volume more than an order of magnitude larger
  than any previous search.  A total of 240 clusters were imaged of
  which 141 had sub-arcsecond image quality.  Our survey has uncovered
  16 new lensing clusters with definite giant arcs, an additional 12
  systems for which the lensing interpretation is very
  likely, and 9 possible lenses which contain shorter arclets or candidate
  arcs which are less certain and will require further observations to
  confirm their lensing origin. The number of new cluster lenses
  detected in this survey is likely $\gtrsim 30$.  Among these new
  systems are several of the most dramatic examples of strong
  gravitational lensing ever discovered with multiple bright arcs at
  large angular separation. These will likely become `poster-child'
  gravitational lenses similar to Abell 1689 and CL0024$+$1654. The
  new lenses discovered in this survey will enable future sysetmatic
  studies of the statistics of strong lensing and its implications for
  cosmology and our structure formation paradigm.
\end{abstract}


\section{Introduction}
\label{sec:intro}

Does the currently favored $\Lambda$CDM cosmological model explain the
detailed distribution of dark matter in galaxy clusters?  Strong
gravitational lensing by clusters is a powerful test of this model,
probing the largest collapsed structures in the Universe, where the
density of dark matter is highest and where mass to light ratios are
sufficiently high ($\sim 10-100$) that baryons are unlikely to
significantly influence the distribution of dark matter. The
statistics of giant arcs -- such as their number counts, redshift
distribution (of both sources and lenses), the distribution of image
separations, and the distribution of angles between arcs in multi-arc
clusters -- provide powerful constraints on the distribution of dark
matter and its evolution over cosmic time.  Detailed modeling of image
positions in individual lens systems can constrain the mass profile of
dark matter halos \citep[e.g.][]{Tyson98,Smith01,Sand04, Broad05a},
and even stronger constraints can be obtained when strongly lensed
image positions are combined with larger scale weak lensing
measurements
\citep{Smith01,Kneib03,Gavazzi03,Broad05b,Oguri05,Dalal06}. Besides
providing cosmological constraints, lensing clusters are natural
gravitational telescopes, whose magnification facilitates the study of
otherwise unobservable faint high redshift background galaxies
\citep{Blain99,Smail02,Metcalfe03,Santos04}.

Surveys for giant arcs in galaxy clusters began just a few years after
they were discovered \citep{LP89}.  \citet{Smail91} conducted the
first homogenous arc survey by imaging 19 rich clusters in the
$V$-band and identifying 20 candidate arcs and arclets.
\citet{Luppino99} conducted an imaging survey ($V<22$) for giant arcs
behind luminous X-ray clusters with $z > 0.15$ and $L_{\rm X} >
2\times 10^{44}~{\rm ergs~s^{-1}}$, selected from the Extended
Medium-Sensitivity Survey (EMSS), following the earlier EMSS imaging
survey of \citet{LeFevre94}.  Luppino et al. found that 8 out of 38
clusters showed arcs with angular separation $\Delta\theta\gtrsim
10\arcsec$ from the brightest clusters galaxies (BCGs). The effective
area of the Luppino et al. survey was $\sim 360~{\rm deg}^2$
\citep{Dalal04}, corresponding to a giant arc abundance of one arc per
$45~{\rm deg}^2$.  \citet{Zaritsky03} discovered two arcs at small
angular separations ($\Delta \theta\lesssim 10\arcsec$) in deep
imaging ($R < 21.5$) of 44 clusters in the redshift range $0.5 < z <
0.7$.  These clusters comprised a subsample of the optically selected
Las Campanas Distant Cluster Survey (LCDCS), and the effective area of
their survey was $69~{\rm deg}^2$. \citet{Gladders03} discovered eight
lensing clusters in the $90~{\rm deg}^2$ imaging ($\mu_{\rm R} <
24~{\rm mag~arcsec^{-2}}$) area of the Red-Sequence Cluster Survey
(RCS).  Oddly, all eight of the Gladders et al. lensing clusters are
at $z > 0.64$ even though the RCS cluster catalog extends down to $z
\sim 0.3$. Although many arc systems have been discovered from
ground-based imaging, the superb image quality delivered by the HST
(FWHM=$0.15\arcsec$) enables the detection of much fainter lower
surface brightness features which would be missed in ground-based
images.  \citet{Sand05} exploited this fact, and mined the
\emph{Hubble Space Telescope} (HST) Wide Field and Planetary Camera 2
(WFPC2) data archive in search for strong lensing.  They presented
$\sim 60$ arcs with image separations of $\Delta \theta \gtrsim
10\arcsec$, which occurred in 32 out of the 128 clusters they
examined\footnote{For ease of comparison with previous surveys we
  quote the length-to-width ratios for arcs with $L\slash W > 10$,
  whereas Sand et al. published all arcs with $L\slash W > 7$.}.

\citet{Bart98} compared the rate of occurrence of giant arcs in a
subsample of the EMSS \citep{LeFevre94} to the theoretical predictions
of ray tracing simulations through CDM clusters, and argued that the
the observed arc abundance exceeded the prediction of the $\Lambda$CDM
cosmological model by an order of magnitude. More recent analyses of
this claimed discrepancy have shown that $\Lambda$CDM is in reasonable
agreement with giant arc statistics
\citep{Oguri03,Wamb04,Dalal04,Li05,Horesh05} at low redshift $z <
0.5$. In particular, \citet{Dalal04} argued that the
order-of-magnitude discrepancy claimed by \citet{Bart98} was likely
due to an underestimate of the EMSS survey volume, an overestimate of
the density of background sources, and a slight overestimate of the
lensing optical depth. \citet{Hennawi06} recently compared the
abundance of giant arcs in \citet{Gladders03} RCS sample to
predictions from ray tracing simulations of $\Lambda$CDM and found
good agreement for the number of $z > 0.6$ lenses; however, a similar
number of $z < 0.6$ lenses were predicted in the RCS area which were
not observed. Although, this putative discrepancy is based on only a
handful of objects, it should be emphasized that the \citet{Hennawi06}
study used $\sigma_8=0.95$, which is discrepant at the 4$-\sigma$
level with the lower most recent WMAP measurement of $\sigma_8
=0.76\pm 0.05$ \citep{Spergel06}, and is $1.5\sigma$ discrepant with
the measurement from the RCS cluster abundance of $\sigma_8
=0.67^{+0.18}_{-0.13}$ \citep{gla06}. A lower value of $\sigma_8$
would significantly reduce the number of high redshift clusters,
making the RCS results even more discrepant with theoretical
predictions.  This putative excess of high redshift lensing clusters
as well as the observation that the proportion of lensing clusters
with arcs from multiple background sources is very high $\sim50$\%
\citep[but see][]{Ho05}, led \citet{Gladders03} to speculate that the
cluster lensing at high redshift is caused by a sub-population of
cluster `superlenses', which have much larger lensing cross section
than the cluster population as a whole.

To date, the three best studied strong lensing clusters are Abell
1689, CL0024$+$1654, and MS2137.3$-$2353.  The presence of multiple
lensed arcs in these systems has enabled detailed modeling of image
positions which strongly constrains the distribution of dark matter,
especially when combined with larger scale weak lensing
measurements. Recent analyses of the mass profiles in these clusters
\citep{Gavazzi03,Kneib03,Broad05a} have uncovered highly concentrated
mass distributions. All three clusters have primary mass components
with NFW concentrations $c_{\rm NFW} > 14$; whereas \citet{Hennawi06}
showed that such high concentrations are expected in${}<2\%$ of
clusters which show giant arcs \citep[see also][]{Oguri05}.  Why
should the three best studied lensing clusters in the Universe all lie
on the tail of the concentration distribution? Is this merely a
selection bias or does it point to a deeper problem with the
$\Lambda$CDM model on cluster scales? A larger sample of similar
multi-arc lensing clusters will be required to determine if the
concentrations of cluster lenses are really discrepant with CDM, which
brings us to the subject of this work.

Recent analyses of strong lensing by galaxy clusters have highlighted
several potential conflicts with the cold dark matter (CDM)
paradigm. But these conclusions are limited by small sample sizes
and/or heterogeneously selected cluster samples. In this paper we
describe a survey for lensing clusters which aims to remedy the
situation by exploiting the $\gtrsim 1~{\rm Gpc}^3$ cosmological
volume of the Sloan Digital Sky Survey (SDSS).  We have carried out a
deep imaging ($\mu_g \lesssim 25.7$) search for giant arcs, conducted on
the Wisconsin Indiana Yale NOAO (WIYN) 4-m telescope and the
University of Hawaii 88-inch telescope (UH88), which targets the
richest clusters identified in the SDSS photometric data. The red
cluster sequence algorithm \citep{GY00,GY05} was applied to the $\sim
8000$ deg$^2$ of SDSS photometric data, resulting in the the largest
catalog of massive galaxy clusters in existence \citep{Gladders06a};
our giant arc survey covers an area 20 times larger than any previous
search for cluster lensing.

The input cluster catalog to our survey is briefly described in
\S~\ref{sec:catalog}.  The deep imaging observations and data
reduction are discussed in \S~\ref{sec:imaging}.  We explain how we
identified giant arcs and present our sample of new lensing clusters
in \S~\ref{sec:sample}. Notes on individual systems are given in
\S~\ref{sec:notes}, and we summarize and conclude in
\S\ref{sec:conclusion}.  Throughout this work we use the best fit
three year WMAP only (maximum likelihood) cosmological model of
\citet{Spergel06}, which has $\Omega_m = 0.24$, $\Omega_\Lambda
=0.76$, $h=0.73$.

\section{Cluster Selection}
\label{sec:catalog}

The Sloan Digital Sky Survey \citep{york00} uses a dedicated 2.5m
telescope and a large format CCD camera \citep{Gunn98,Gunn06,tucker06}
at the Apache Point Observatory in New Mexico to obtain images in five
broad bands \citep[$u$, $g$, $r$, $i$ and $z$, centered at 3551, 4686,
  6166, 7480 and 8932 \AA, respectively;][]{Fuku96,Stoughton02} of
high Galactic latitude sky in the Northern Galactic Cap.  The imaging
data are processed by the astrometric pipeline \citep{Astrom} and
photometric pipeline \citep{Photo,lupton06}, and are photometrically
calibrated to a standard star network
\citep{Hogg01,Smith02,ivezic04}. Based on this imaging data,
spectroscopic targets chosen by various selection algorithms (i.e.,
quasars, galaxies, stars, serendipity) are observed with two double
spectrographs producing spectra covering \hbox{3800--9200 \AA} with a
spectral resolution ranging from 1800 to 2100 (FWHM
$\simeq150-170\kms$).  Details of the spectroscopic observations can
be found in \citet{Castander01} and \citet{Stoughton02}.  Additional
details on the SDSS data products can be found in \citet{DR1,DR2,DR3}
and \citet{DR4}.

The total imaging area of the SDSS Data Release 5 \citep[DR5][]{DR5}
is $\sim 8000~{\rm deg}^2$.  We applied the RCS selection algorithm
\citep{GY00,GY05} to the SDSS photometric galaxy catalog, and produced
a sample of $\sim 2\times 10^4$ massive galaxy clusters over the
redshift range $0.05 \lesssim z \lesssim 0.6$, which is the largest
catalog of massive clusters in existence. The RCS algorithm exploits
the fact that all clusters have a red sequence of early-type galaxies,
and a cluster is localized as an overdensity in position, magnitude
and color simultaneously.  This method has been demonstrated to be
robust and effective in the RCS survey \citep{GY00,GY05}, and has been
tested on both real and mock data.  Using photometric redshift
techniques, the RCS algorithm yields cluster redshifts which are
accurate to $\lesssim 0.05$ by fitting for the location of the red
sequence of the cluster galaxies in the color-magnitude plane. These
initial approximate redshifts were used as the input to our follow-up
imaging survey.  For those clusters which we imaged at WIYN or UH88,
we re-computed more accurate 5-band photometric redshifts $\sigma
\lesssim 0.02$, restricting attention to only those galaxies which
satisfied the Luminous Red Galaxy (LRG) color-selection criteria
\citep{Eisenstein01,Paddy05}, because these galaxies have been
demonstrated to have more robust photometric redshifts, 
accurate to $\sigma_z \sim 0.03$ in the redshift range $0.2\lesssim z
\lesssim 0.55$.  We employed the photometric redshift algorithm
described in \citet{Paddy05}.

Only clusters with redshifts $z\gtrsim 0.1$ were targeted for
follow-up imaging, since strong lensing is significantly less likely
behind lower redshift clusters \citep[see e.g.][]{Hennawi06}. The
cosmological volume accessible to our survey is thus $1.9~{\rm h^{-1}
  Gpc^{3}}$ ($0.1 \lesssim z \lesssim 0.6$; $\sim 8000~{\rm deg}^2$).
The clusters were assigned priorities based on a combination of
richness, detection significance, and a parameter that describes the
concentration of the galaxy distribution, with the richest, most
significant, most highly concentrated clusters receiving a higher
priority.  The redshift range $z=0.1-0.6$ was divided into five equal
bins of width $dz=0.1$, and priorities were assigned based on the
relative rank of the cluster in that bin, and higher priority clusters
were targeted first.  We did not exclude previously known lensing
clusters from our imaging survey, although they tended to receive a
lower priority.  Full details of the SDSS-RCS cluster catalog
\citep{Gladders06a}, and the selection criteria for our cluster lens
survey \citep{Gladders06b} will be presented in the future.

In addition to the clusters selected by the RCS algorithm, we also
imaged a handful of candidates for which visual inspection of the SDSS
imaging suggested strong lensing.  One of these objects is the lensing
cluster SDSS~J0146$-$0929, which was serendipitously discovered in the
SDSS imaging by Pat Hall \citep{Allam06}. The others were candidate 
cluster lenses identified via visual inspection of the SDSS coadded
southern strip imaging \citep{Allam06}, for which the lensing
interpretation was ambiguous because the SDSS imaging was too shallow. 

\section{Deep Imaging Survey}
\label{sec:imaging}

\subsection{WIYN 3.5 Imaging}

The WIYN 3.5m telescope imaging observations took place during three
runs on 10-12 June 2004, 7-12 May 2005, and 24-27 May 2006, for which
conditions were photometric.  An additional run on 27-28 June 2006 was
completely lost to poor weather. With the exception of the first run,
we used the Orthogonal Parallel Transfer Imaging Camera
\citep[OPTIC][]{OPTIC}, which consists of two Lincoln Lab CCID28
2K$\times$x4K chips, resulting in a 4K$\times$4K imager. For 10-12
June 2004 run, we used the WIYN Mini-Mosaic Imager
\citep[Minimo][]{minimo}, which is also a 4K$\times$4K imager, but
consisting of two SITe 4096$\times$2048 CCDs.  Both OPTIC and MIMO 
have 15 $\mu m$ pixels subtending $0.14\arcsec$, and giving rise to a
total field of view of 9.6$\times$9.6 arcminutes.  The readout time
for OPTIC is 25s, whereas Minimo takes 182s to read out the full
frame. The combined median seeing of all of our WIYN runs was
$0.90\arcsec$, measured in the $g$-bandpass.

Our primary observing strategy was to conduct a fast $g$-band survey
consisting of a 600s total integration on each cluster, which was
broken into two 300s exposures. The majority of the clusters were
observed in this mode. We chose the SDSS $g$ bandpass both because
giant arcs are known to typically be blue and to exploit the fact that
the night sky is darker in $g$ than in a redder bandpass. The 10-12
June 2004 imaging using Minimo was also conducted in $g$, however we
performed longer 2400s integrations. We chose longer 600s individual
exposures with Minimo because of the considerable readout time (182s)
of this detector.

Each image was bias subtracted using the overscan region and median
bias frames, and flat-fielded using a median twilight sky flat. Final
images were produced by registering and shifting the images, and
cleaning them of cosmic ray hits.

\subsection {UH 88-inch Imaging}

The UH 88-inch telescope was employed during two runs on 6-7 May 2005
and 24-25 April 2006, using the University of Hawaii 8k wide field
imager (UH8k), which is a mosaic CCD camera consisting of eight
2048$\times$4096 pixel CCDs, resulting in a total of 8192$\times$8192
pixels. The plate scale is $0.235\arcsec$, giving rise to a 32$\times$32
arcminute field of view.  The median seeing for the May 2005 and April
2006 runs was $0.86\arcsec$ and $1.13\arcsec$, respectively, measured
in the $V$-bandpass. Conditions were not photometric for the April
2006 run. One of our serendipitous lensing cluster targets
(SDSS~J0146$-$0922; see \S~\ref{sec:sample}) was observed on the
earlier date 17 December 2004 as part of a different observing
program.

Our observing strategy at the UH 88-inch differed for the two
runs. During the 6-7 May 2005 run, we imaged clusters in the
$V$-bandpass for total integration times of 1800s. The total
integrations were shortened to 900s for the run of 24-25 April
2006. The $V$-band was chosen because because this was the bluest
filter available for the UH8k camera at the time of observing. The
individual exposure times for the UH8k were 180s, which were chosen
because guiding is not available for this instrument, and longer
integrations would have resulted in image distortions. 

Each image was bias subtracted using a median bias frame, and
flat-fielded using a median dome flat. Final images were produced by
registering and shifting the images, and averaging the individual exposures. 

\subsection{Photometric Calibration and Seeing Measurements}

Astrometric solutions were obtained for our images by using the SDSS
astrometry of stars identified in each image.  Images were then
photometrically calibrated to the SDSS photometry using stars and
galaxies.  Because the UH88 images were taken in the V-band,  the photometric
transformations given in \cite{Fuku96} were used to produce estimated
V-band magnitudes for SDSS objects within these images, and hence a
nominal V-band calibration.  For the WIYN/OPTIC data (600s
integrations), our $1\sigma$ surface brightness limit in a $1\arcsec$
aperture is typically $\mu_g = 25.7\pm 0.1$; whereas the deeper
(2400s) WIYN/Minimo images from the 10-12 June 2004 run have a fainter
limit of $\mu_g = 26.5\pm 0.1$. For the UH88 images from the 6-7 May
2005 (1800s exposures) the surface brightness limit was $\mu_V=26.9\pm
0.1$. Seeing was measured for each image by hand by fitting a Gaussian
PSF to several stars in each image.

\section{Lensing Cluster Sample}
\label{sec:sample}
\begin{deluxetable*}{lcccccl}
\tablecolumns{7}
\tablewidth{0pc}
\tablecaption{Parameters for Definite Lensing Clusters\label{table:cutout1}}
\tablehead{Cluster Name &    RA         &     Dec       &  $z$     & Seeing    & Telescope/   & Notes \\
                        &     (2000)    &    (2000)     &          &($\arcsec$)& Instrument             &}
\startdata
SDSS~J0146$-$0929      &  01:46:56.01  & $-$09:29:52.5  & 0.444    &    0.97   &  UH88/8k     & known lens\\
SDSS~J1115$+$5319      &  11:15:14.85  & $+$53:19:54.6  & 0.466    &    1.02   &  WIYN/OPTIC  & complex morphology\\
NSC~J115347+425155     &  11:53:49.49  & $+$42:50:43.1  & 0.327    &    0.58   &  WIYN/OPTIC  & arc near cD\\
SDSS~J1217$+$3641      &  12:17:31.94  & $+$36:41:12.3  & 0.364    &    0.71   &  WIYN/OPTIC  & confusion from nearby galaxy\\
\smallskip
Abell 1703             &  13:15:05.24  & $+$51:49:02.6  & 0.281    &    0.97   &  WIYN/Minimo & two arcs\\
GHO~132029$+$315500    &  13:22:48.77  & $+$31:39:17.8  & 0.307    &    1.00   &  WIYN/OPTIC  & definite giant arc\\
RX~J1327.0$+$0211      &  13:27:01.01  & $+$02:12:19.5  & 0.260    &    0.64   &  WIYN/OPTIC  & high surface-brightness arc\\
NSC~J134610$+$030555   &  13:46:03.53  & $+$03:09:31.0  & 0.232    &    0.66   &  WIYN/OPTIC  & faint arc\\
Abell 1835             &  14:01:02.07  & $+$02:52:42.5  & 0.252    &    1.06   &  WIYN/Minimo & known lens\\
\smallskip
Abell 1914             &  14:25:56.67  & $+$37:48:59.3  & 0.170    &    0.62   &  WIYN/OPTIC  & known lens\\
Abell 1942             &  14:38:21.87  & $+$03:40:13.2  & 0.225    &    0.79   &  WIYN/OPTIC  & known lens\\
SDSS~J1446$+$3032      &  14:46:34.02  & $+$30:32:58.2  & 0.47\phn &    0.63   &  WIYN/OPTIC  & multiple bright arcs\\
Abell 2070             &  15:24:19.51  & $+$35:15:59.3  & 0.253    &    0.80   &  WIYN/OPTIC  & definite bright arc\\
SDSS~J1531$+$3414      &  15:31:10.60  & $+$34:14:25.0  & 0.335    &    1.20   &  WIYN/OPTIC  & Multiple high surface brightness arcs\\
\smallskip
Abell 2141             &  15:57:42.40  & $+$35:30:29.8  & 0.159    &    0.63   &  WIYN/OPTIC  & Bright arc with knots.\\
SDSS~J1557$+$2131      &  15:57:44.84  & $+$21:31:49.3  & 0.427    &    0.54   &  WIYN/OPTIC  & CB58's?\\
SDSSJ1602$+$4038       &  16:02:24.68  & $+$40:38:50.1  & 0.387    &    0.58   &  WIYN/OPTIC  & length/center uncertain\\
Abell 2219             &  16:40:19.81  & $+$46:42:41.5  & 0.234    &    0.69   &  WIYN/OPTIC  & known lens\\
RX~J1652.2$+$4449$^\dagger$      &  16:52:14.59  & $+$44:49:23.9  & 0.175    &    0.67   &  WIYN/OPTIC  & Arc near BCG\\
\\
\smallskip
Abell 2261$^\ast$              &  17:22:27.21  & $+$32:07:55.1  & 0.224    &    0.66   &  WIYN/OPTIC  & known lens \\
SDSS~J1747$+$5428      &  17:47:21.76  & $+$54:28:13.9  & 0.31\phn &    1.01   &  WIYN/OPTIC  & arc near BCG\\
SDSS~J2111$-$0115      &  21:11:19.34  & $-$01:14:23.5  & 0.68\phn &    0.74   &  WIYN/Minimo & multiple high surface brightness arcs\\
\enddata
\tablecomments{\footnotesize Parameters for definite lensing clusters
  shown in Figure~\ref{fig:cutout1}.  The coordinates are those of the
  brightest cluster galaxy as identified by our cluster finding
  algorithm. For cases where the arcs surround a secondary
  concentration of galaxies, the coordinates of a cluster member in
  this concentration are substituted. Spectroscopic redshifts, either
  from the SDSS or from the literature, are quoted to three
  significant digits; while, photometric cluster redshift are
  quoted to two digits. The seeing of the arc discovery image is
  indicated, as well as the telescope and instrument used. Brief notes
  about the lensing features for each clusters are included. For more
  detailed notes including the references for known lenses see \S~\ref{sec:notes}.
  Clusters which are labeled as SDSS are new discoveries in the RCS
  cluster catalog \citep{Gladders06a}. Previously known clusters are indicated by their 
  names. Abell clusters are from the compilation of \citet{Abell}.  Clusters labeled NSC were 
  previously discovered in the optical catalog of \citet{NSC}. GHO indicates 
  clusters from the photographic distant survey of \citet{GHO}. Clusters labeled 
  RX are X-ray clusters from the NORAS catalog of \citet{Bohringer00}.\\
  $^\dagger$ Spectroscopic redshift from the NORAS catalog \citep{Bohringer00}.\\
  $^\ast$ Spectroscopic redshift from \citet{Crawford95}.
}
\end{deluxetable*}

\begin{deluxetable*}{lcccccl}
\tablecolumns{7}
\tablewidth{0pc}
\tablecaption{Parameters for Tentative Lensing Clusters\label{table:cutout2}}
\tablehead{Cluster Name &    RA         &   Dec         &  $z$    &  Seeing   & Telescope/ & Notes \\
                        &   (2000)      &  (2000)       &         &($\arcsec$)& Instrument          &}
\startdata
SDSS~J1040$+$3405       &   10:40:19.72 &  $+$34:05:19.8  & 0.445    & 1.10 & UH88/8k       & short arc\\
SDSS~J1131$+$3554       &   11:31:10.08 &  $+$35:54:19.0  & 0.49\phn & 1.10 & WIYN/OPTIC  & confusion from bright star\\ 
Abell 1351$^\dagger$              &   11:42:23.02 &  $+$58:30:45.0  & 0.322    & 0.97 & UH88/8k       & known lens \\
SDSS~J1150$+$0650       &   11:50:30.29 &  $+$06:50:19.1  & 0.301    & 1.11 & WIYN/OPTIC  & arclet\\
\smallskip
Abell 1550              &   12:29:02.53 &  $+$47:36:56.0  & 0.262    & 1.30 & WIYN/OPTIC  & faint arc\\
SDSS~J1258$+$4702       &   12:58:02.09 &  $+$47:02:54.2  & 0.32\phn & 1.39 & WIYN/OPTIC  & arc near BCG\\
Abell 1758              &   13:32:38.42 &  $+$50:33:35.7  & 0.279    & 0.85 & WIYN/OPTIC  & known lens\\
SDSS~J1406$+$3945       &   14:06:32.72 &  $+$39:45:46.2  & 0.427    & 0.89 & UH88/8k       & short arc\\
SDSS~J1414$+$2703       &   14:14:39.13 &  $+$27:03:10.5  & 0.46\phn & 0.63 & WIYN/OPTIC  & many suggestive features\\
\smallskip
SDSS~J1527$+$0652       &   15:27:45.83 &  $+$06:52:33.6  & 0.40\phn & 0.67 & WIYN/OPTIC  & very high surface brightness\\
SDSS~J1537$+$3926       &   15:37:52.22 &  $+$39:26:09.9  & 0.444    & 0.57 & WIYN/OPTIC  & faint resolved arc\\
Abell 2136              &   15:53:18.86 &  $+$51:07:24.9  & 0.229    & 0.87 & WIYN/Minimo & multiple arclets\\
GHO~155414$+$405306       &   15:55:57.95 &  $+$40:44:14.8  & 0.393    & 0.52 & WIYN/OPTIC  & ambiguous\\
RXC~J1749.3$+$4245$^\dagger$      &   17:49:18.05 &  $+$42:46:38.4  & 0.230    & 0.63 & WIYN/OPTIC  & short blobby arc \\
\enddata
\tablecomments{\footnotesize Same as Table~\ref{table:cutout1}, but for the tentative 
  lensing clusters shown in Figure~\ref{fig:cutout2}.\\
  $^\dagger$ Spectroscopic redshift from the NORAS catalog \citep{Bohringer00}.}
\end{deluxetable*}

\begin{deluxetable*}{lcccccl}
\tablecolumns{7}
\tablewidth{0pc}
\tablecaption{Parameters for Possible Lensing Clusters\label{table:cutout3}}
\tablehead{Cluster Name &    RA         &   Dec            &   $z$ &  Seeing   & Telescope/   & Notes \\
                        &   (2000)      &  (2000)          &       &($\arcsec$)& Instrument            &}
\startdata
SDSS~J0821$+$2519        &   08:21:21.27 &  $+$25:19:03.0  & 0.268    &    1.05   &  UH88/8k       & possible faint arc\\ 
SDSS~J1008$+$4529        &  10:08:54.85  & $+$45:29:13.2  & 0.479    &    0.81    &  UH88/8k       & possible pair of faint pair\\
NSC~J113554$+$40012      &   11:35:59.06 &  $+$40:05:17.9  & 0.295    &    0.88   &  UH88/8k       & possible faint arc\\
SDSS~J1240$+$4250        &   12:40:29.98 &  $+$42:50:08.1  & 0.406    &    0.73   & WIYN/OPTIC    & geometry ambiguous\\
\smallskip
Abell 1926               &   14:30:28.62 &  $+$24:40:19.3  & 0.135    &    1.33   & WIYN/OPTIC   & arclet?\\
Abell 1934$^{\dagger\dagger}$               &   14:33:21.85 &  $+$29:27:01.3  & 0.220    &    0.63   & WIYN/OPTIC   & possible faint arc\\
SDSSJ1513$+$0525         &   15:13:46.27 &  $+$05:25:38.6  & 0.49\phn &    0.69   & WIYN/OPTIC   & high SB images at large separation\\
NSC~J163346$+$243312     &   16:33:48.85 &  $+$24:32:37.3  & 0.195    &    0.54   & WIYN/OPTIC   & faint arclets around secondary group\\
Abell 2224               &   16:43:28.83 &  $+$13:21:53.9  & 0.126    &    0.50   & WIYN/OPTIC   & pair of arclets around secondary group\\
\enddata
\tablecomments{\footnotesize Same as Table~\ref{table:cutout1}, but for the possible 
  lensing clusters shown in Figure~\ref{fig:cutout3}.\\
  $^{\dagger\dagger}$ Spectroscopic redshift from \citet{SR87}.}
\end{deluxetable*}

\begin{figure}
  \centerline{\epsfig{file=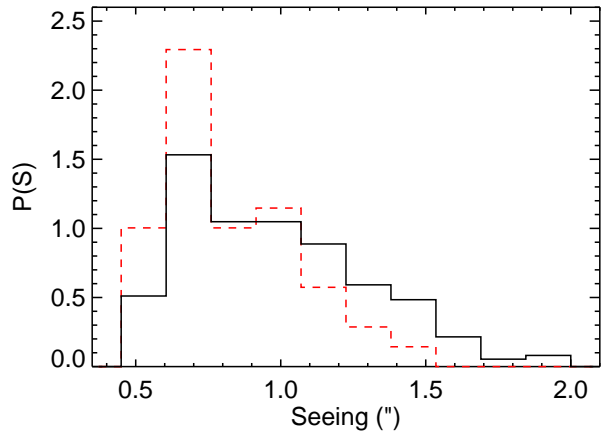,width=0.5\textwidth}}
  \caption{ Seeing distributions for the clusters imaged in our
    survey. The solid (black) histogram shows the seeing distribution
    for the 240 clusters imaged in our survey; whereas, the dashed
    (red) histogram is the seeing distribution for the 45 clusters
    which were identified as candidate strong lenses (see
    \S~\ref{sec:sample}). We imaged 141 clusters in sub-arcsecond
    seeing, The median seeing for the total survey is $0.94\arcsec$,
    and 141 clusters were imaged in sub-arcsecond seeing. The median
    seeing for the lenses was $0.74\arcsec$.\label{fig:seeing}}
\end{figure}

\begin{figure}
  \centerline{\epsfig{file=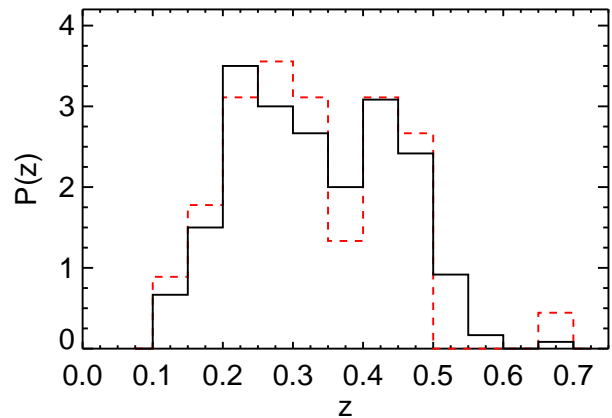,width=0.5\textwidth}}
  \caption{ Redshift distributions of the clusters imaged in our
    survey. The solid (black) histogram is the redshift distribution
    of the 240 clusters imaged in our survey; whereas, the dashed
    (red) histogram is the seeing distribution for the 45 clusters
    which were identified as candidate strong lenses (see
    \S~\ref{sec:sample}).\label{fig:zdist}}
\end{figure}

The final best-seeing image for each cluster was visually inspected by
two of the authors (J. Hennawi and M. Gladders) independently. Possible
lensing features were assigned an integer grade of 1-3 (1 being
`possible', 2 `likely', and 3 'definite'), and pixel positions of
each feature were noted. The intent of these rankings is such that a
well-detected giant arc with appropriate curvature and orientation
would receive a grade of 3, putative arclets would receive a grade of
1, and grade 2 features typically are apparent giant arcs at the sky
limit, or brighter apparent arcs with less plausible or ambiguous
image geometries, etc. The complete lists of each examiner were then
matched within a radius of $10\arcsec$, and the rankings summed,
producing a final candidate list of lensing features with rankings
between 1 and 6. We report in this paper on three sub-samples derived
from this list. The first and most definite sample contains objects
with a total rank of 4 or greater (which at the lower limit were
either assigned a grade of 3 by at least one of the examiners, or more
typically a 2 by both examiners). This is the core sample of the paper
which we refer to as our `Definite' lensing cluster sample. These
represent the clusters for which we have the greatest confidence in a
strong lensing interpretation. Next are all objects with a total rank
equal to 3; which we consider as `Likely' strong lensing features.
Our final sample of `Possible' cluster lenses, consists of objects
which received a score of 2 from one examiner but were missed by the
other.  These features were typically missed by one of us because they
are short arcs/arclets at the sky limit or they were brighter features but
were located well away from the cluster center. These possible lensing
clusters will require further imaging or spectroscopic observations to 
confirm their lensing origin. 

The three samples are summarized in
Tables~\ref{table:cutout1}-\ref{table:cutout3}, which give coordinates
and redshifts for the clusters, the seeing of the arc discovery image,
the telescope and instrument used for the observation, and brief notes
on the lensed features and clusters. Cutout images of the clusters in
the respective samples are shown in Figures~\ref{fig:cutout1}-\ref{fig:cutout3}.
Strong lensing features are indicated by lines and a number which is the 
total score of the identified strong lensing feature.

The cluster redshifts are derived either from SDSS spectroscopy, from
a photometric redshift analysis of the SDSS photometry of cluster
members, or for previously known clusters which did not have an SDSS
spectroscopic redshift, from the literature. For clusters with SDSS
spectra, the redshift reported is that of the early-type galaxy
nearest to the cluster center which had a redshift consistent with the
photometrically derived cluster redshift. For the photometric
redshifts, we examined the photo-$z$ distribution of all galaxies
within a projected aperture of 0.5~$\hMpc$ from the cluster center
which also satisfied the LRG color-cuts of \citet{Paddy05}, which are
based on the initial SDSS LRG color selection described in
\citet{Eisenstein01}.  We extended the LRG flux limits $0.7$
magnitudes fainter than those used by \citet{Paddy05} to obtain more
LRG cluster members and better statistics on the cluster
redshift. This amounted to a flux limit of $r < 20.4$ for LRGs in the
redshift range $0.2 \lesssim z \lesssim 0.4$, and $i < 20.7$ in the
redshift range $0.4 \lesssim z \lesssim 0.6$ (see Padmanabhan et
al. 2005 for details). We took the cluster redshift distribution to be
the sum of a set of Gaussians with means and dispersions given by the
photo-z and the photo-z error of the individual galaxies. The mode of
this distribution was then taken to be the cluster photometric
redshift. The average number of LRGs within the 0.5~$\hMpc$ aperture
was 14 for the 240 clusters which we imaged. Of these, $65\%$ had SDSS
spectroscopic redshifts, for which the average difference between the
spectroscopic redshift and the cluster photometric redshift was
$\langle |z_{\rm phot}-z_{\rm spec}|\rangle = 0.019$, and the median of
the same quantity was $0.013$.

In summary, a total of 240 clusters were imaged (195 from WIYN and 45
from UH88), out of which we found 22 definite cluster lenses (6
previously known), 14 likely lenses (2 previously known), and 9
possible systems. Because our imaging runs have only been in the
Spring months of April-June, we have primarily covered the range of
RA$= 10-17$ hours, and our survey of high priority clusters is roughly
half finished, so that a comparable number of systems are to be
expected after the completion of the survey.  The solid (black) curve
in Figure~\ref{fig:seeing} shows the seeing distribution for our
imaging survey and the dashed (red) curve is the seeing distribution
for the 45 clusters listed in
Tables~\ref{table:cutout1}-\ref{table:cutout3}. The median seeing of
all of our imaging is $0.94\arcsec$; while for the lenses it is
$0.74\arcsec$, illustrating that we were more likely to identity
lensing features with better image quality. Figure~\ref{fig:zdist}
compares the redshift distribution of the clusters we imaged (solid,
black) to that of the lenses (dashed, red). The cutoff in the
histograms for $z \lesssim 0.2$ reflects the fact that we were less
likely to observe lower redshift clusters. At $z\gtrsim 0.5$, only the
richest massive clusters are being identified in the SDSS photometric
survey, which explains the high redshift cutoff
\citep[see][]{Gladders06a}.

\section{Notes on Cluster Lenses}
\label{sec:notes}

In this section we briefly describe the cluster lenses published in
Tables~\ref{table:cutout1}-\ref{table:cutout3}. Each cluster in the
definite and likely samples is discussed, as well as a few of the more
notable clusters in the possible lens sample. Previously known lenses
which we re-discovered in our search are mentioned as well as known
lensing clusters which were imaged but missed. We matched all of the
clusters discussed below to the Northern ROSAT All-Sky (NORAS) galaxy
cluster sample \citep{Bohringer00}, the ROSAT-ESO Flux Limited X-ray
(REFLEX) galaxy cluster sample \citep{Bohringer04}, and also the ROSAT
All Sky Survey - SDSS cluster sample
\citep[RASS-SDSS;][]{Popesso04}. For the clusters detected, we note
X-ray luminosities in the $0.1-2.4~{\rm keV}$ energy band, converted
to the $\Lambda$CDM cosmology used in this paper.

\subsection{Definite Lenses}

\noindent {\bf SDSSJ~0146$-$0929} This lens at $z=0.444$ was
serendipitously discovered in the SDSS imaging by Pat
Hall \citep{Allam06}. Our UH88/8k $V$-band image shows three high surface
brightness arcs at $\Delta \theta \sim 13\arcsec$ from the central
galaxy. The morphology of the arcs are that of a classic quad
configuration, which suggests that these are three images of the same
multiply imaged source, as does the comparable surface brightness and
similar colors of the images.

\noindent {\bf SDSSJ~1115$+$5319} Three candidate arcs were identified
in our WIYN image (seeing $= 1\arcsec.02$) of this cluster at $z=0.466$,
which is one of the richest $z > 0.4$ clusters in our catalog.  The
most compelling feature is the arc situated $\Delta\theta = 31\arcsec$
east of the BCG (see Figure~\ref{fig:cutout1}), which received a score
of 6. This arc threads between two cluster members, suggesting that
the galaxies help in boosting the lensing potential. Two other arcs
were identified situated to the southeast at $\Delta\theta =
35\arcsec$ and $\Delta\theta = 57\arcsec$, and received scores of 3
and 4, respectively. The furthest of these two arcs also flanks a
cluster member. The morphology of the cluster is very irregular with
the galaxy distribution highly elongated along the east-west
direction.

\noindent {\bf NSC~J115347$+$425155} This cluster at $z=0.327$ has a
complex morphology. There are several bright early type galaxies at
the same redshift with comparable magnitudes, making the determination
of the cluster center somewhat arbitrary.  One of these secondary mass
concentrations extends as far as $\sim 2.5^\prime$ to the north of the
putative cluster center. Three candidate strong lensing features are
identified, two of which straddle the brightest early type galaxy
at the southern end of the cluster, at separations of $\Delta
\theta = 4\arcsec$ and $\Delta\theta = 6\arcsec$, respectively. The
third is an arclet situated $\Delta\theta \sim  50\arcsec$ to the north of 
this features among a group of galaxies which are members of the cluster. 

\noindent {\bf SDSS~J1217$+$3641} An obvious arc-like feature was
identified $\Delta \theta= 33\arcsec$ from the cluster center;
however, the gravitational lensing interpretation in this system is
complicated by the presence of a foreground elliptical galaxy ($z_{\rm
  phot}= 0.12$) at a separation of $\sim 10\arcsec$ from the candidate
arc. Although the feature we identified is most likely an arc, it
could also be a tidal tail or debris associated with the foreground
object. Broad band colors could be used to distinguish between these
two possibilities, but an image to comparable depth in another filter
(like $i$-band) is required.

\noindent {\bf Abell 1703} Abell 1703, a massive X-ray cluster with
$L_{\rm X} = 5.3\times 10^{44}~{\rm erg~s^{-1}}$ \citep{Bohringer00},
is one of the most dramatic lensing clusters in our sample. A long
giant arc at $\Delta\theta =35\arcsec$ stretches across the southeast
end of the cluster. A second arc is identified at $\Delta \theta =
13\arcsec$ from the BCG, with very high surface brightness knots at
each end, which suggests a merging pair of images of the same
background source. The large difference in angular separations of the
two arcs is unlikely to be caused by very different source redshifts,
especially considering that the two arcs have comparable surface
brightnesses. Furthermore, the radius of curvature of the smaller
separation arc as well as its merging pair morphology, suggest that it
could be a minor axis tangential arc, rather than an arc at very
different redshift or a radial arc \citep[see e.g.][]{Dalal04}. If
this interpretation is correct, the tangential critical curves in this
lens will be highly flattened, with the wide arc demarcating the major
axis, and the smaller arc indicating the minor axis. Such flattened
cigar-shaped critical curves arise naturally from the shallow radial
profiles and high ellipticity \citep{DK03,Dalal04} characteristic of
clusters in $\Lambda$CDM.

\noindent {\bf GHO~132029$+$315500} This cluster at $z=0.307$ was
discovered in the photographic high-redshift cluster survey of
\citet{GHO}, and has an X-ray luminosity $L_{\rm X} = 4.4\times
10^{44}~{\rm erg~s^{-1}}$\citep{Bohringer00}. Even with the modest 
seeing of our WIYN image (seeing $=1\arcsec.00$), we identify a
definite giant arc at angular separation $\Delta \theta = 21\arcsec$
to the east of the cluster center.

\noindent {\bf RX~J1327.0$+$0211} This cluster at $z=0.260$ has an
X-ray luminosity of $L_{\rm X} = 7.7\times 10^{44}~{\rm erg~s^{-1}}$
\citep{Popesso04}. An extremely high surface brightness arc is
detected $4\arcsec$ away from a cluster galaxy and curves slightly
around it.  This arc is at an angular separation of $\Delta =
93\arcsec$ to northeast of the cluster center, which is clearly
defined by a very luminous BCG, making it one of the largest
separation giant arcs ever discovered, although the nearby galaxy is
likely responsible for boosting the gravitational potential at such a
large distance from the cluster center.  The arc feature was bright
enough to be detected in the SDSS photometry and it has magnitudes
$(u,g,r,i,z) = (21.3,21.8,21.3,20.9,20.6)$.

\noindent {\bf NSC~J134610$+$030555} A very faint arc is detected in
our WIYN  image (seeing $=0.66$) $\Delta \theta = 19\arcsec$
from the center of this cluster at $z=0.232$.

\noindent {\bf Abell 1835} This cooling-flow cluster with $L_{\rm X} =
19.5\times 10^{44}~{\rm erg~s^{-1}}$\citep{Bohringer00} is a well
known gravitational lens and has a prominent giant arc $31\arcsec$
from the cluster center, as well as several other candidate lensed
features \citep{Schmidt01,Smith05,Sand05}. The giant arc is clearly
identified in our WIYN image in Figure~\ref{fig:cutout1} (seeing =
$1.06\arcsec$), and it received the highest possible score of 6.

\noindent {\bf Abell 1914} The arc candidates in this bright X-ray
cluster $L_{\rm X} = 8.9\times 10^{44}~{\rm erg~s^{-1}}$
\citep{Bohringer00} were first noted in the weak lensing study of
\citet{Dahle02}. \citet{Sand05} identify a faint tangential arc
$\Delta \theta = 28\arcsec$ to the southwest of the BCG, but no
feature at this location was noted by either examiner of our WIYN
image, which had a seeing of $0.62\arcsec$. However, we identified an
arc oriented tangentially around a secondary concentration of galaxies
which is located $\sim 1.5\arcmin$ to the North and $\sim 1\arcmin$ to
the East of the cluster BCG. The arc is at separation $\Delta \theta
34\arcsec$ from the brightest galaxy in the secondary concentration.
\citet{Sand05} do not show the image of Abell 1914, but considering
that the WFPC2 has an effective area of $134\arcsec \times 134\arcsec$
but with an ``L''-shaped-field-of-view, it is conceivable that the arc
which we detected in our WIYN image was not imaged by the WFPC2,
explaining its absence from the \citet{Sand05} compilation.

\noindent {\bf Abell 1942} The arc in this cluster was previously
noted in the $V$-band imaging arc survey of survey of \citet{Smail91}.

\noindent {\bf SDSS~J1446$+$3032} This cluster at $z_{\rm phot} =
0.47$ one the most dramatic examples of strong gravitational lensing
ever discovered. Five extended high surface brightness arcs are
arranged about the center of the cluster in an ellipse with minor axis
$\Delta \theta = 13\arcsec$ and major axis $\Delta \theta
=22\arcsec$. Additional $i$-band and $r$-band images of this cluster
indicate that the arcs are very blue and that they all have similar
colors, but it is not clear that whether they are images of the same
source.

\noindent {\bf Abell 2070} A bright arc is detected $\Delta \theta =
15\arcsec$ from the center of this cluster at $z=0.253$. 

\noindent {\bf SDSS~J1531$+$3414} This cluster at $z=0.335$ is another
poster child gravitational lens, similar to SDSS~J1446$+$3032, with a
series of definite arcs arranged in a ellipse, with minor axis $\Delta
\theta =11\arcsec$ and major axis $\Delta \theta = 15\arcsec$. The
arcs surround the very prominent BCG, but there is a secondary
grouping of galaxies $1\arcmin$ to the southeast. These high sufrace
brightness arcs are very wide and they are thus resolved, even in the
poor seeing of our WIYN $g$-band image (seeing = $1.2\arcsec$).

\noindent {\bf Abell 2141} Abell 2141 is an X-ray cluster with $L_{\rm
  X} = 1.8~{\rm erg~s^{-1}}$ \citep{Bohringer00} at $z=0.159$. We
identify an arc at $\Delta \theta = 22\arcsec$ from the BCG. Four
distinct blobs can be seen in the arc, which suggests a merging quad of
multiple images of the same source.

\noindent {\bf SDSS~J1557$+$2131} This cluster at $z=0.427$ has a
complex morphology with a secondary mass concentration centered on a
bright early-type galaxy $\Delta \theta = 49\arcsec$ to the
southwest. We identify three candidate lensing features in this
cluster. The first is located $\Delta\theta = 58\arcsec$ to the
southwest of the BCG, close to the secondary mass concentration. The
other two, located at $\sim 5\arcsec$ from a double-BCG, have
extremely high surface brightness, and we anticipate that these could be
highly magnified radial arcs. One of the images is more highly
distorted than the other (the northernmost image) which is why it
received a higher score (4 versus 2). It is especially intriguing that
there is a double BCG: if the total mass distribution traces the
orientation of these galaxies then the tangential critical curve would
be elongated along the north-south direction, and the images which we
identify would lie along the minor axis -- which is exactly where one
expects radial arcs to be located \citep{Dalal04,Sand05}.  Both of the
candidate radial arcs were detected in the SDSS photometry, and they
have $(u,g,r,i,z)$ magnitudes of $(22.0,21.9,21.9,21.8,22.6$) and
$(25.3,21.8,20.9,20.5,20.2)$, for the northernmost and southernmost
images, respectively. The differences in color suggest that these are
not the images of the same source, but distinct features. If either of
these images are confirmed to be at high redshift, then they are highly
magnified and will be useful for studies of high-redshift
galaxies similar to the famous lensed image of the high redshift
galaxy MS~1512$-$CB58 \citep{Yee96}. 

\noindent {\bf SDSSJ1602$+$4038} This cluster at $z=0.387$ has a faint
arc $\Delta \theta = 7\arcsec$ from the brightest cluster galaxy. There
are hints of other suggestive features at the sky limit making this a good 
target for deeper imaging. 

\noindent {\bf Abell 2219} The two giant arcs in Abell 2219 were
discovered by \citet{Smail95}. \citet{Smith05} and \citet{Sand05}
summarize the locations and redshifts of other candidate arclets and
multiply imaged features. In our WIYN image (seeing = $0.69\arcsec$),
we identified the giant arc at $z=2.73$ \citep{Smith05} as two
features which received a score of 6 (northwest) and 5 (west),
respectively (see Figure~\ref{fig:cutout1}). The brighter arc at $z=
1.070$ \citep{Smith05} to the southeast received a score of 5.

\noindent{\bf RX~J1652.2$+$4449} This X-ray cluster at $z=0.175$ with
$L_{\rm X} = 2.3\times 10^{44}~{\rm erg~s^{-1}}$ \citep{Bohringer00}
has a bright arc $\Delta \theta = 10\arcsec$ from its BCG. The arc is
resolved in the $0.67\arcsec$ seeing of our WIYN image.

\noindent {\bf Abell 2261} The arc in Abell 2261 to the southwest of
the BCG was first noted in the weak lensing study of
\citet{Dahle02}. This arc is clearly identified in our WIYN 
image (seeing = $0.66\arcsec$) and it received a score of 6. The \citet{Sand05}
summary identifies a second arc to the northeast with a much
smaller length-to-width ratio $L\slash W=7.7$ (compared to $L\slash
W=25.5$ for the southwesterly arc), but this feature was not
identified in our WIYN image.

\noindent {\bf SDSSJ1747$+$5428} A small separation arc 
$\Delta\theta =5\arcsec$ east of the BCG is detected in this 
cluster with photometric redshift $z_{\rm phot} = 0.31$. 

\noindent {\bf SDSSJ2111$-$01115} This cluster at $z_{\rm phot}= 0.68$
was targeted for imaging at WIYN after it was identified as an arc
candidate in the visual inspection search of the SDSS coadded southern
strip imaging \citep{Allam06}. Our WIYN image shows two very
high surface brightness arcs south of the BCG with angular separations
of $\Delta \theta =11\arcsec$ and $\Delta \theta =16\arcsec$, for the
longer and shorter arc, respectively. Additional images in $r$-band and
$i$-band were taken, and both arcs are very blue and have similar colors, 
suggesting that they are images of the same
source. Furthermore, three bright high surface brightness knots can be
identified in both arcs. Based on the additional imaging, we identify 
a counter-image candidate, which is a high surface brightness blue 
feature $\Delta \theta = 28\arcsec$ north of the BCG. If all three of these
features are images of the same source, they are not very well centered on the
BCG, which could indicate that the BCG is offset significantly from the 
center of the dark matter potential well. 

\subsection{Likely Lenses}

\noindent {\bf SDSS~J1040$+$3405} A short arc $\Delta \theta =
26\arcsec$ from the BCG is identified in our UH88 $V$-band image (seeing $=
1.10$) of this cluster at $z=0.445$, which is one of the richest
members of our cluster catalog with $z > 0.4$.

\noindent {\bf SDSS~J1131$+$3554} This cluster at $z_{\rm phot}=0.49$
shows a likely giant arc $\Delta \theta = 12\arcsec$ from the double-BCG
which defines the cluster center. However, the interpretation is
complicated by the superposition with a bright star and the relatively
poor seeing of our WIYN image ($1.10\arcsec$).

\noindent {\bf Abell 1351} The weak gravitational lensing study of
\citet{Dahle02} noted a ``bright red gravitational arc offset from the
cluster light center'' in this bright X-ray cluster at $z=0.322$,
which has $L_{\rm X} = 5.2\times 10^{44}~{\rm
  erg~s^{-1}}$\citep{Bohringer00}. In our UH88 $V$-band image (seeing
= $0.97\arcsec$), we identify an arc-like feature $29\arcsec$ to the
southwest of the cluster center, which received a score of 3.

\noindent {\bf SDSS~J1150$+$0650} The arclet identified south of the 
BCG in this cluster cluster at $z=0.301$ is notable for its large
separation $\Delta\theta = 25\arcsec$. 

\noindent {\bf Abell 1550} Despite the poor seeing ($1.30\arcsec$) of
our WIYN image of Abell 1550, we identify a faint arc $\Delta\theta=
16\arcsec$ northeast of the BCG. This cluster is at $z=0.262$ and has
$L_{\rm X} = 3.4\times 10^{44}~{\rm erg~s^{-1}}$ \citep{Bohringer00}.

\noindent {\bf SDSS~J1258$+$4702} The WIYN image of this cluster at
$z_{\rm phot}=0.32$ has the worst seeing ($1.39\arcsec$) of any the
candidate cluster lenses we identified, but we still identified a
small separation arc $\Delta \theta = 3\arcsec$ from the BCG. 

\noindent {\bf Abell 1758} In our WIYN image (seeing $= 0.85\arcsec$),
Abell 1758 appears as a double cluster, which is possibly in the
process of merging. Both subclumps are concentrated around a bright
early type galaxy, and the total X-ray luminosity is $L_{\rm X} =
11.2\times 10^{44}~{\rm
  erg~s^{-1}}$\citep{Bohringer00}. \citet{Dahle02} noted the presence
of a blue arc associated with the northwest mass clump, which received
a score of 3 in our arc search.

\noindent {\bf SDSS~J1406$+$3945} A short arc $\Delta\theta =22\arcsec$ 
south of the BCG is identified in our UH88 $V$-band image of 
this cluster at $z=0.427$. 

\noindent {\bf SDSS~J1414$+$2703} This cluster at $z_{\rm phot}=0.46$
shows a likely arc $\Delta \theta = 14\arcsec$ northeast of the
double-BCG, which defines the cluster center. There are many other features
at or near the sky limit which are highly suggestive of strong lensing, making
this an excellent candidate for deeper multicolor imaging. 

\noindent {\bf SDSS~J1527$+$0652} This cluster at $z_{\rm phot}=0.40$
is one of the richest clusters in our catalog. An extremely high
surface brightness blue arc candidate is identified $\Delta \theta
=17\arcsec$ south of the BCG. This arc candidate was detected in the
SDSS imaging and has magnitudes $(u,g,r,i,z) =
(24.8,20.9,20.7,20.9,21.1)$, making it an easy target for
spectroscopic follow-up. This arc was classified as likely rather than
definite: its high surface brightness led us to wonder whether it
might be a foreground edge-on spiral galaxy.

\noindent {\bf SDSS~J1537$+$3926} We detect a faint arc which is
optically resolved and very near the sky limit of our WIYN image in 
this cluster at $z=0.444$. The arc is located $\Delta \theta  = 17\arcsec$ 
from a bright cluster galaxy located in a secondary concentration of
galaxies $1.5^\prime$ north of the cluster center. 

\noindent {\bf Abell 2136} Abell 2136 at $z=0.229$ has an X-ray
luminosity of $L_{\rm X} = 1.7\times 10^{44}~{\rm
  erg~s^{-1}}$\citep{Bohringer00}. Two very likely arcs are identified
to the northwest and southeast, at separations of $\Delta \theta =
14\arcsec$ and $\Delta \theta = 21\arcsec$, respectively.

\noindent {\bf GHO~155414$+$40530} An ambiguous arclike feature overlaps
a cluster galaxy at $\Delta \theta = 40\arcsec$ northeast of the BCG
in this cluster at $z=0.393$.

\noindent {\bf RXC~J1749.3$+$4245} This cluster at $z=0.230$ has an
X-ray luminosity $L_{\rm X} = 1.9\times 10^{44}~{\rm
  erg~s^{-1}}$\citep{Bohringer00}.  A short arc is identified $\Delta
\theta =22\arcsec$ west of the cluster center.

\subsection{Possible Lenses}

\noindent {\bf SDSS~J1008$+$4529} This pair of arcs was missed by one
of the examiners because the lensing features are centered on a galaxy
which is several arcminutes from the galaxy identified as the BCG by
our cluster finding algorithm. However, this $z=0.479$ system has a
complex morphology and hence the location of the cluster center is
somewhat ambiguous. Two arc-like features are identified at separations $\Delta
\theta = 8\arcsec$ (eastern) and $\Delta \theta = 10\arcsec$
(northwest) of the cluster galaxy.

\noindent {\bf SDSS~J1240$+$4250} This cluster at $z=0.406$ is
$2.9\arcmin$ southwest of the known galaxy cluster NSC
J124039$+$425228 with published redshift $z=0.3955$ \citep{NSC}. The
similarity of the redshifts suggests a centering error in the NSC
cluster catalog. This cluster has several very suggestive high surface
brightness arc candidates. The orientation/geometry of these features
does not obviously suggest a lensing interpretation, although the 
cluster has a complex morphology.

\noindent {\bf SDSSJ1513$+$0525} A very high surface brightness arc
candidate is identified $\Delta \theta = 38\arcsec$ to the north of
the BCG of this cluster at $z_{\rm phot} = 0.49$. Because of the the
large angular separation and the fact that the feature is so bright
$(u,g,r,i,z) = (22.8,21.5,20.2,19.7,19.4)$, we were not certain of the
lensing interpretation and classified this system as a possible lens.


\subsection{Known Lenses which were Missed}

\noindent {\bf Abell 773} \citet{Smith05} identified several arcs and
many candidate multiple image systems in HST Wide Field Planetary
Camera 2 (WFPC2) imaging (F$702W$ filter) of this bright X-ray cluster
($L_{\rm X} = 7.0\times 10^{44}~{\rm erg~s^{-1}}$). Our WIYN image of this cluster
has a seeing of $0.79\arcsec$ and the counterparts to the majority of
the lensed features from Smith et al. can be identified when comparing
the HST and WIYN images side by side. But only those features with the
largest length-to-width ratios could be convincingly identified as
arcs or arclets given our image quality. Five features were given a
score of 1 when inspected by one examiner; whereas the second examiner
found only the most conspicuous of these and gave it a score of
1. Thus the highest score of any feature was 2, with a score of 1 from
each examiner, which is not sufficient to make it into our lens samples.

\noindent {\bf Z3146} \citet{Sand05} identified a single faint
(integrated magnitude F$606 = 23.55$) tangential arc at
$\Delta\theta=26\arcsec$ to the southeast of the center of this well
known cooling flow cluster.  An arclet like feature can be seen at
roughly this location in our WIYN image (seeing $= 1.09\arcsec$),
however it is near the sky limit and does not appear very tangentially
distorted. Neither of the image examiners identified this feature.

\noindent {\bf Abell 1763} \citet{Smith05} list several candidate
multiple-image systems in this cluster and \citet{Sand05} identified
one of them as a tangential arc with length-to-width ratio $L/W=8.7$
and magnitude F$702=24.87$. However, this feature is extremely faint and
the lensing interpretation is not very compelling, even in the HST
image. No candidate lensing features were identified in our WIYN image
(seeing = $0.64\arcsec$) of this cluster.

\section{Summary and Conclusions}
\label{sec:conclusion}

We have conducted a systematic deep imaging survey for lensed arcs and
arclets in a large, homogeneous, optically selected sample of distant
clusters. A total of 240 clusters were imaged (195 from WIYN and 45
from UH88), of which 141 had sub-arcsecond image quality. Our survey
uncovered 22 definite lensing clusters (6 previously known), 14
likely lenses (2 previously known), and 9 possible lensing
clusters. It is probable that $\gtrsim 50\%$ of our likely and
possible samples are indeed cluster lenses, so that the number of new
cluster lenses discovered here is $\sim 30$. This is a substantial
contribution to the total number of lensing clusters known, which was
previously $\sim 50$.  Among these new systems are some of the most
dramatic examples of gravitational lensing ever discovered. Clusters
such as SDSS~J1115$+$5319, Abell 1703, SDSS~J1446$+$3032, and
SDSS~J1531$+$3414, which have multiple bright arcs at large angular
separation, will likely become `poster-child' gravitational lenses
similar to Abell 1689 and CL0024$+$1654. The selection function of our
cluster lens survey will be presented in the future
\citep{Gladders06b}. Considering that our imaging campaign is
conducted from the ground with a median seeing of $0.90\arcsec$, our
completeness for identifying lensed images is surely lower than a
survey from space, such as the WFPC2 archive search of
\citet{Sand05}. However, 8 out of the 11 previously known lenses which
were imaged in our survey were recovered, and most of the arcs/arclets
which we missed had magnitudes close to or below our surface
brightness limit.

There are several advantages to our ground-based survey strategy.
First and foremost, the cosmological volume accessible to our survey
is $\sim 2~{\rm h^{-1} Gpc^{3}}$ ($0.1 \lesssim z \lesssim 0.6$; $\sim
8000~{\rm deg}^2$), which is more than an order of magnitude larger
than any previous search. This enables us to characterize the lensing
cross-sections of a more generic population of galaxy clusters;
whereas most previous arc searches have been dominated by nearby
bright X-ray clusters \citep[i.e.][]{LeFevre94,Luppino99, Sand05}.
Second, inspection of Figures~\ref{fig:cutout1}-\ref{fig:cutout3}
reveals that our survey primarily discovers high-surface brightness
arcs which are resolved in ground-based image quality. Indeed, several
of the arcs published here have $g\lesssim 21$, making them among the
brightest giant arcs ever discovered.  Spectroscopic observations of
the bright arcs in our sample, which are required both to confirm the
lensing hypothesis and to construct accurate mass models, will be much
easier than for arcs discovered in HST images, which tend to be $\gtrsim
2$ magnitudes fainter.

The cosmological applications of this sample of $\sim 30$ new lensing
clusters are many.  The sample size and redshift coverage ($z \lesssim
0.6$; see Figure~\ref{fig:zdist}) will allow us to characterize the
strong lensing properties of a statistical sample of low redshift
clusters, which is important in light of the recent suggestion by
\citet{Gladders03} that most cluster strong lensing occurs at
high redshift.  Our low redshift lensing survey is thus a necessary
complement to ongoing high redshift giant arc searches such as the RCS
and RCS2 \citep{GY00, Gladders03, GY05}, as well as the Massive Cluster
Survey (MACS) \citep{MACS01}. The abundance and statistics of giant
arcs will be addressed in a future paper \citep{Gladders06b}. HST
imaging of the most dramatic lenses in our sample will likely uncover
many more arcs and candidate multiple images. Using these image
positions, detailed models can measure the distribution of dark matter
in each cluster \citep[e.g.][]{Tyson98,Smith01,Sand04, Broad05a} and
around the cluster galaxies \citep{Priya04}, and stronger constraints
can be obtained if the strongly lensed image positions are combined
with larger scale weak lensing measurements
\citep{Kneib03,Gavazzi03,Smith05,Broad05b,Priya06,Dalal06}. This will
provide much improved statistics on the properties of the dark matter
distribution in clusters such as ellipticity, radial profile,
concentrations, and substructure, enabling statistical comparisons to
theoretical predictions \citep{Hennawi06,Priya06}. These comparisons
are of particular interest considering that the large ($c_{\rm NFW} >
14$) concentrations recently measured in cluster lenses
\citep{Gavazzi03,Kneib03,Broad05a} are significantly higher than the
expectations from CDM \citep{Oguri05,Hennawi06,Dalal06}.

Our survey for clusters lenses is only half complete, and we expect to
find a comparable number of lenses as the $\sim 30$ published here upon its
completion. By conducting the largest giant arc search to date, we
will help transform strong lensing by galaxy clusters from the study
of a handful of rare systems, into a powerful statistical probe of the
formation of structure in the Universe.


\begin{figure*}
  \vskip -0.1cm
  \centerline{\epsfig{file=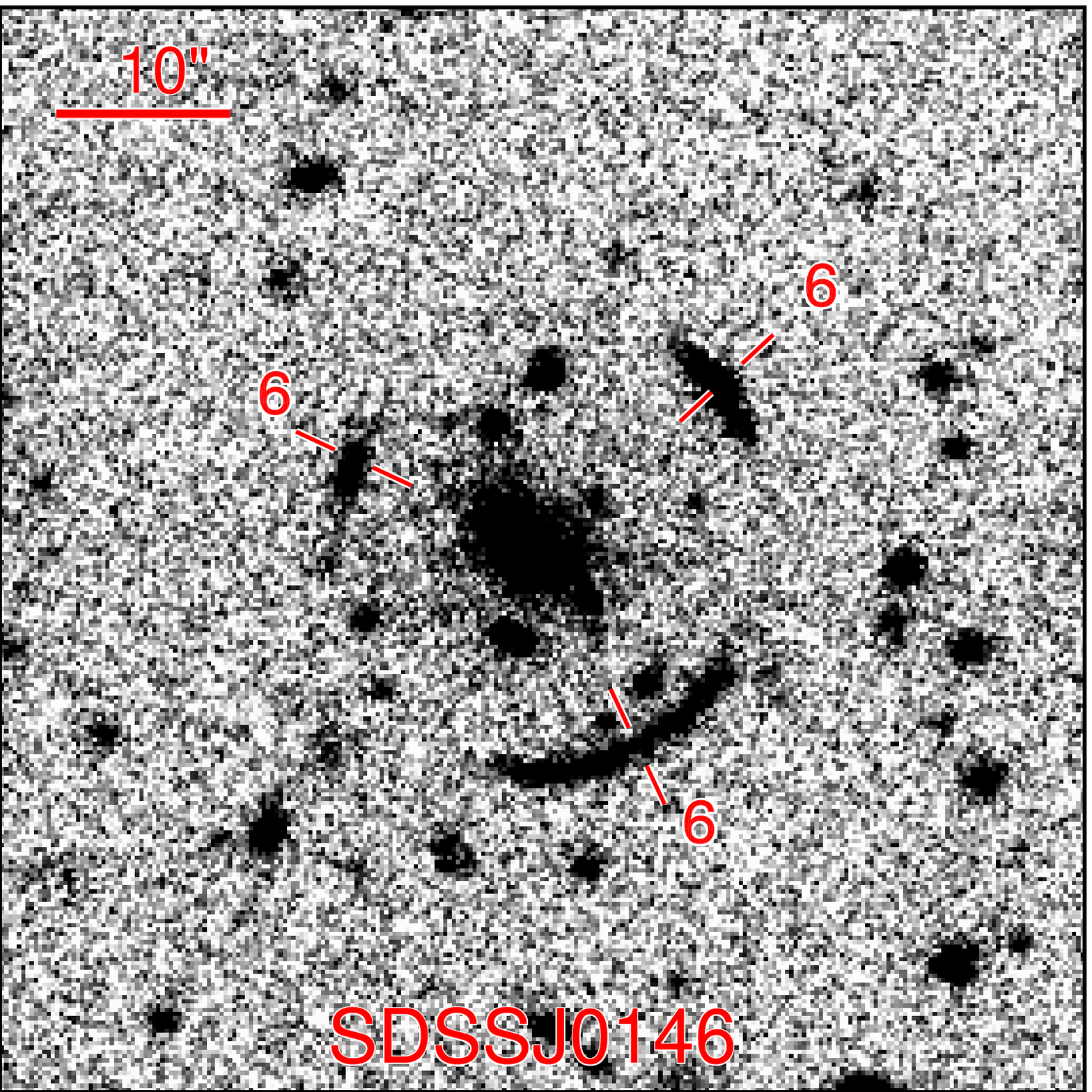,width=0.45\textwidth}
    \epsfig{file=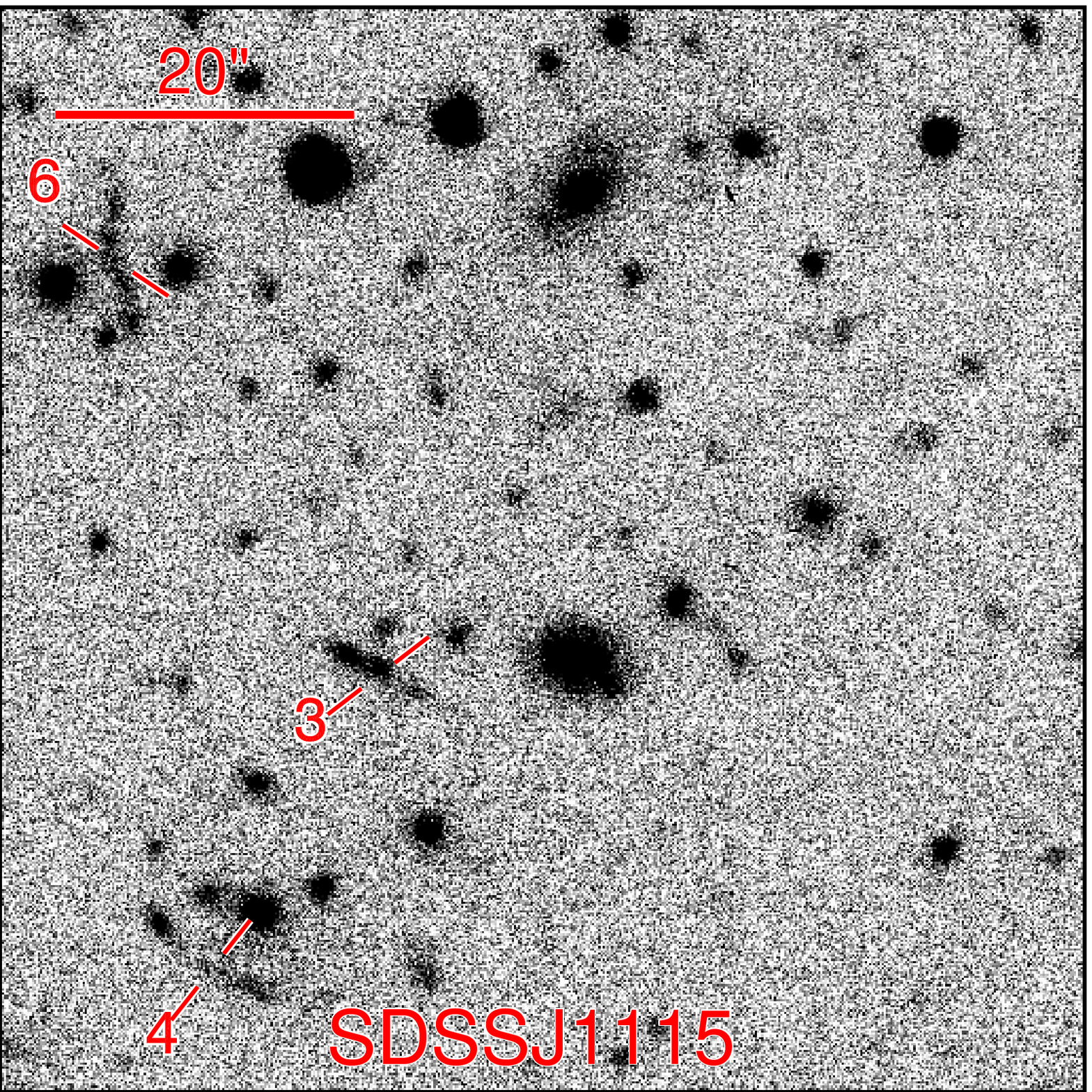,width=0.45\textwidth}}
  \vskip 0.05cm
  \centerline{\epsfig{file=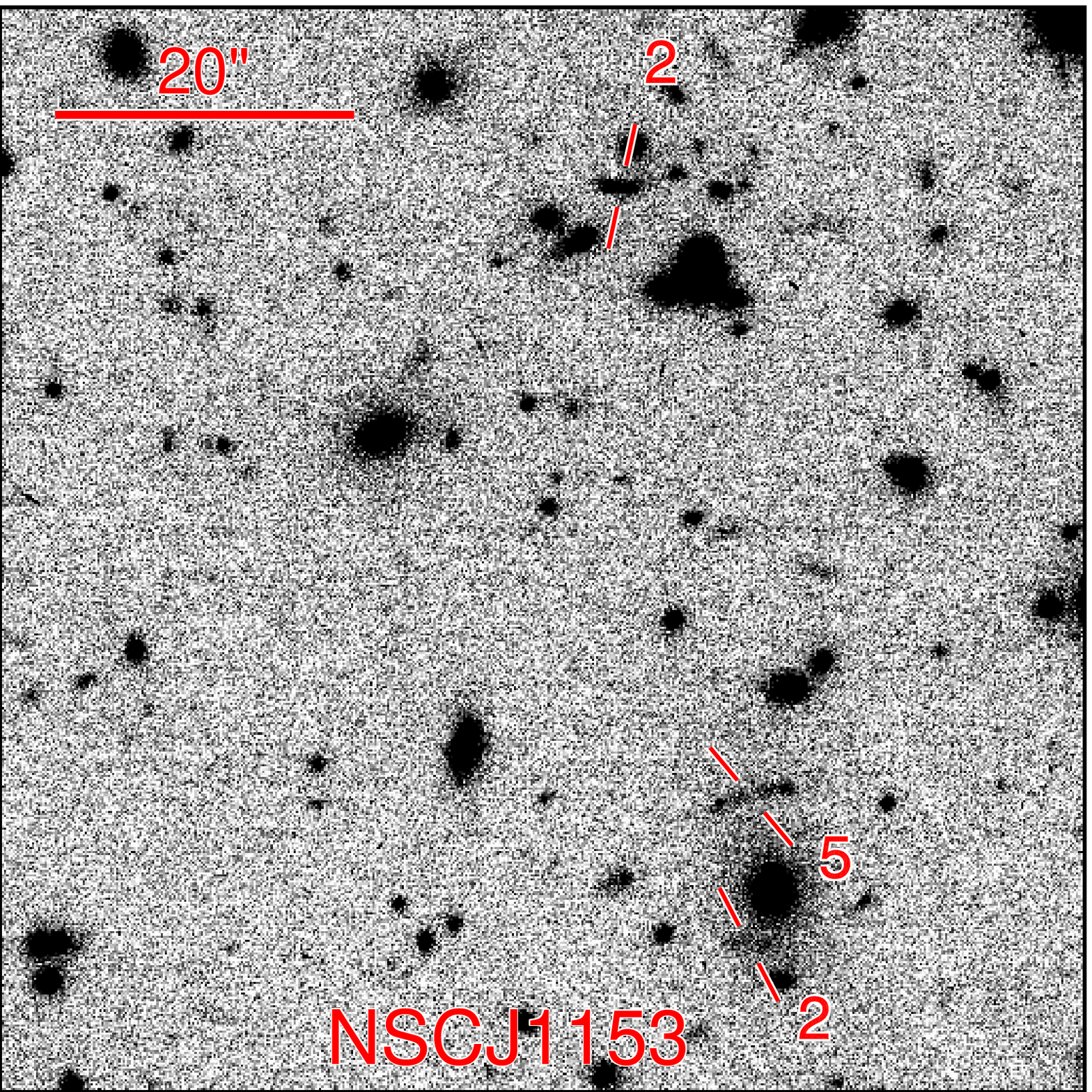,width=0.45\textwidth}
    \epsfig{file=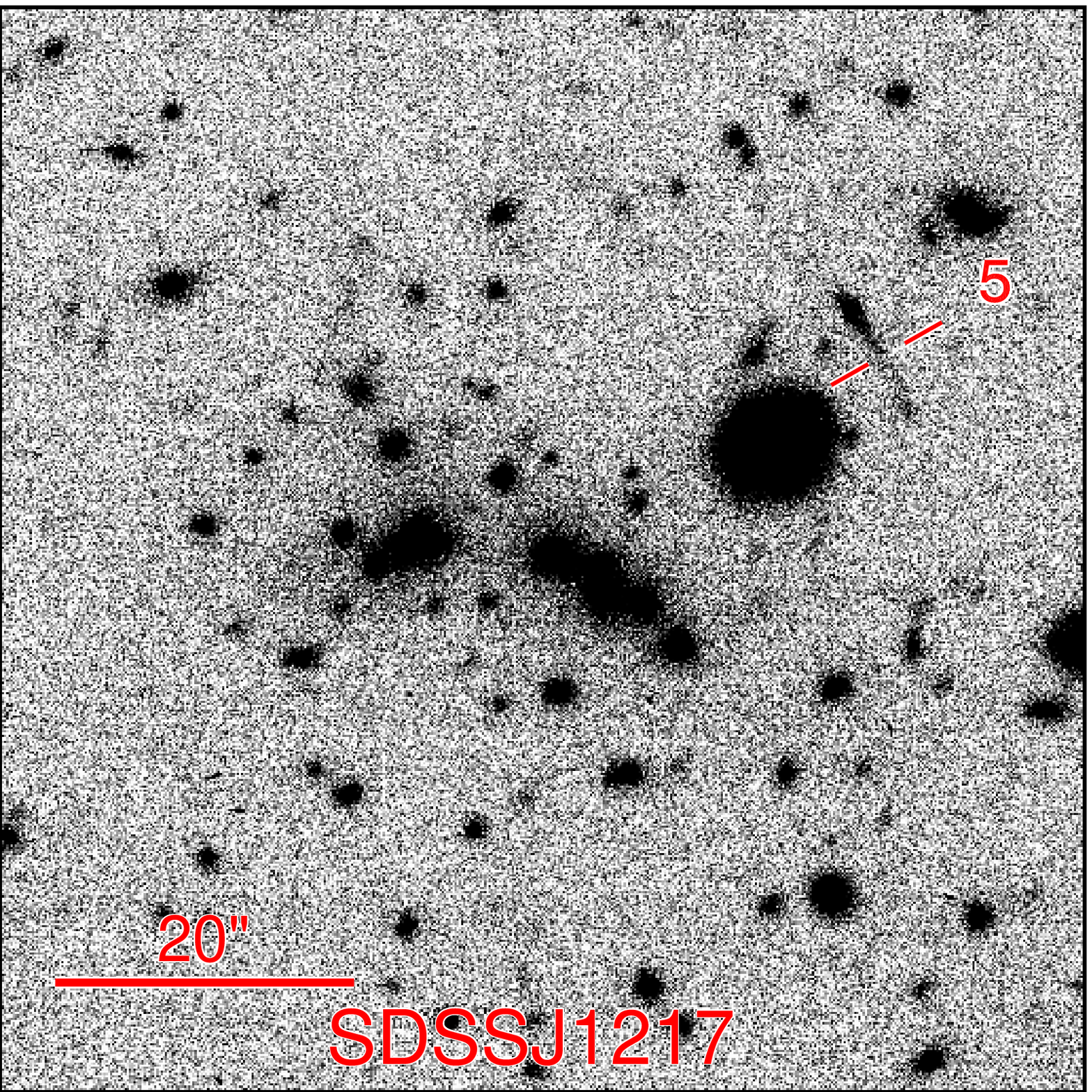,width=0.45\textwidth}}
  \vskip 0.05cm
  \centerline{\epsfig{file=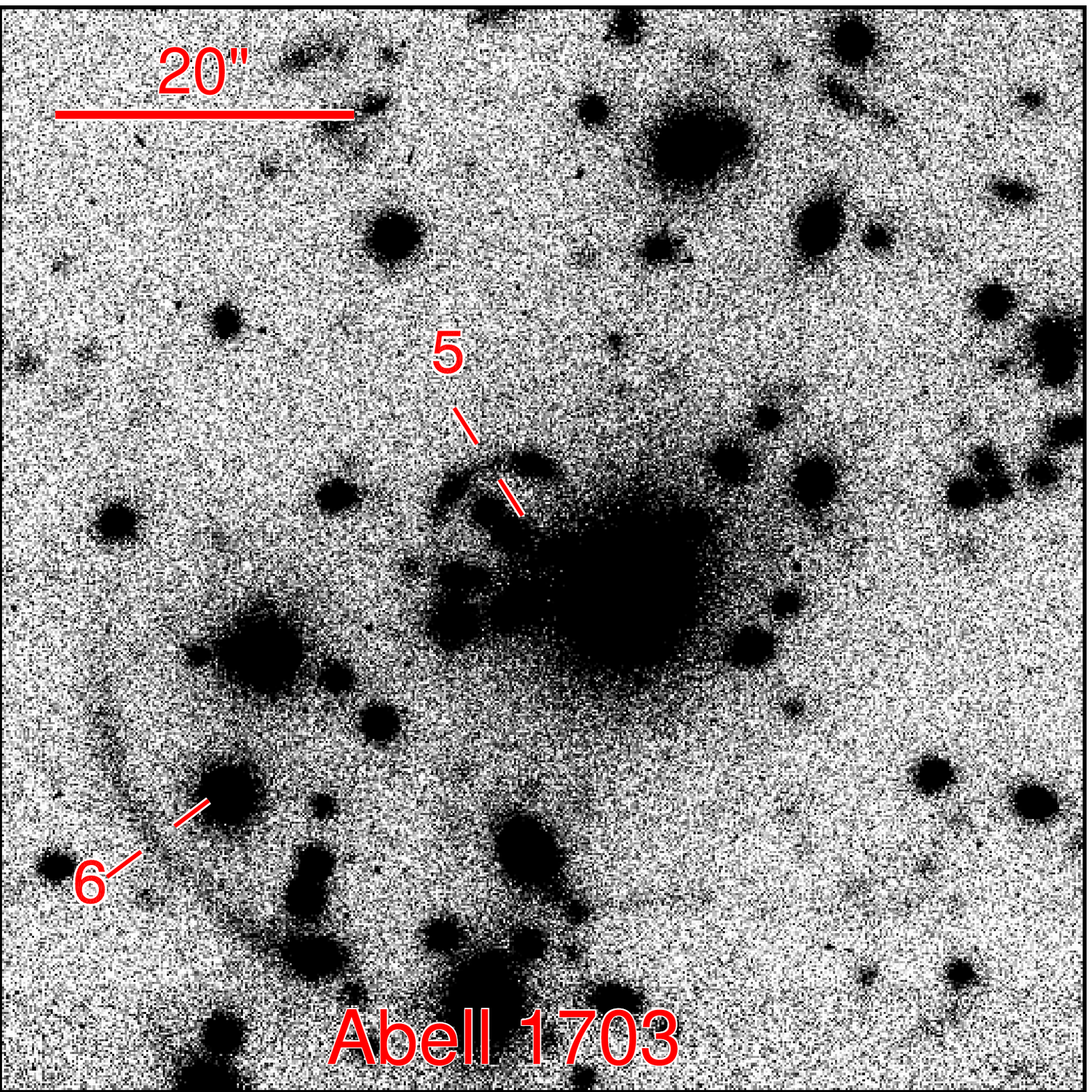,width=0.45\textwidth}
    \epsfig{file=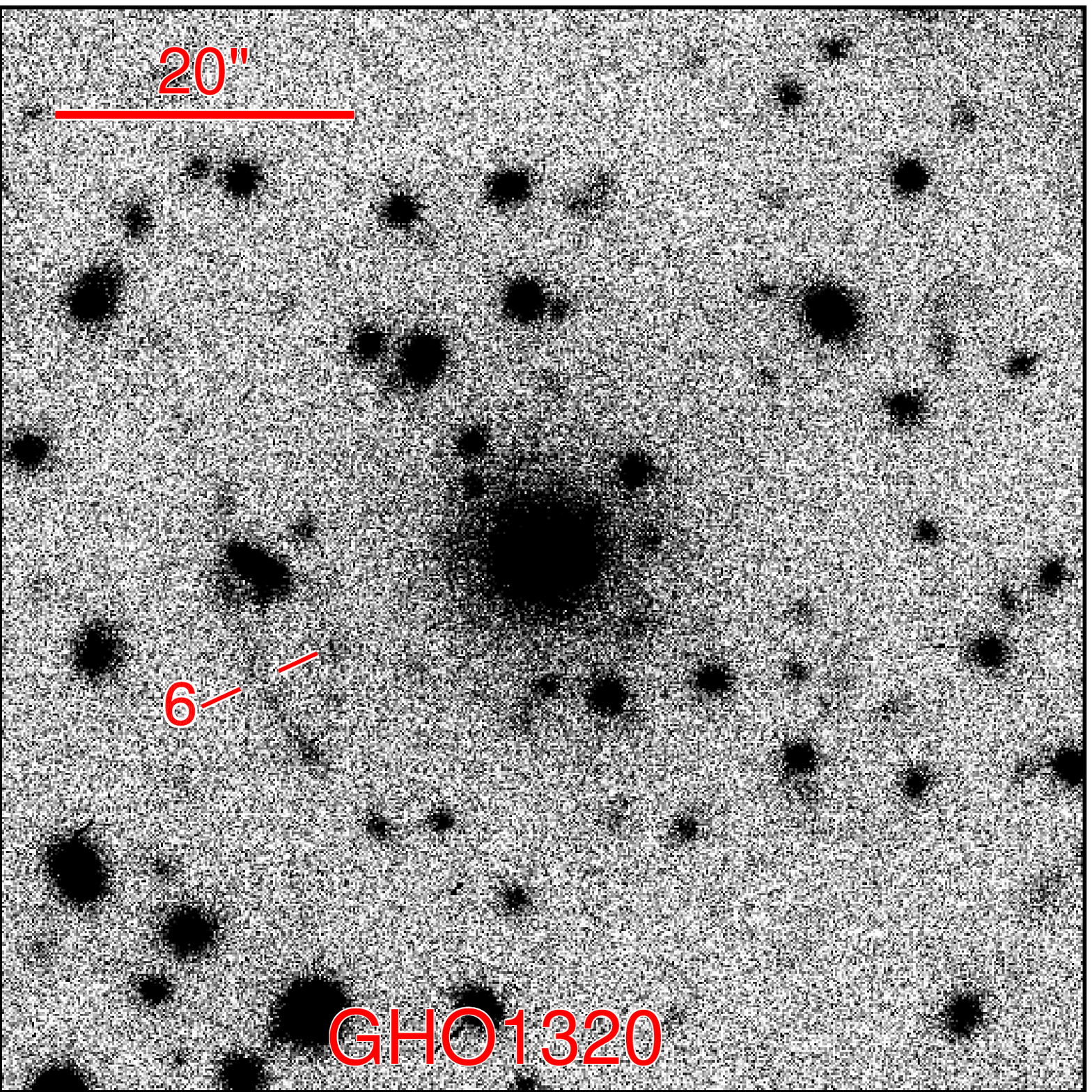,width=0.45\textwidth}}
  \caption{ Cutout images of the definite lensing clusters listed in
    Table~\ref{table:cutout1}. Strong lensing features are indicated
    by lines and a number which is the score of the feature. North is
    up and east is to the left. The scalebar indicates the size of the
    cutouts and varies from $10-30\arcsec$, depending on the angular
    separation of the arcs. The name of the cluster is indicated at
    the bottom of each cutout.\label{fig:cutout1}}
\end{figure*}
\addtocounter{figure}{-1}
\begin{figure*}
  \centerline{\epsfig{file=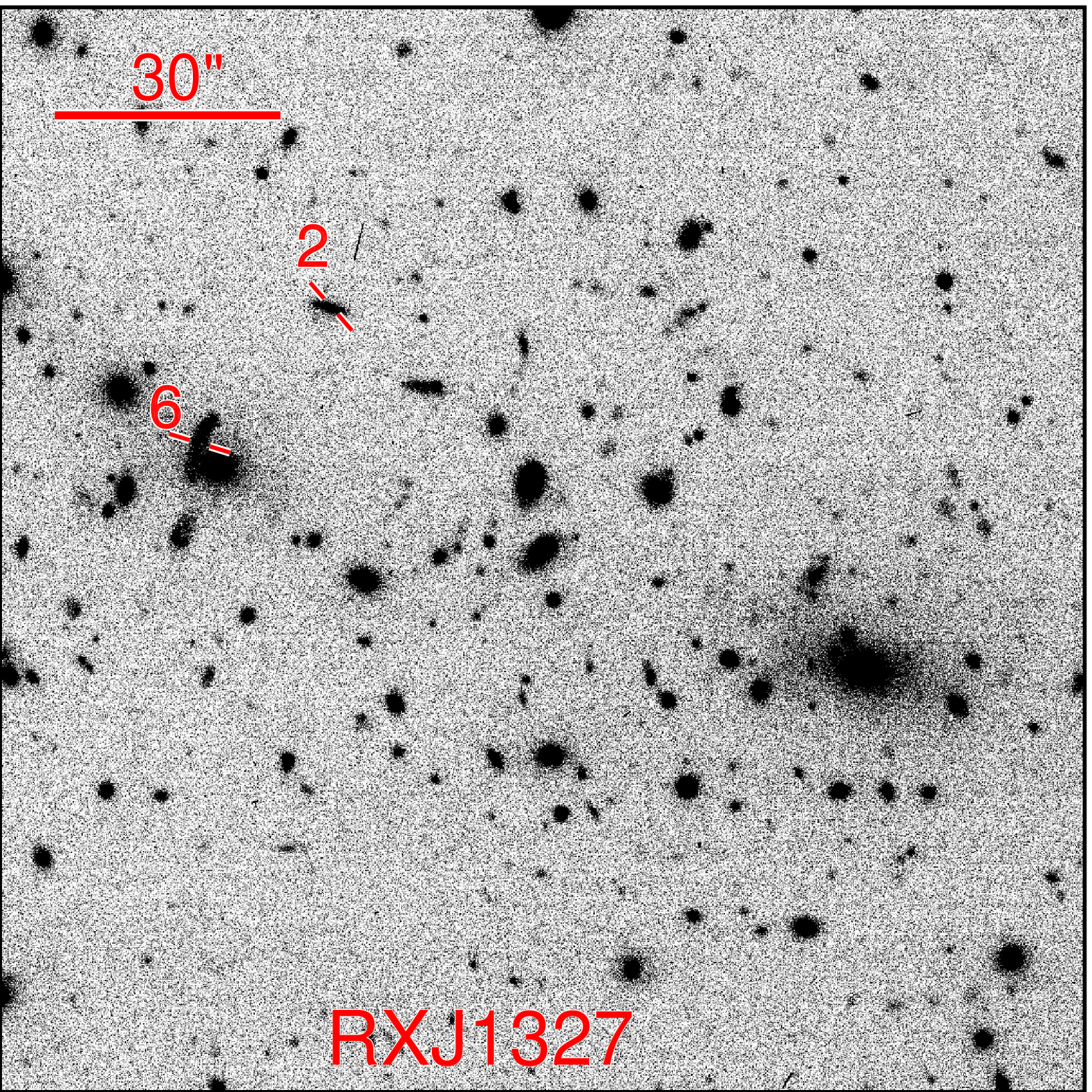,width=0.45\textwidth}
    \epsfig{file=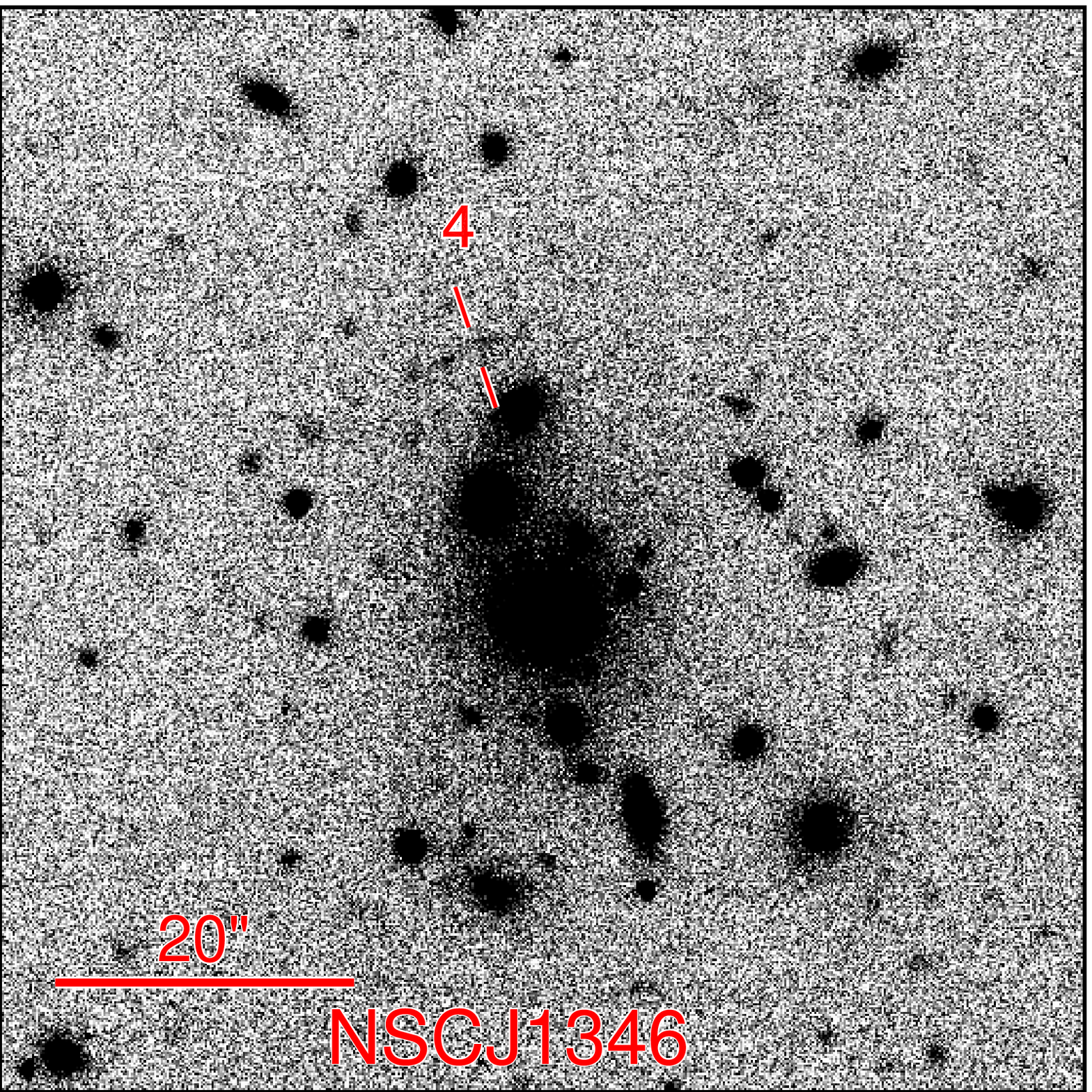,width=0.45\textwidth}}
  \vskip 0.05cm
  \centerline{\epsfig{file=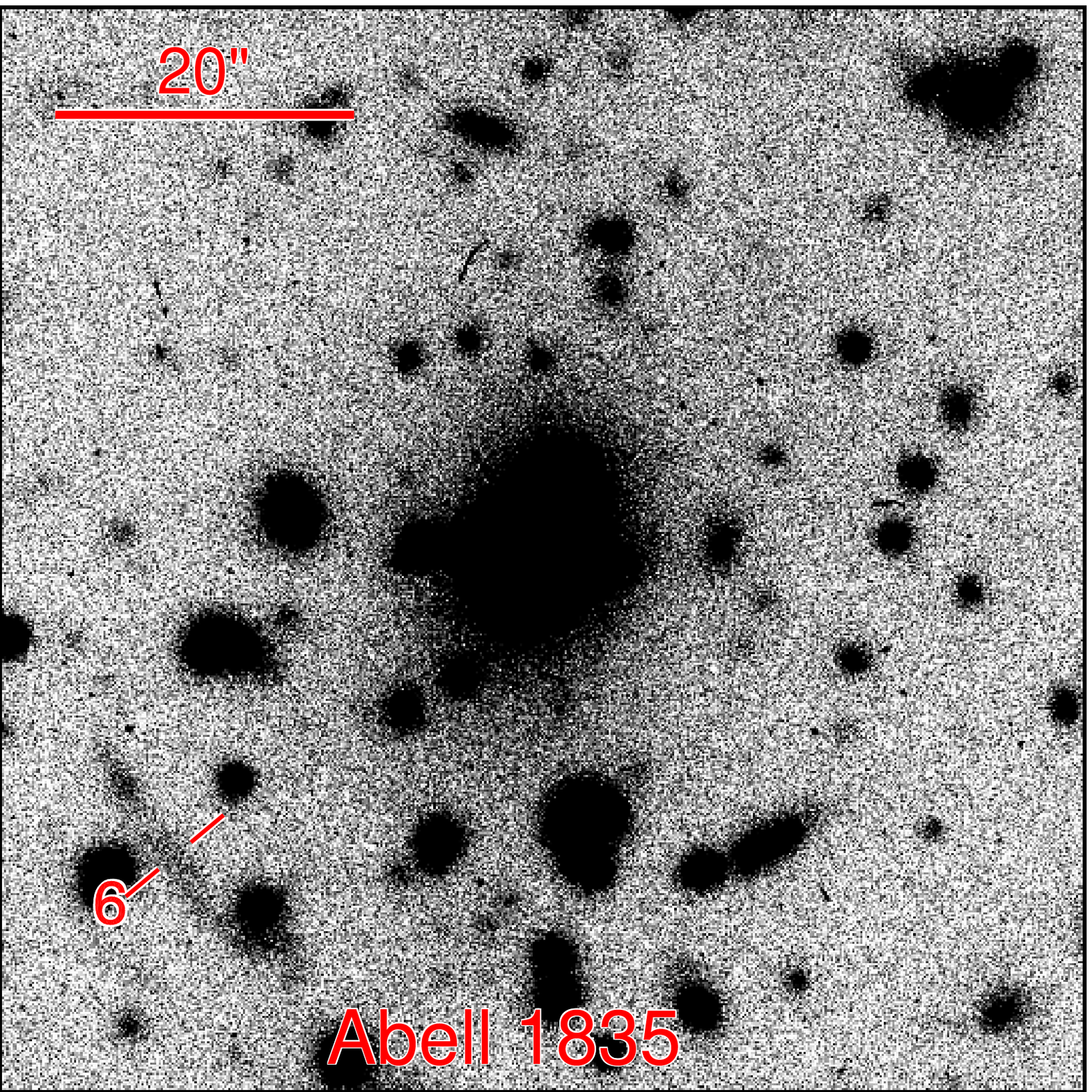,width=0.45\textwidth}
    \epsfig{file=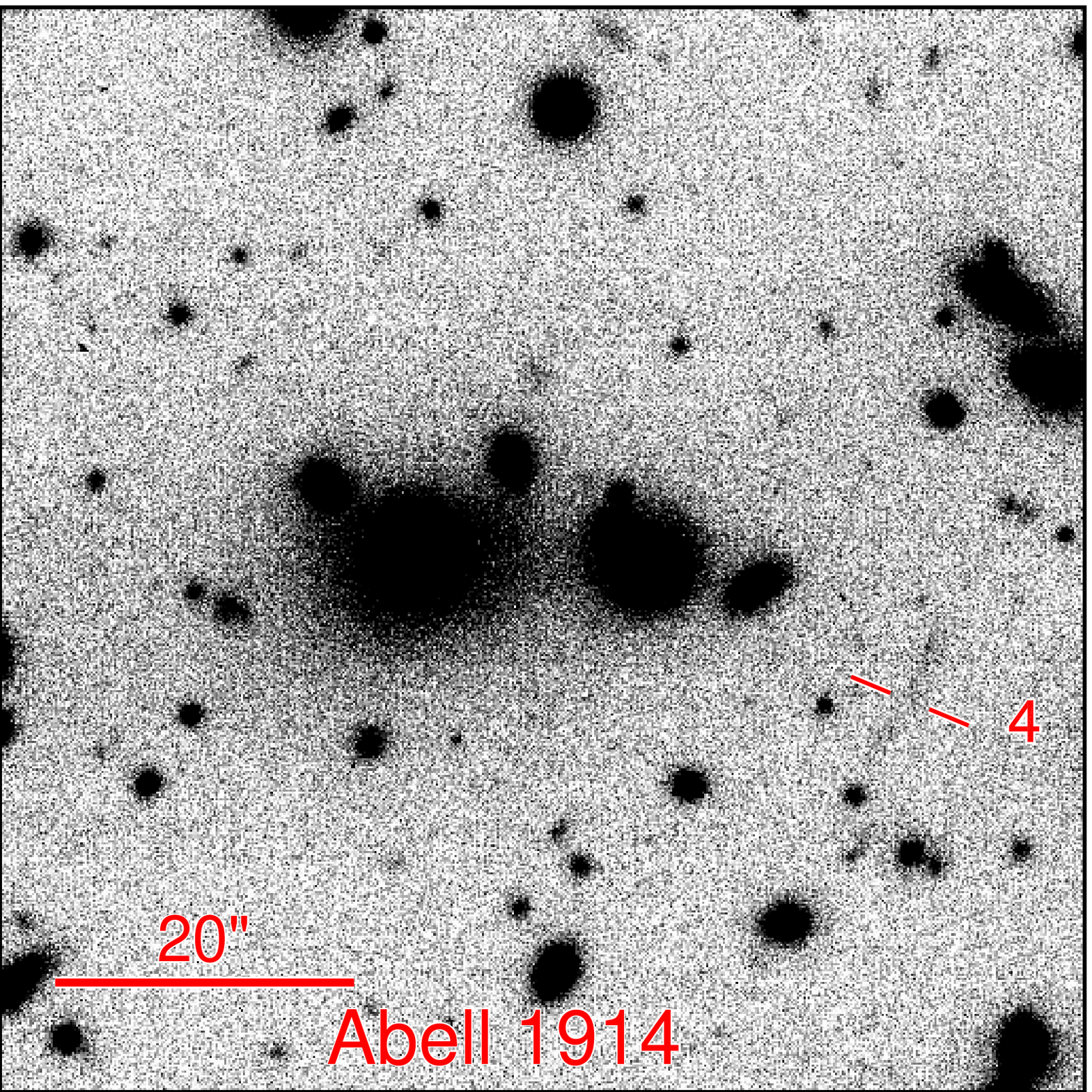,width=0.45\textwidth}}
  \vskip 0.05cm
  \centerline{\epsfig{file=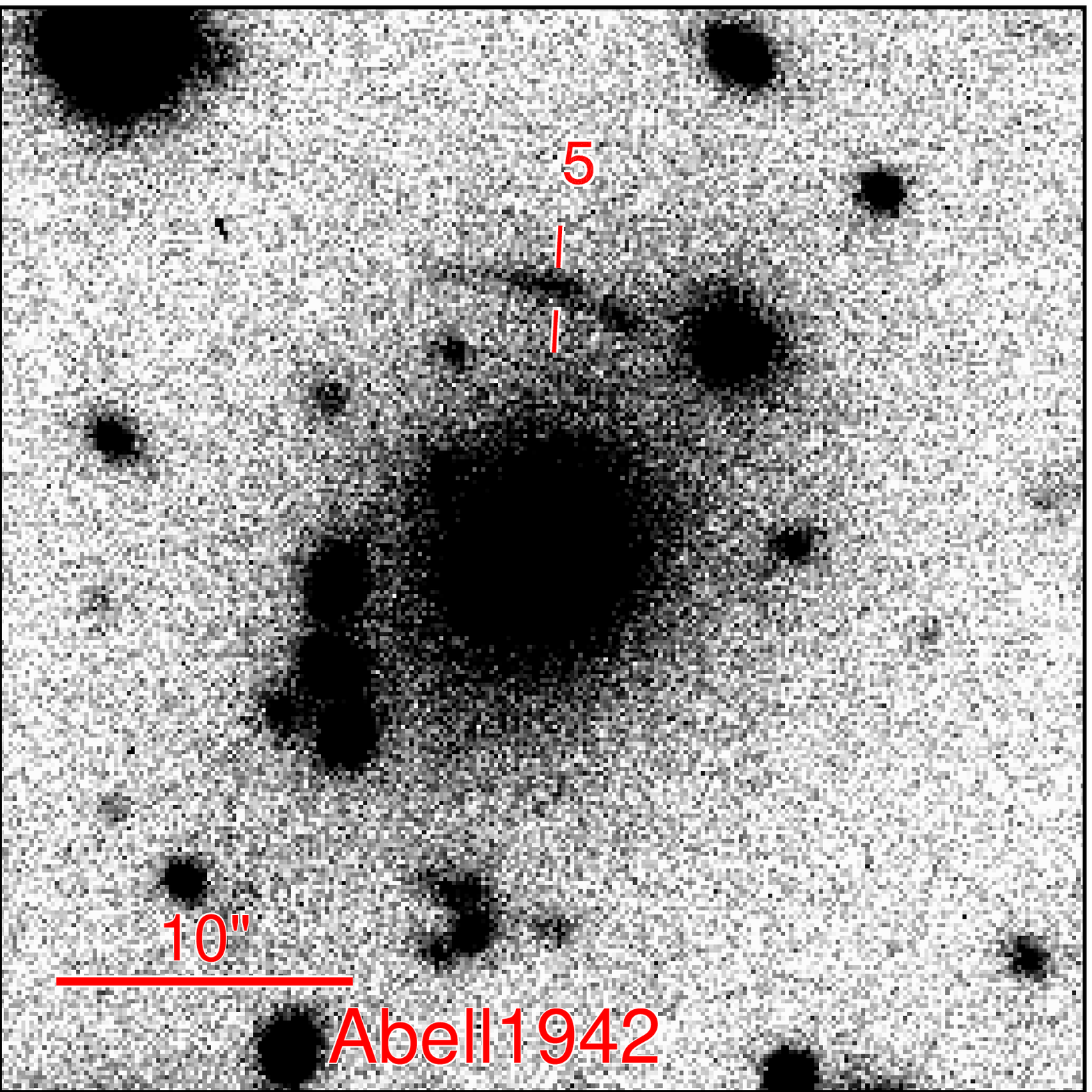,width=0.45\textwidth}
    \epsfig{file=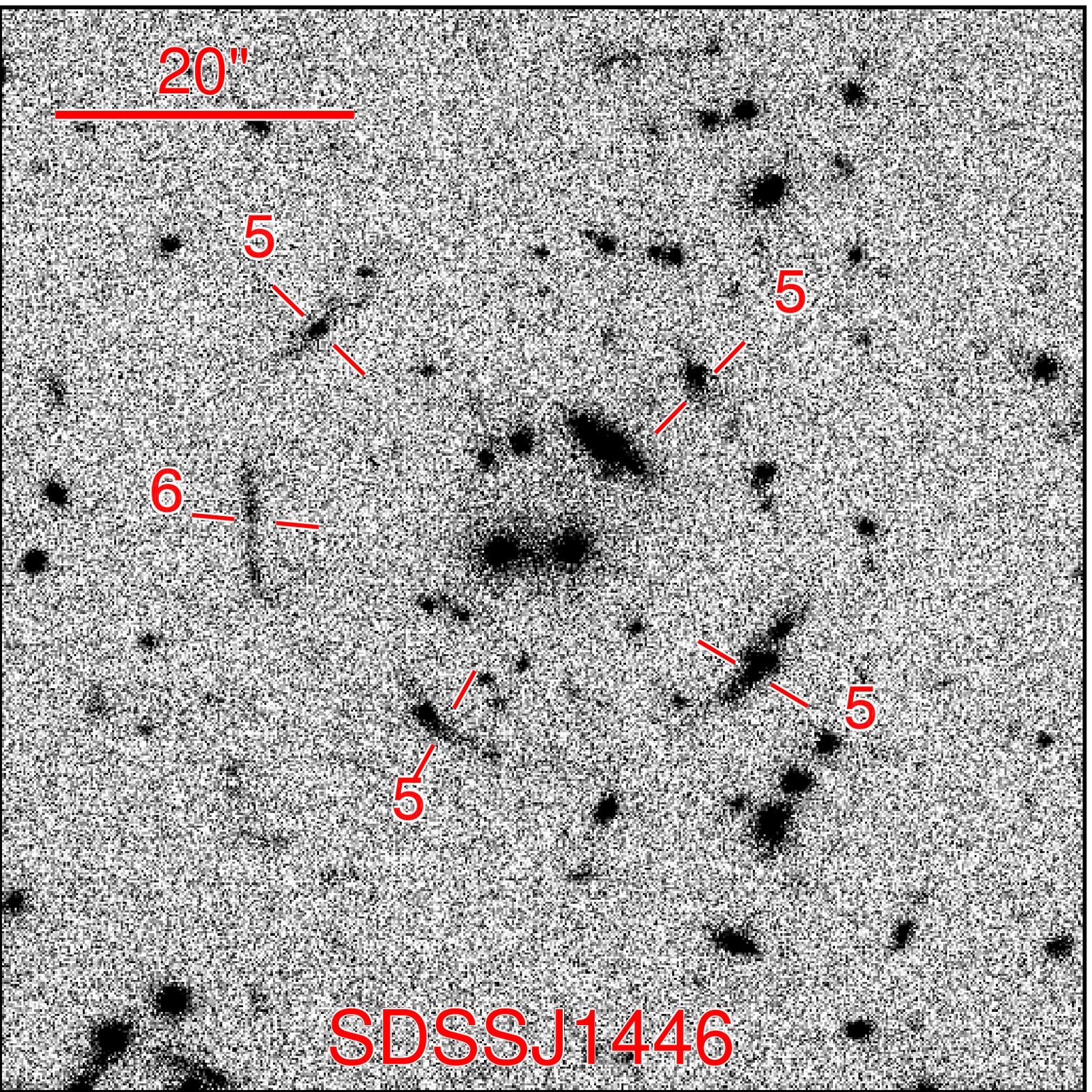,width=0.45\textwidth}}
  \caption{ continued.}
\end{figure*}
\addtocounter{figure}{-1}
\begin{figure*}
  \vskip -0.1cm
  \centerline{\epsfig{file=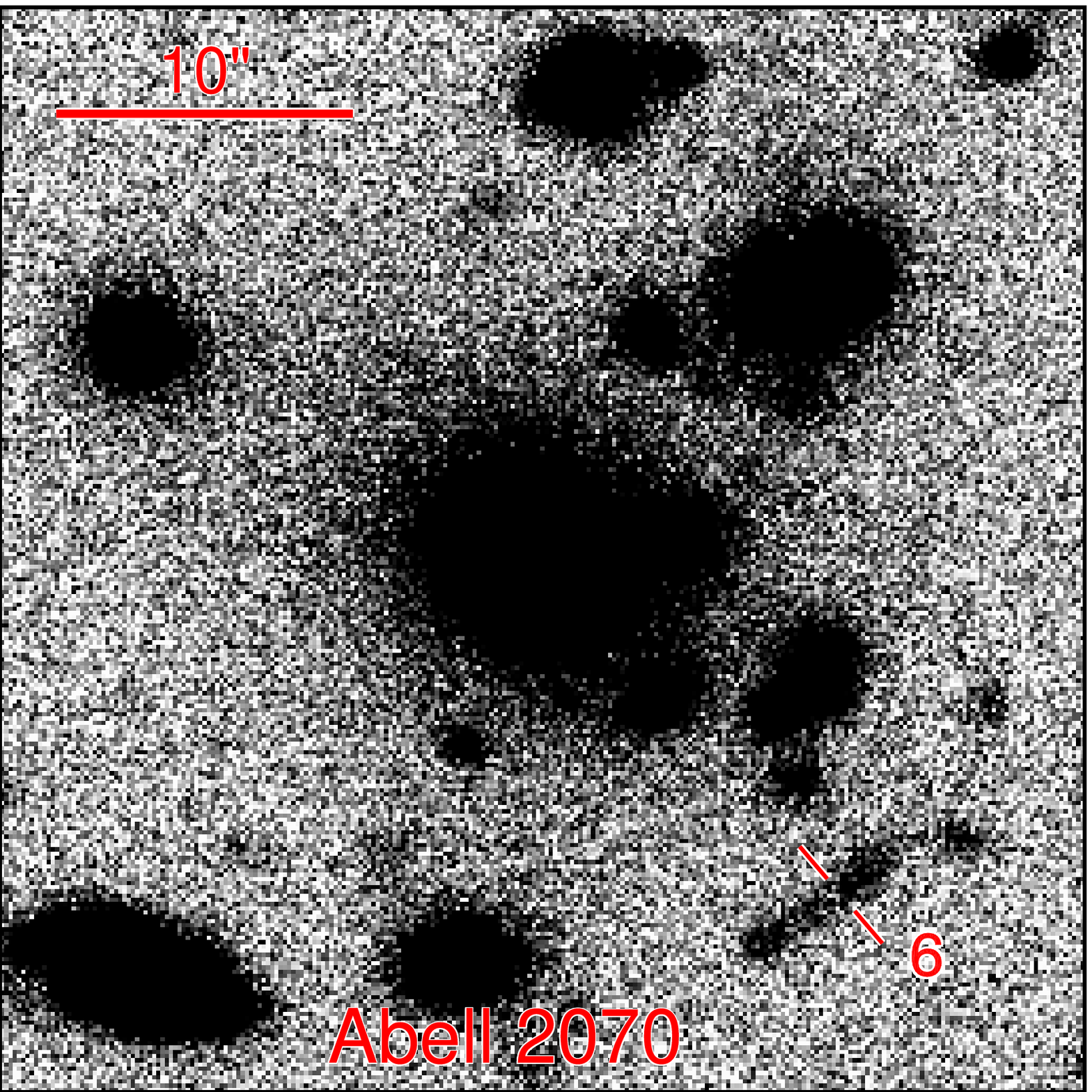,width=0.45\textwidth}
    \epsfig{file=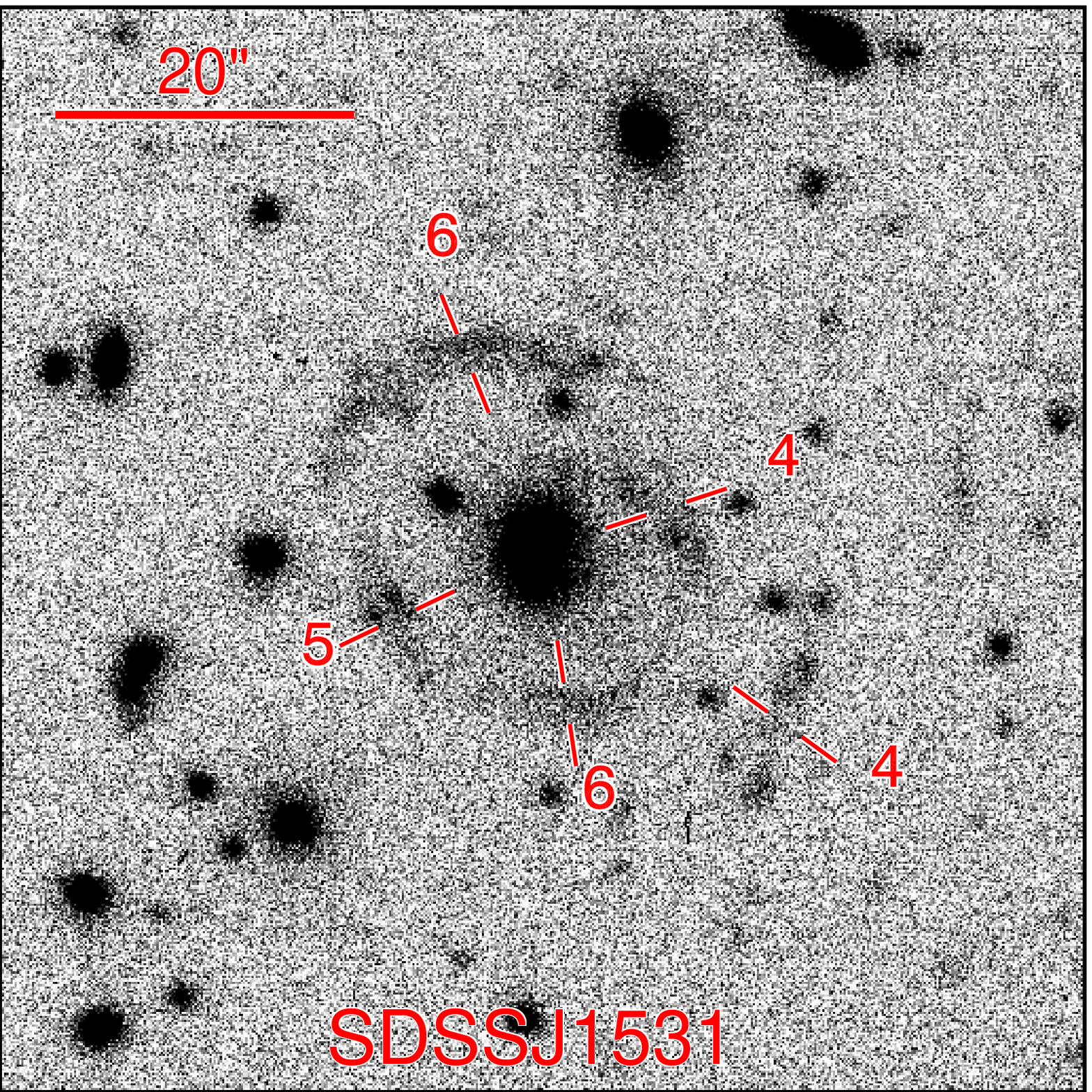,width=0.45\textwidth}}
  \vskip 0.05cm
  \centerline{\epsfig{file=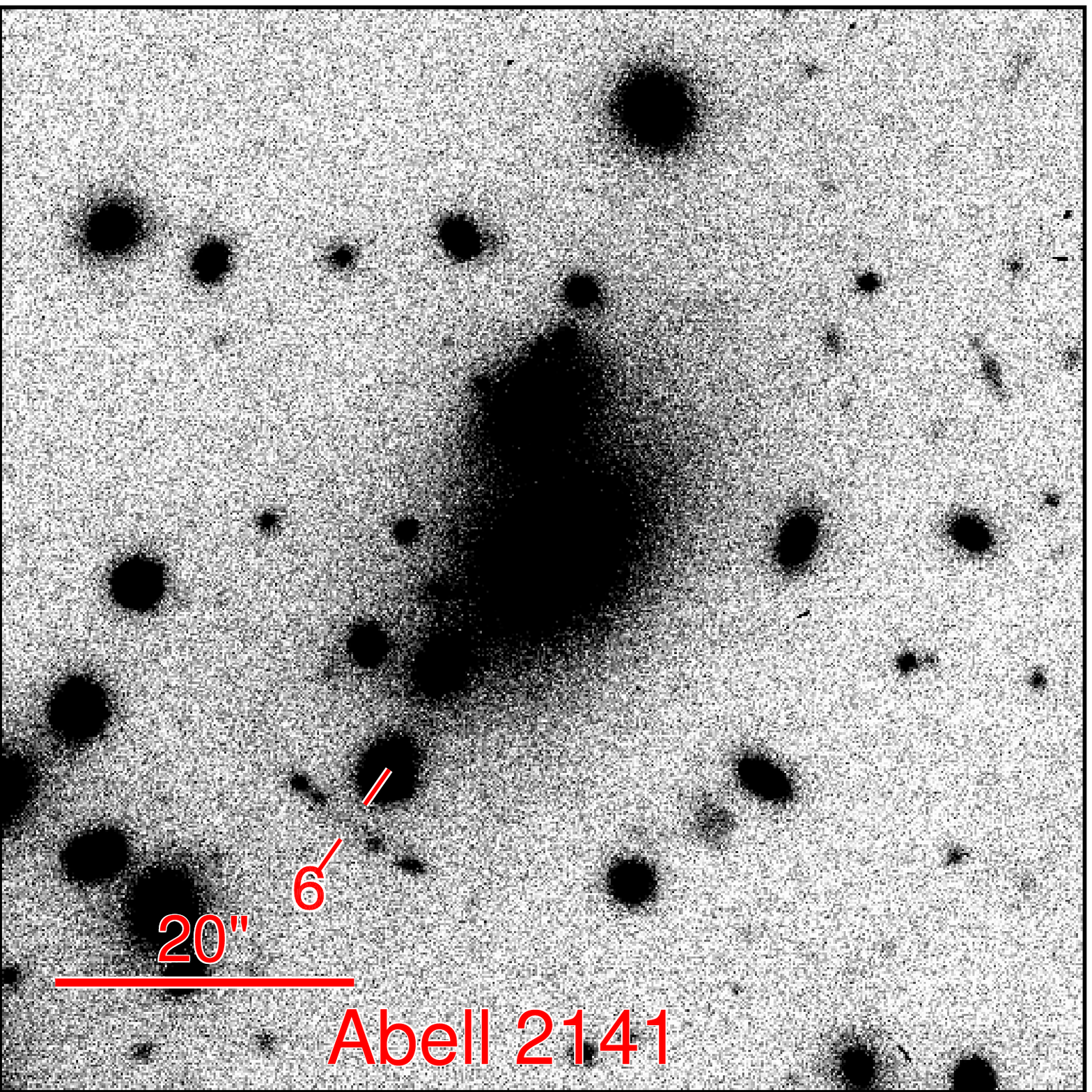,width=0.45\textwidth}
    \epsfig{file=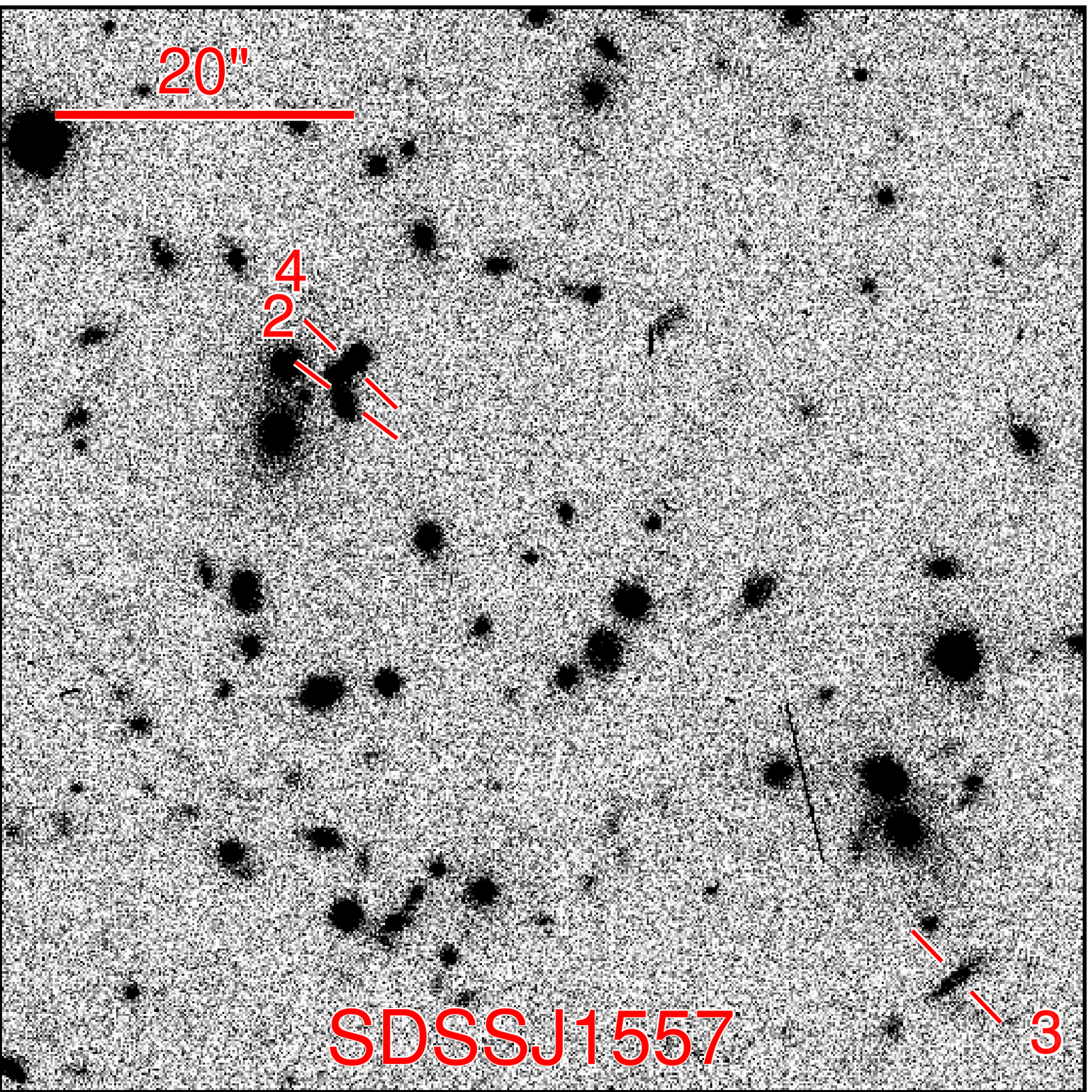,width=0.45\textwidth}}
  \vskip 0.05cm
  \centerline{\epsfig{file=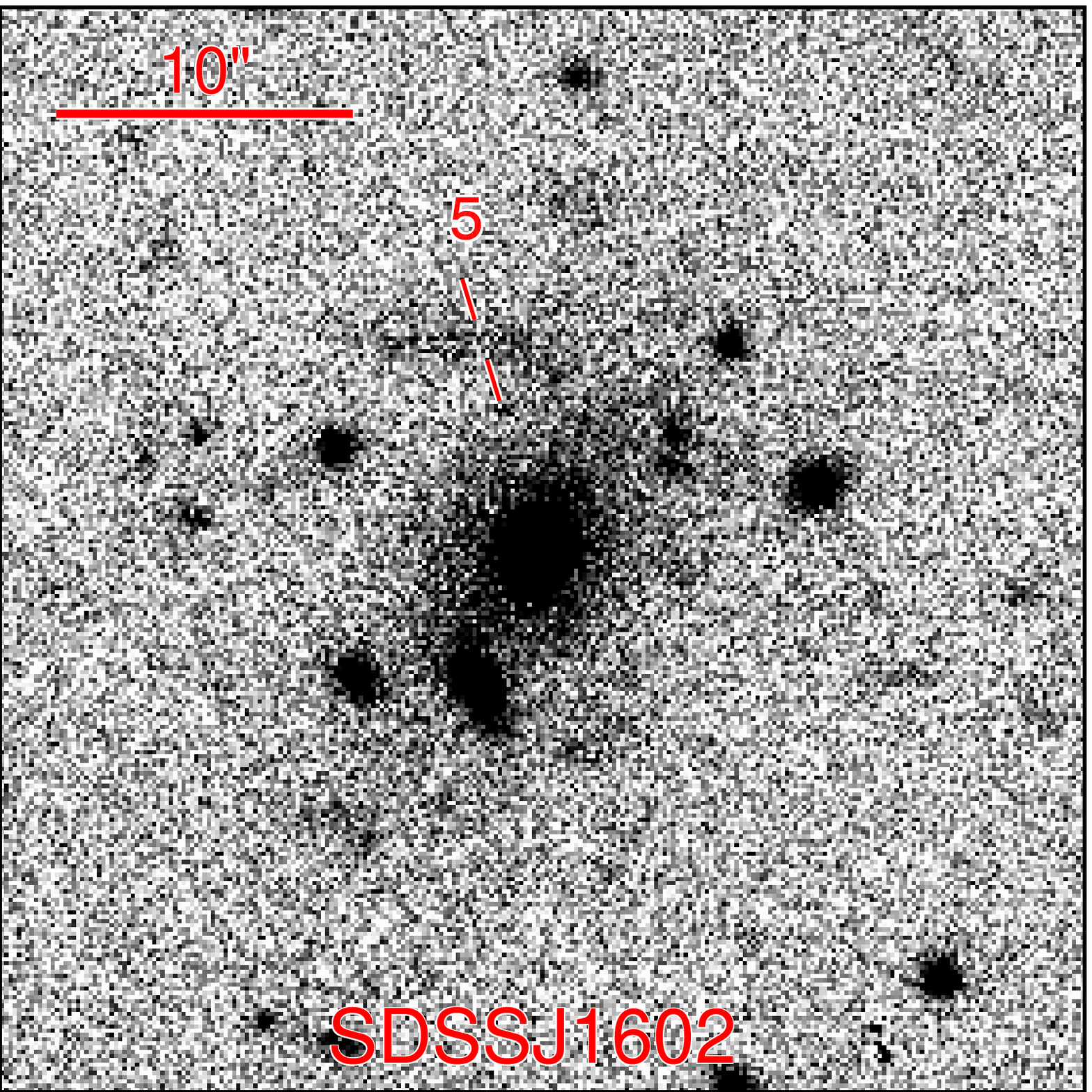,width=0.45\textwidth}
    \epsfig{file=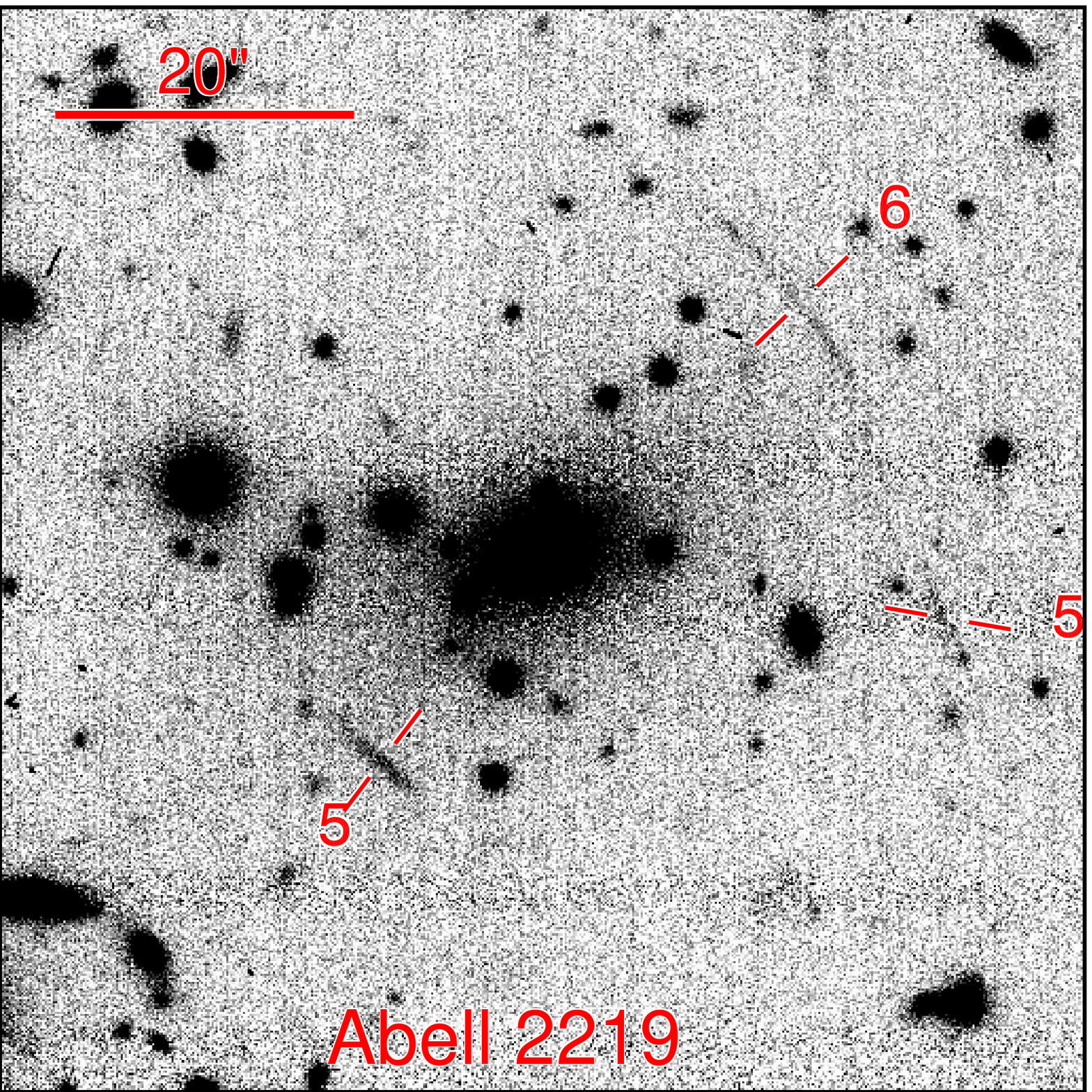,width=0.45\textwidth}}
  \caption{ continued.}
\end{figure*}
\addtocounter{figure}{-1}
\begin{figure*}
  \vskip -0.1cm
  \centerline{\epsfig{file=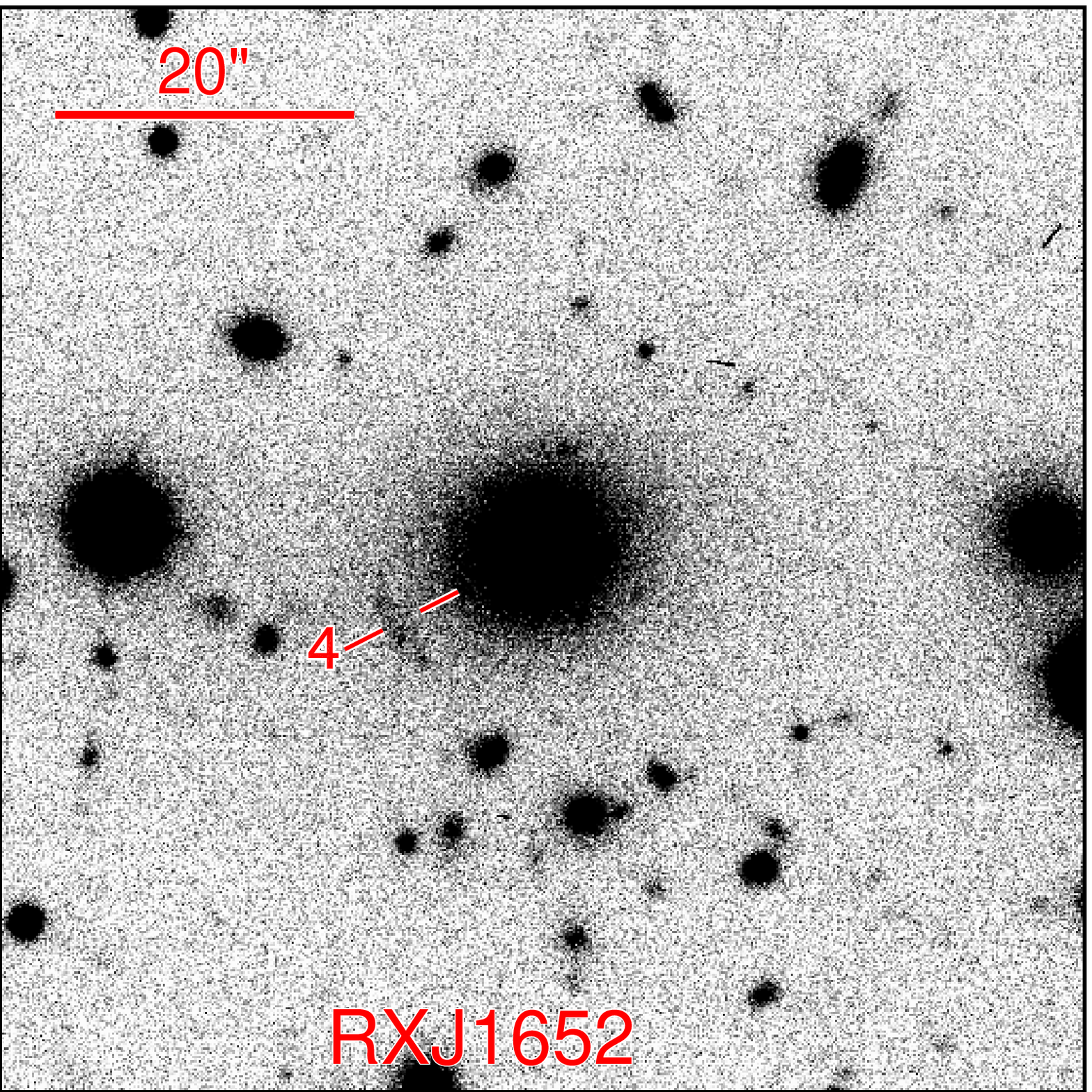,width=0.45\textwidth}
    \epsfig{file=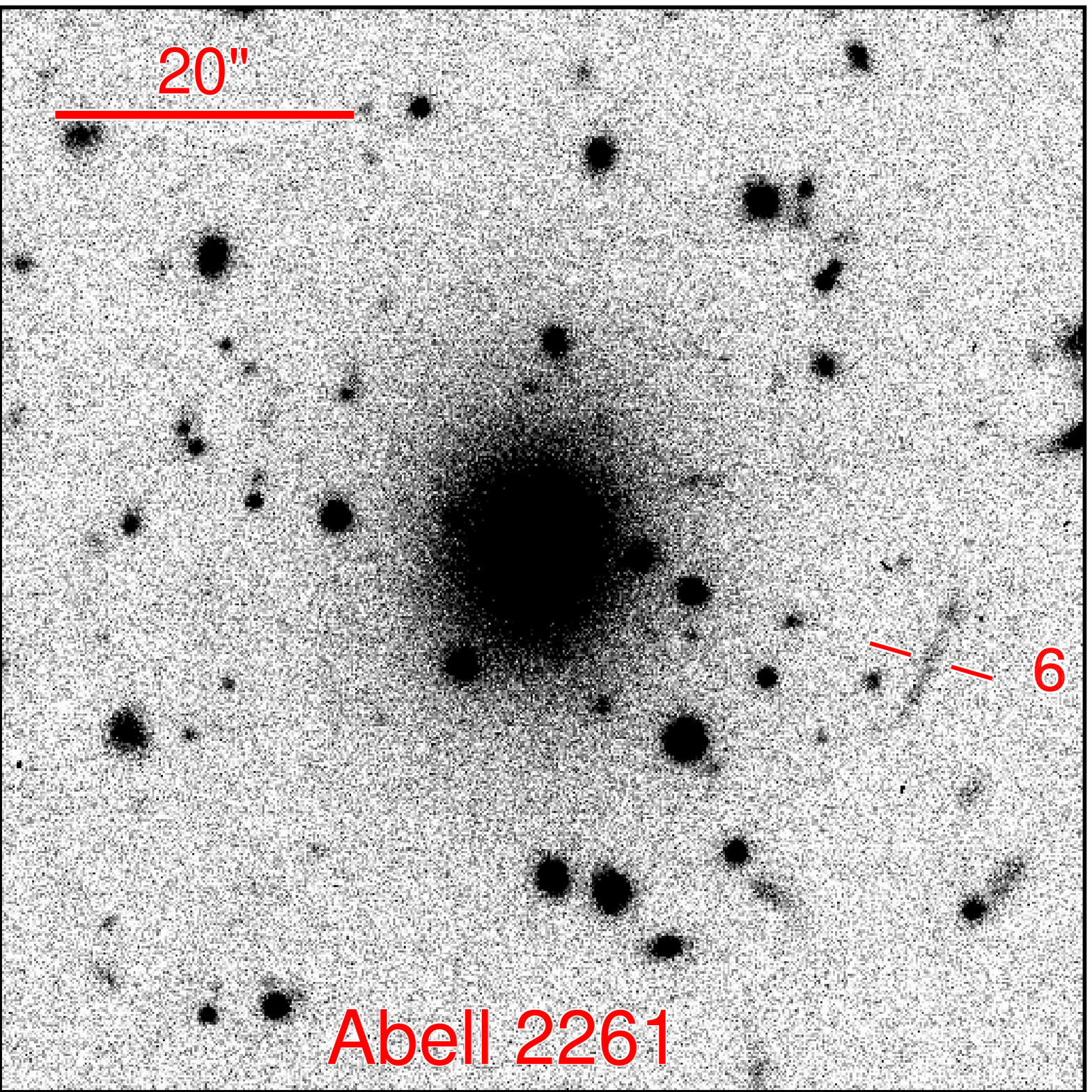,width=0.45\textwidth}}  
  \vskip 0.05cm
  \centerline{\epsfig{file=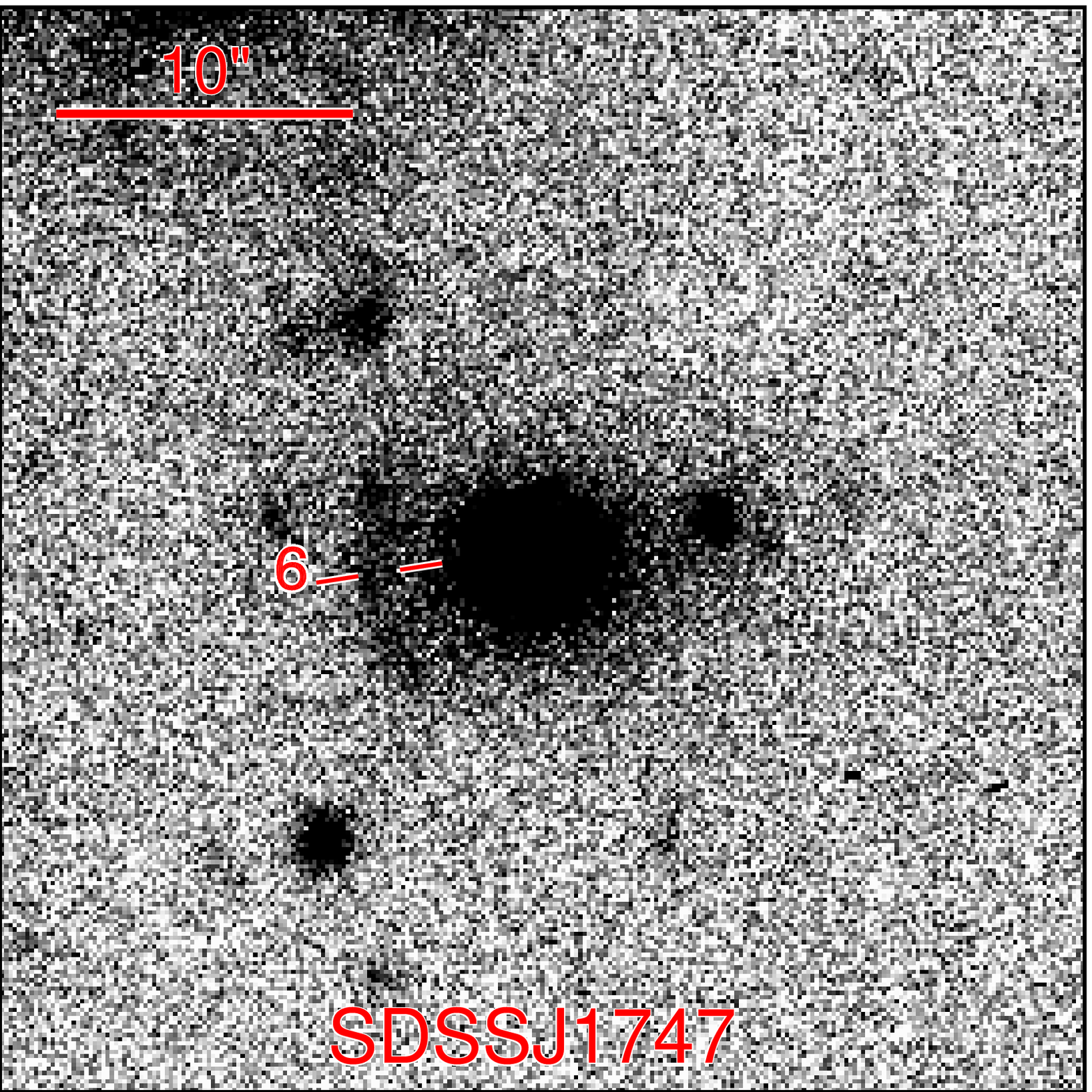,width=0.45\textwidth}
    \epsfig{file=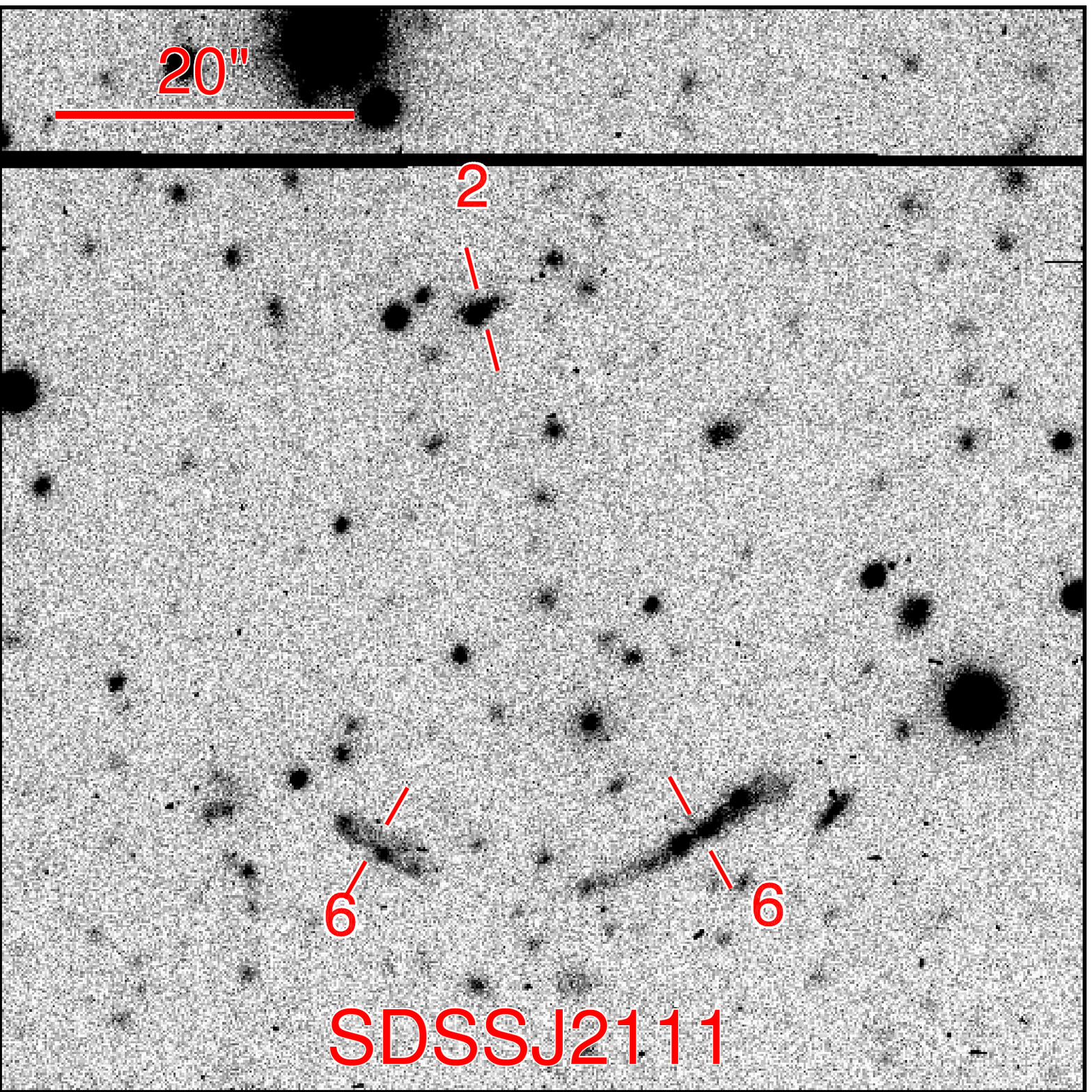,width=0.45\textwidth}}
  \caption{ continued.}
\end{figure*}

\begin{figure*}
  \vskip -0.1cm
  \centerline{\epsfig{file=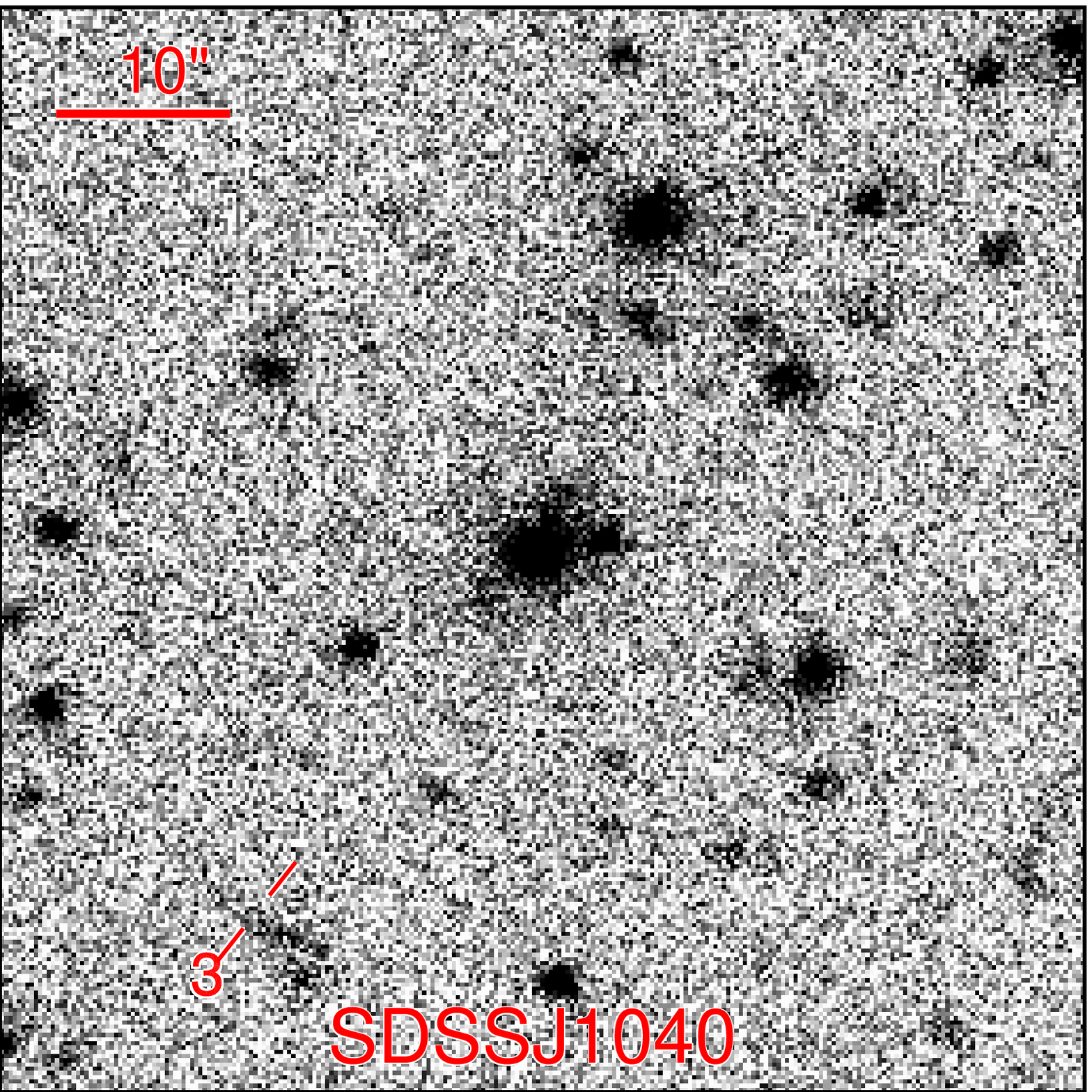,width=0.45\textwidth}
    \epsfig{file=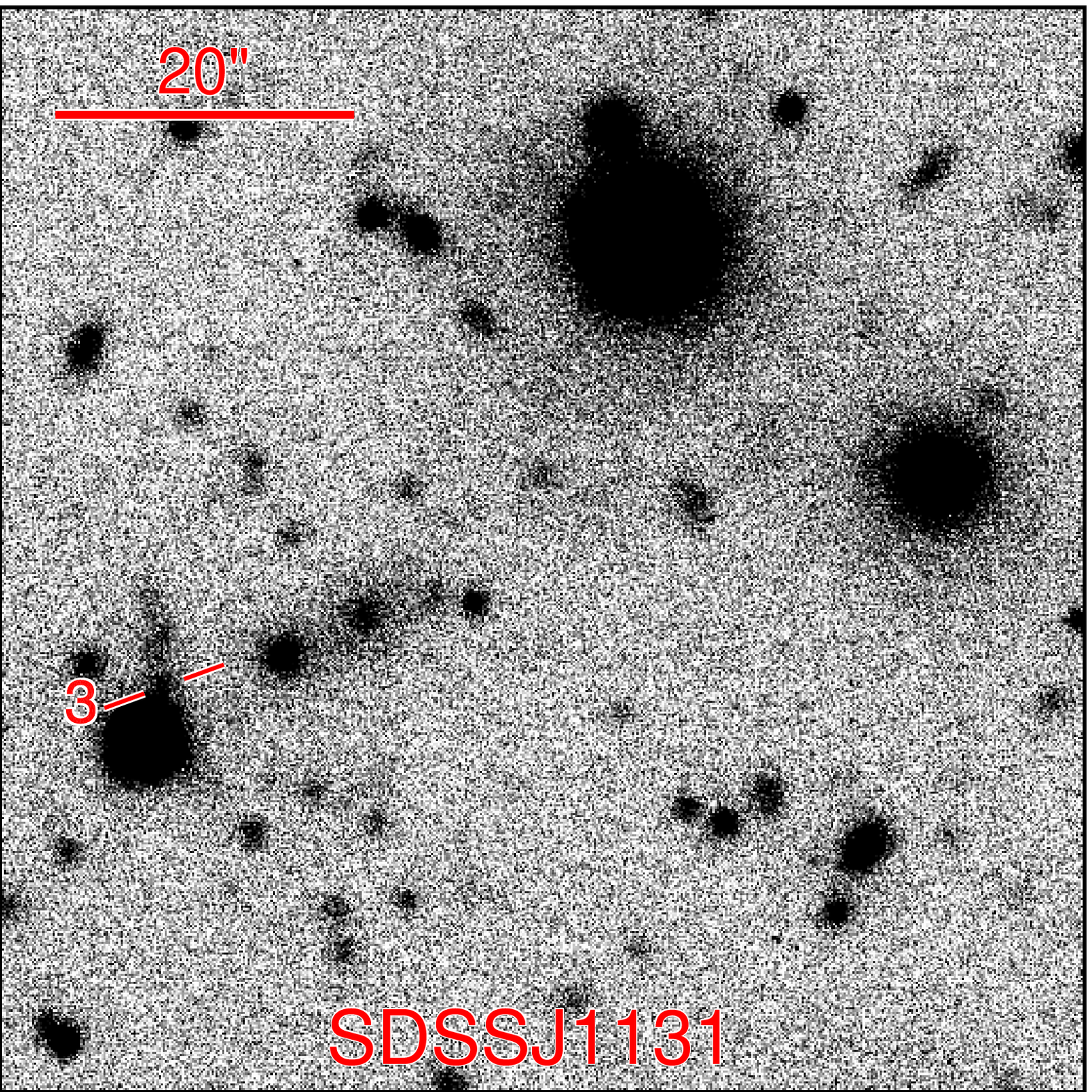,width=0.45\textwidth}}
  \vskip 0.05cm
  \centerline{\epsfig{file=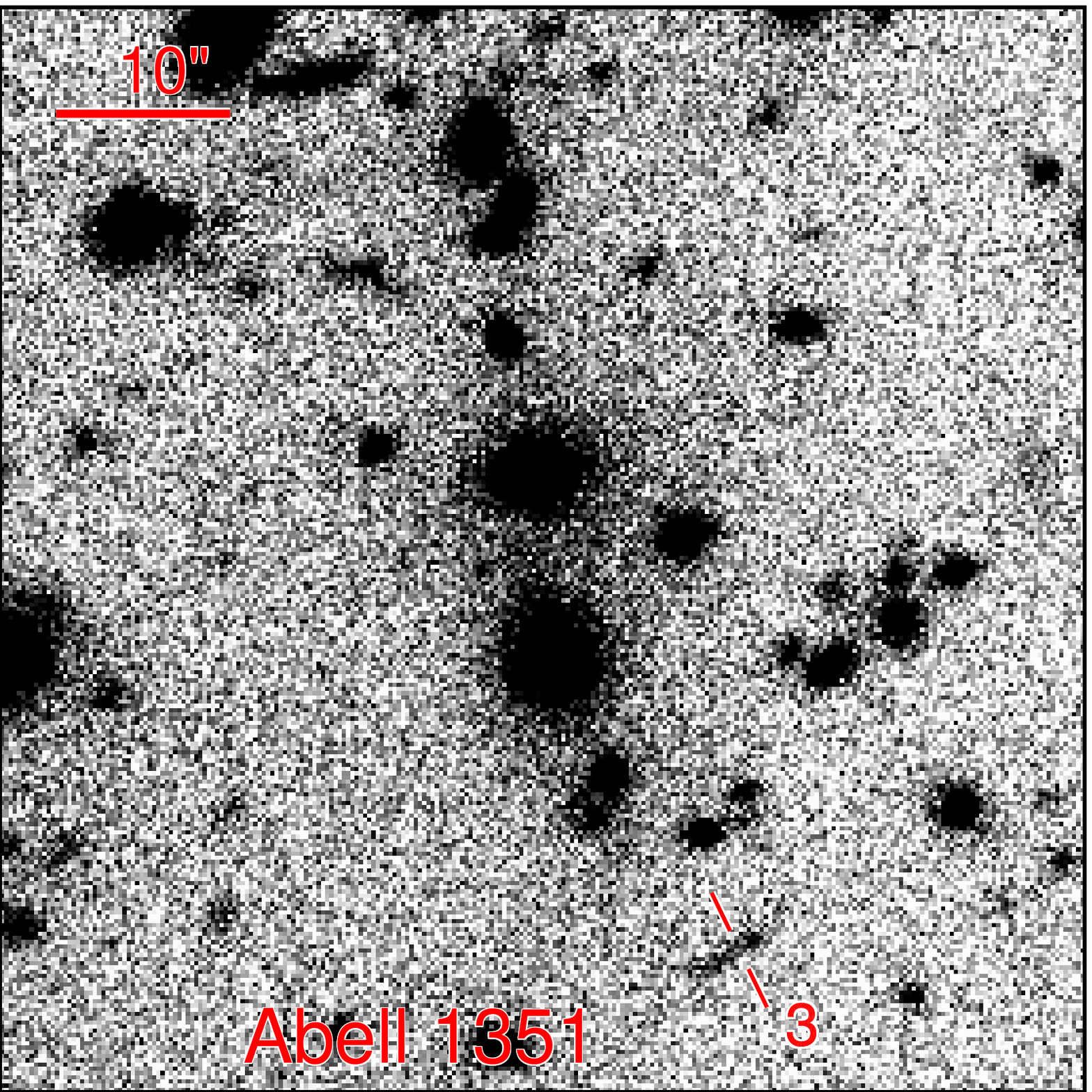,width=0.45\textwidth}
    \epsfig{file=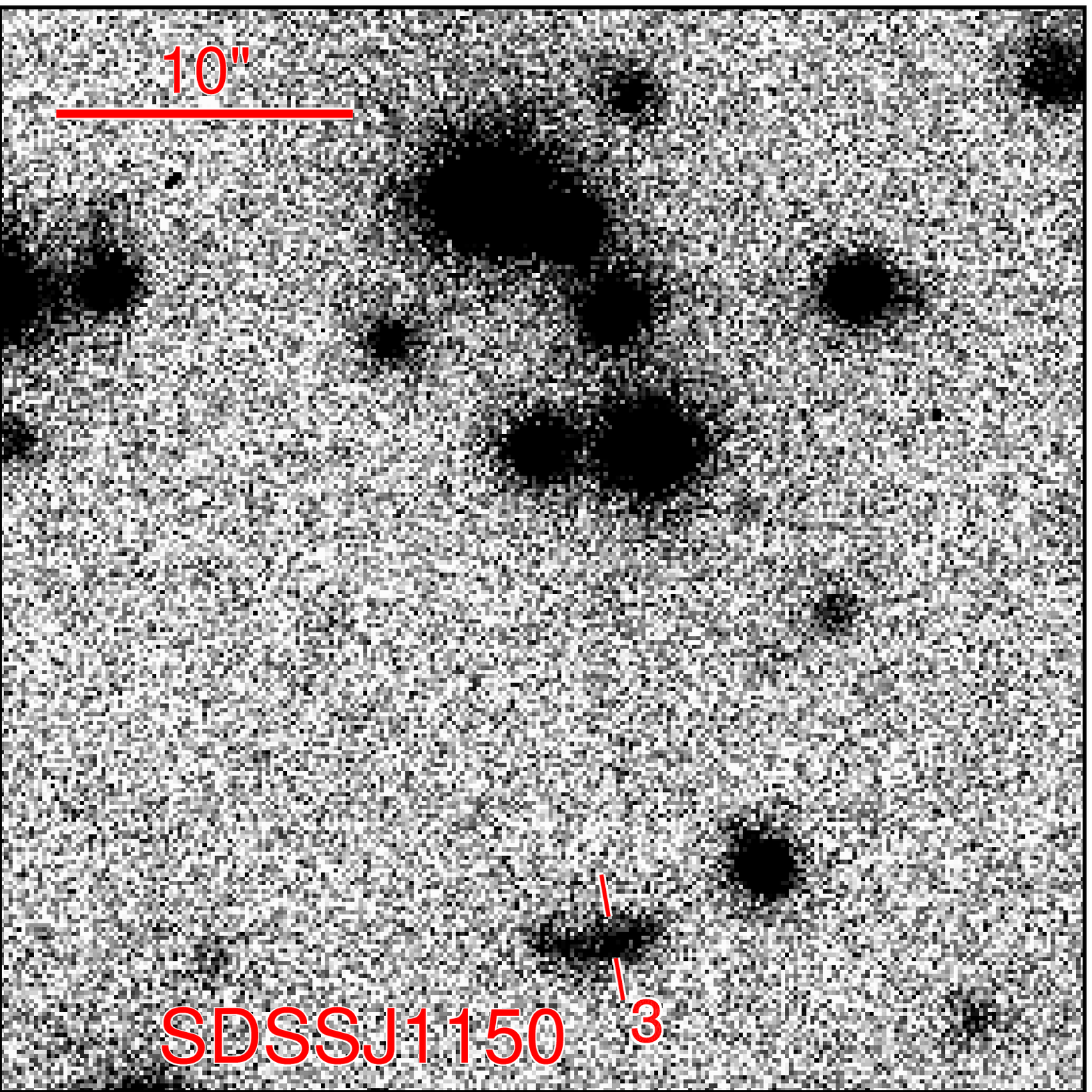,width=0.45\textwidth}}
  \vskip 0.05cm
  \centerline{\epsfig{file=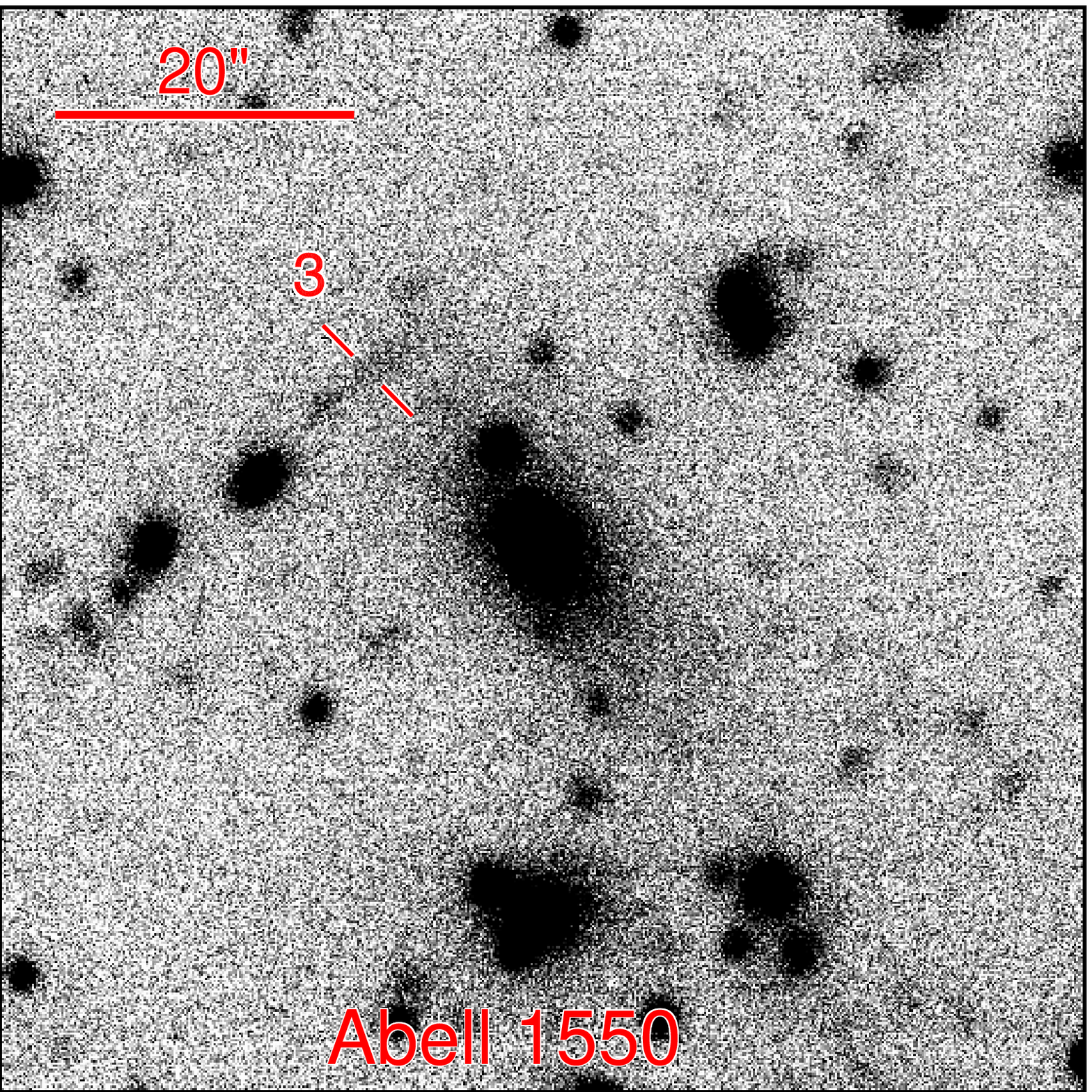,width=0.45\textwidth}
    \epsfig{file=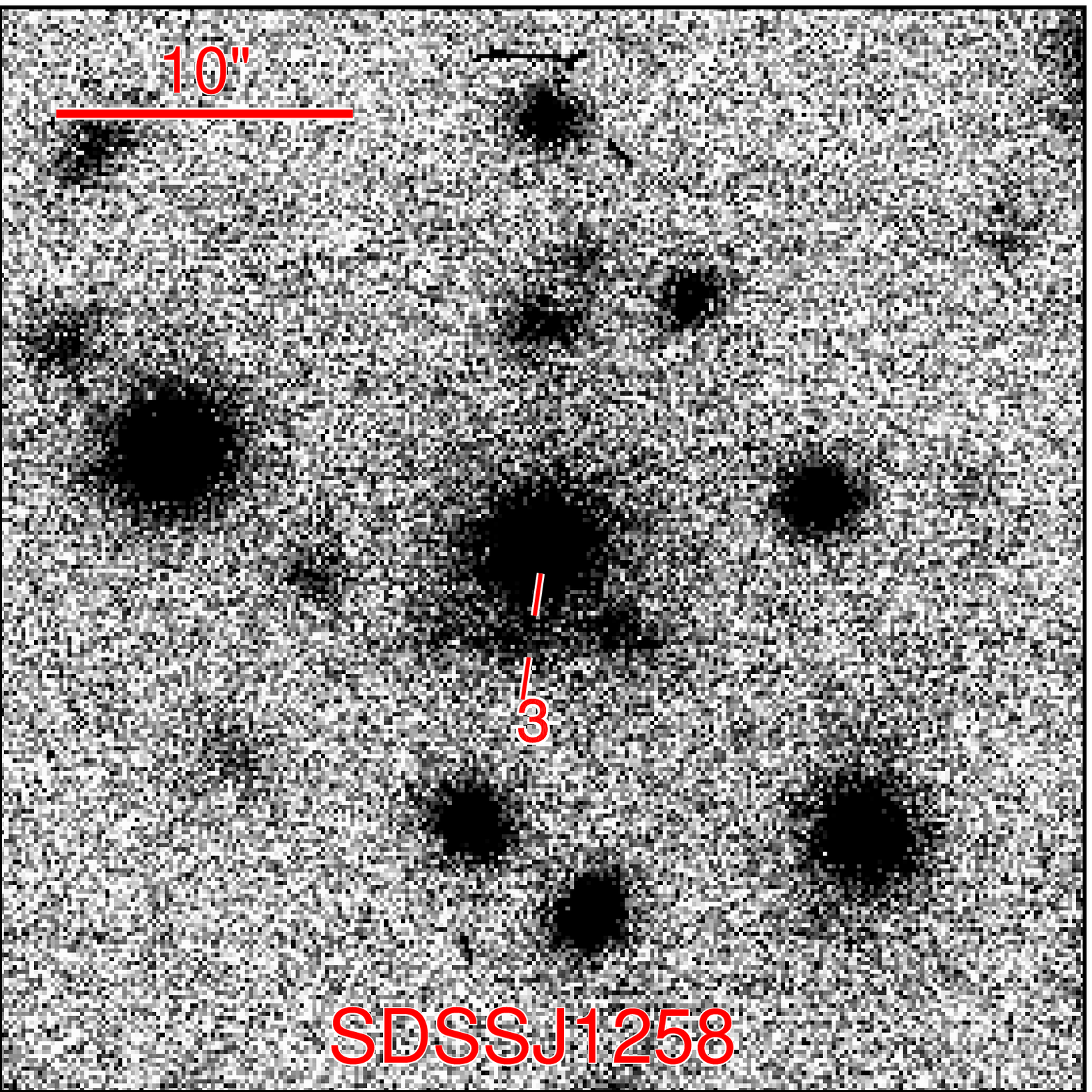,width=0.45\textwidth}}
  \caption{Same as Figure~\ref{fig:cutout1}, but for the likely lensing 
    clusters listed in Table~\ref{table:cutout2}.\label{fig:cutout2}}
\end{figure*}
\addtocounter{figure}{-1}
\begin{figure*}
  \vskip -0.1cm
  \centerline{\epsfig{file=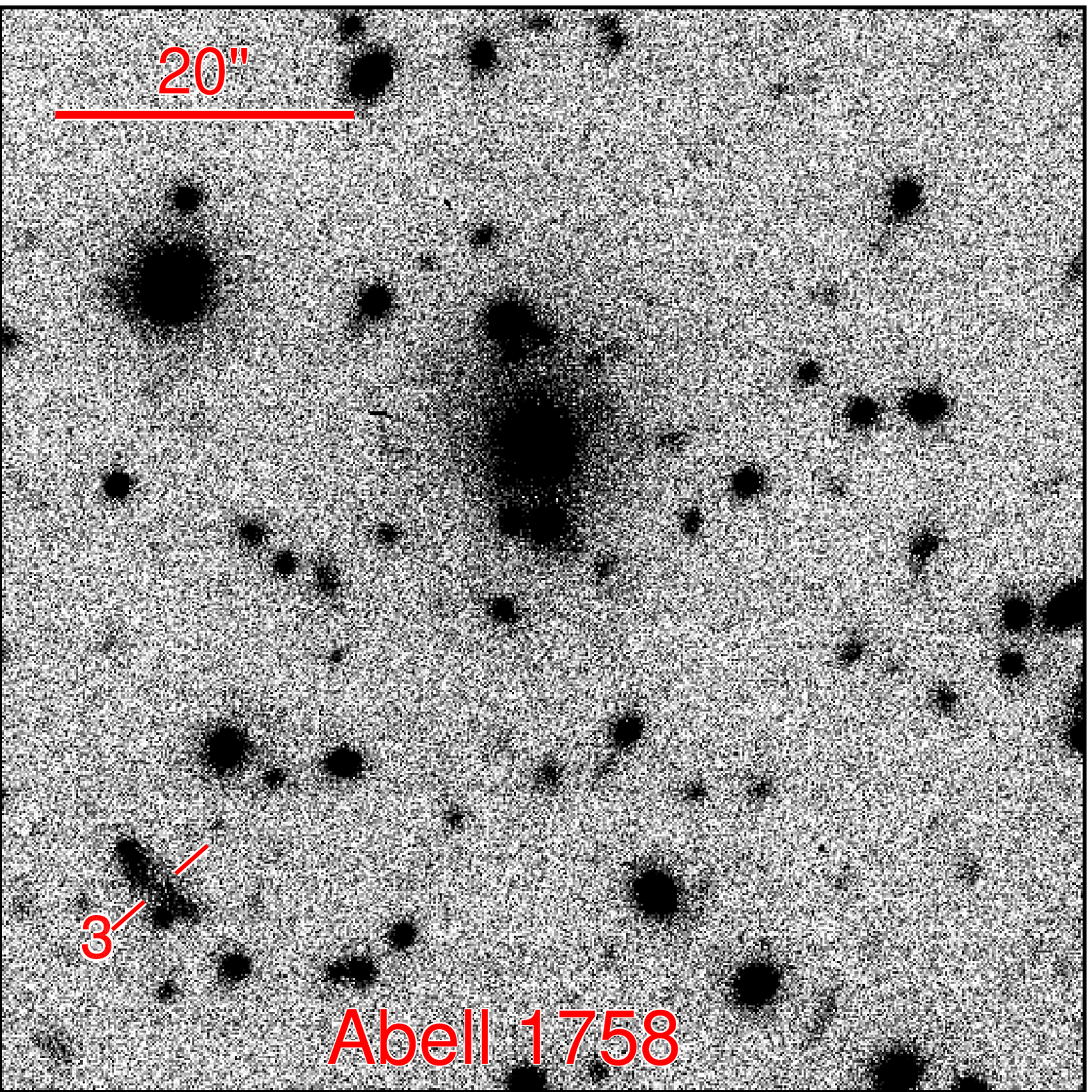,width=0.45\textwidth}
    \epsfig{file=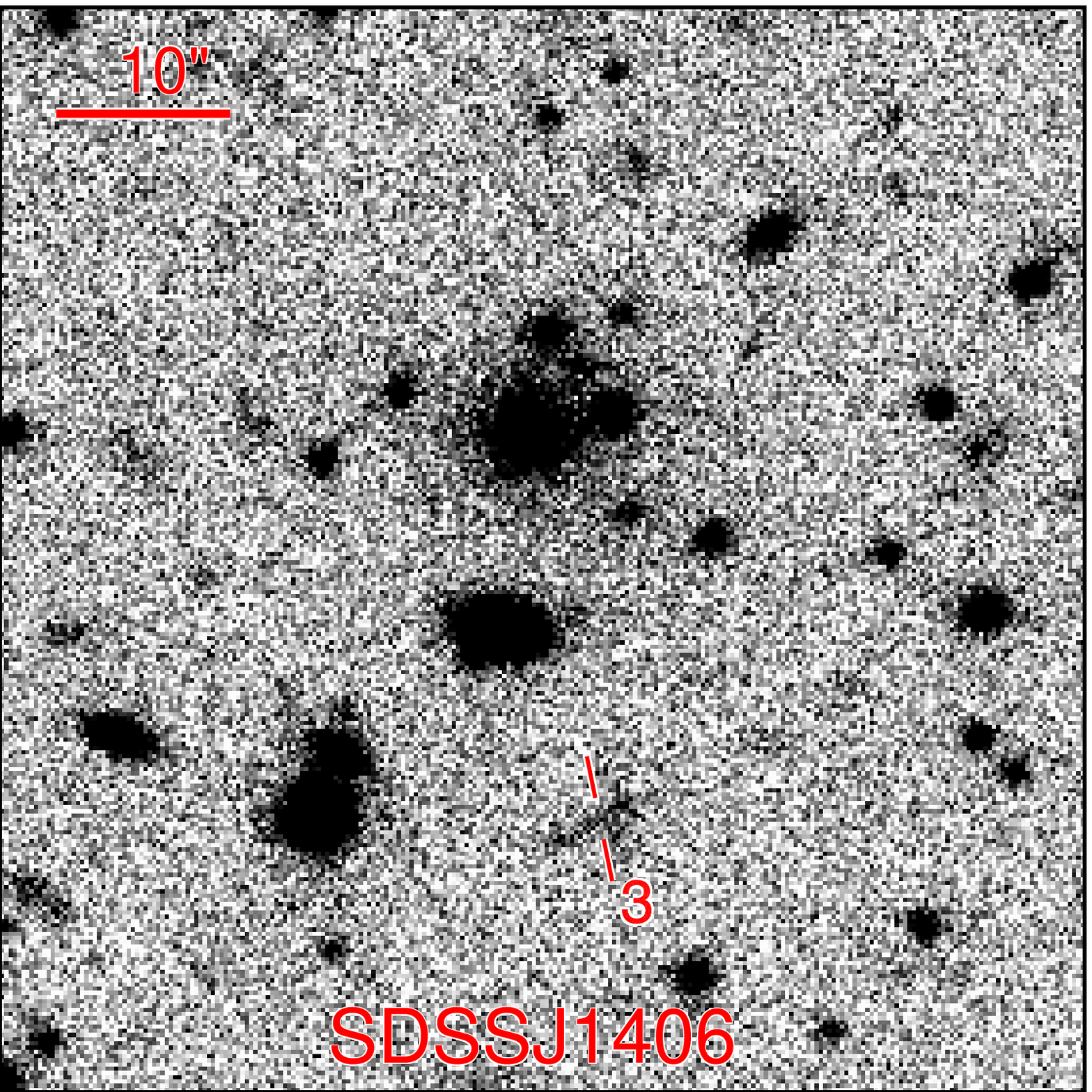,width=0.45\textwidth}}
  \vskip 0.05cm
  \centerline{\epsfig{file=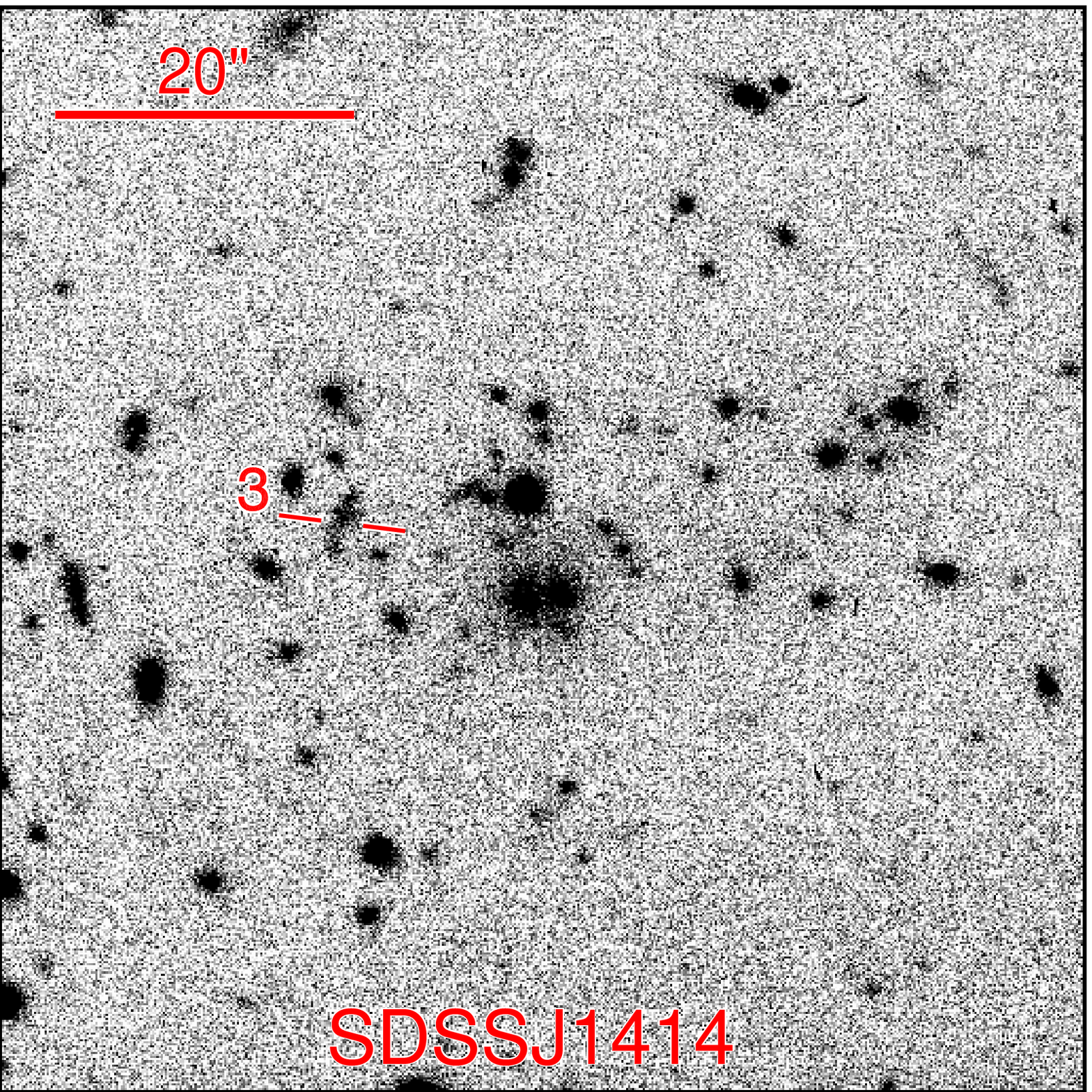,width=0.45\textwidth}
    \epsfig{file=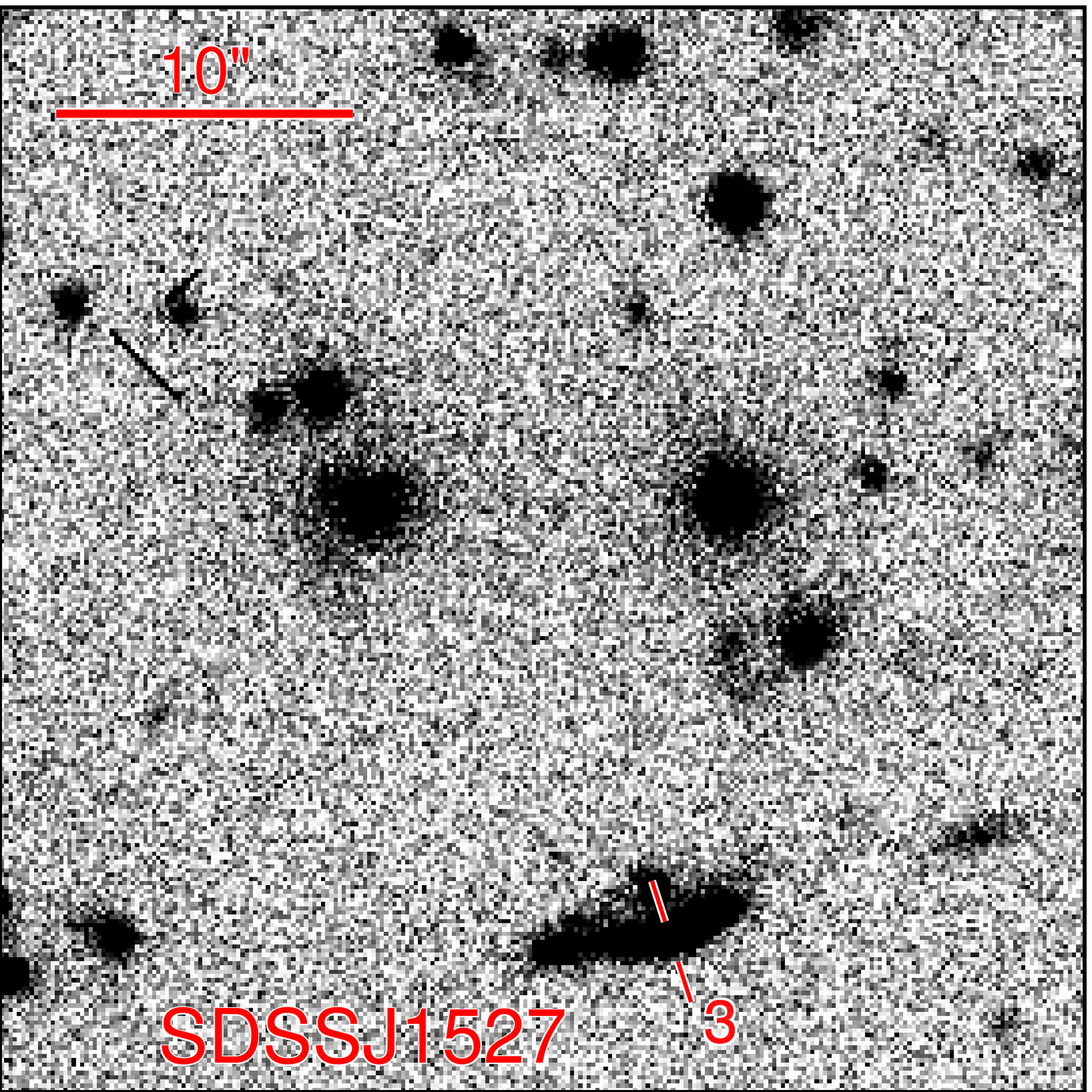,width=0.45\textwidth}}
  \vskip 0.05cm
  \centerline{\epsfig{file=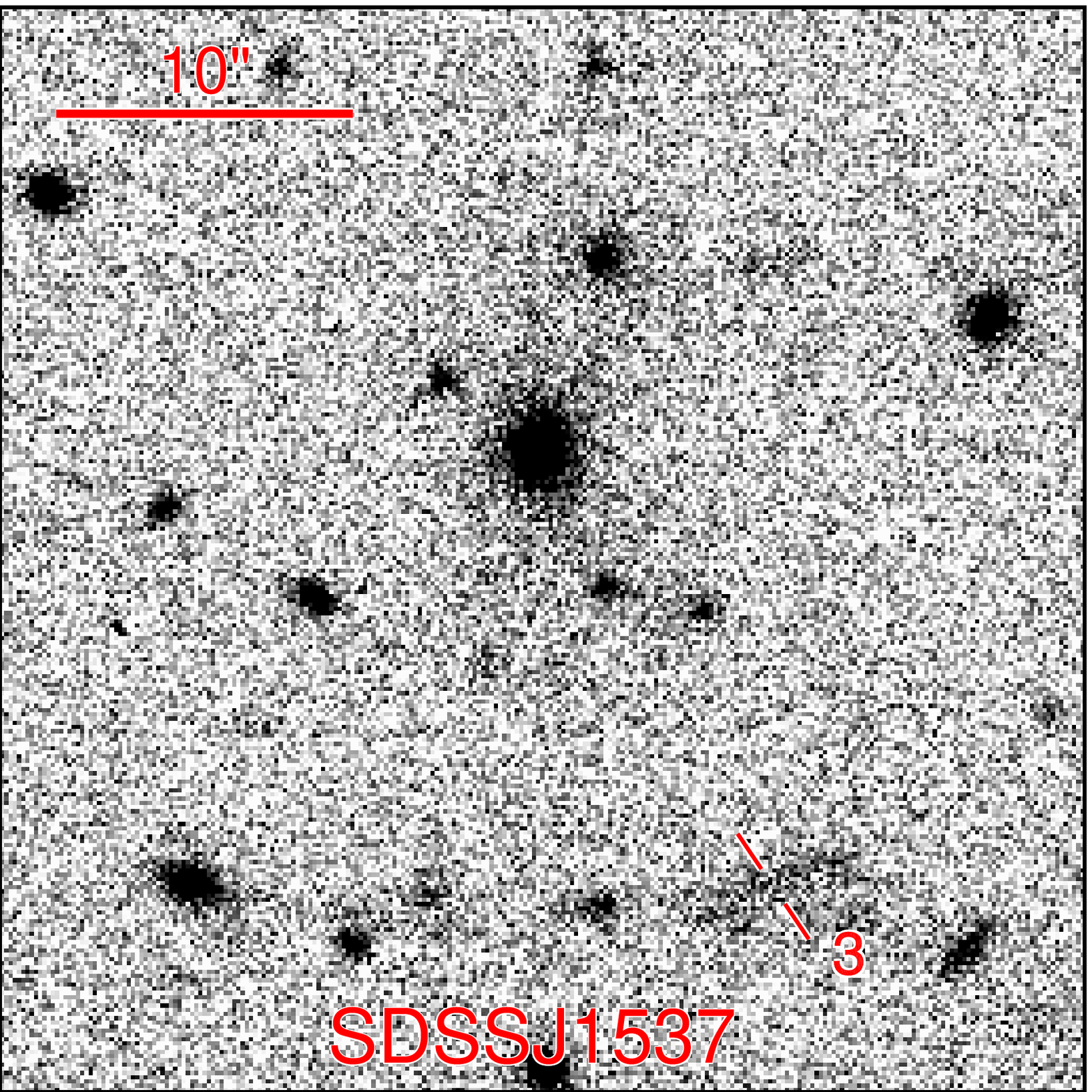,width=0.45\textwidth}
    \epsfig{file=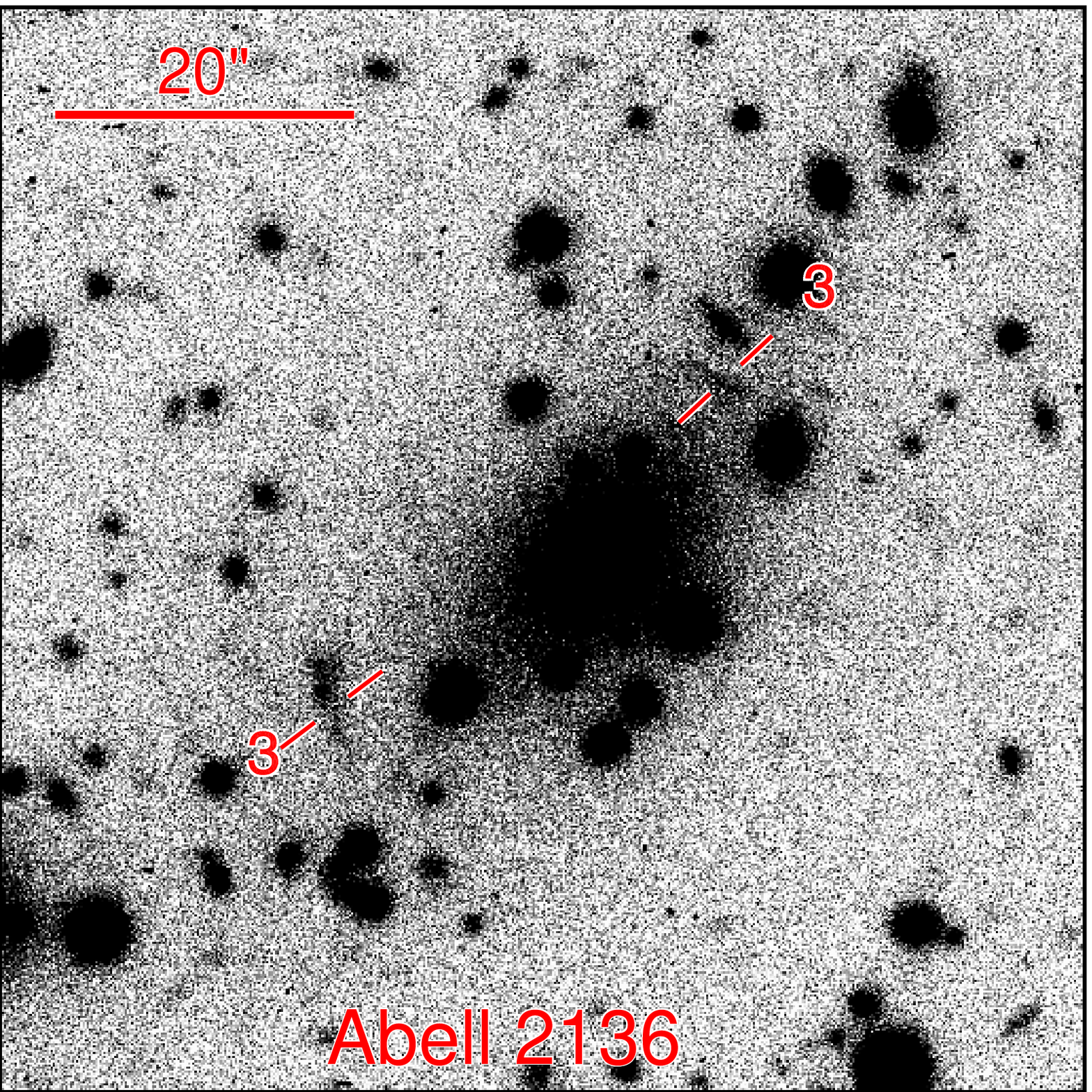,width=0.45\textwidth}}
  \caption{ continued.}
\end{figure*}
\addtocounter{figure}{-1}
\begin{figure*}
  \vskip -0.1cm
  \centerline{\epsfig{file=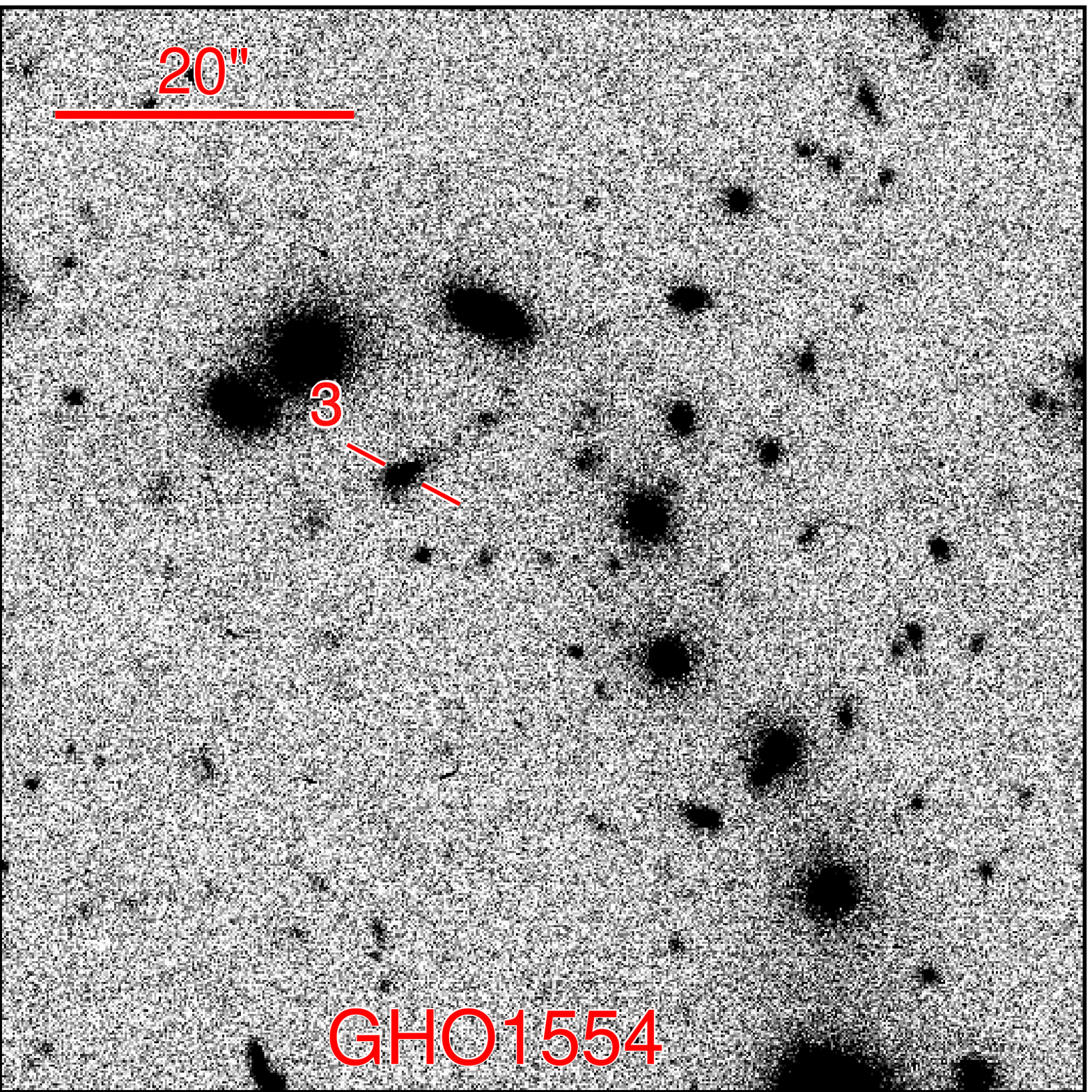,width=0.45\textwidth}
    \epsfig{file=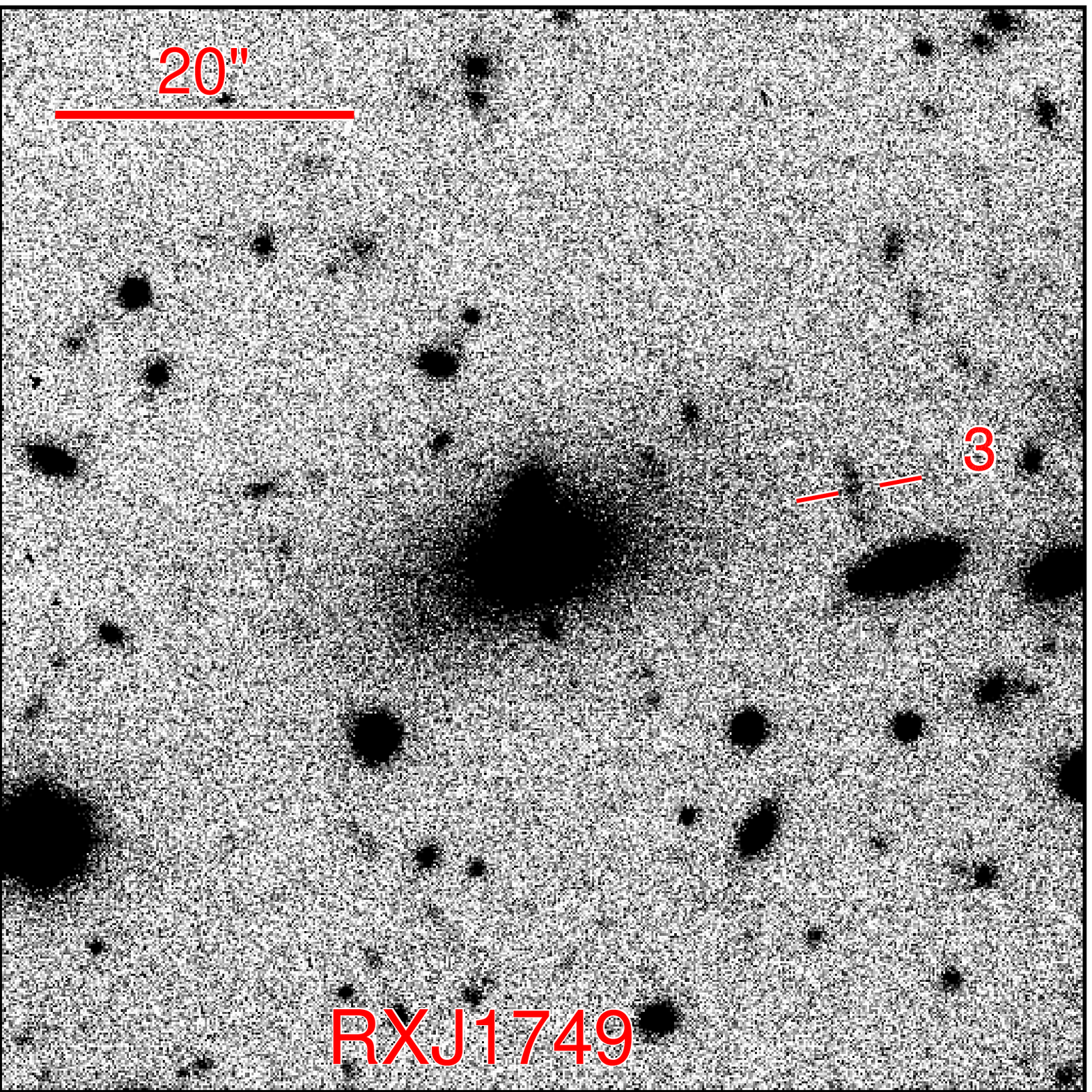,width=0.45\textwidth}}
  \caption{ continued.}
\end{figure*}

\begin{figure*}
  \vskip -0.1cm
  \centerline{\epsfig{file=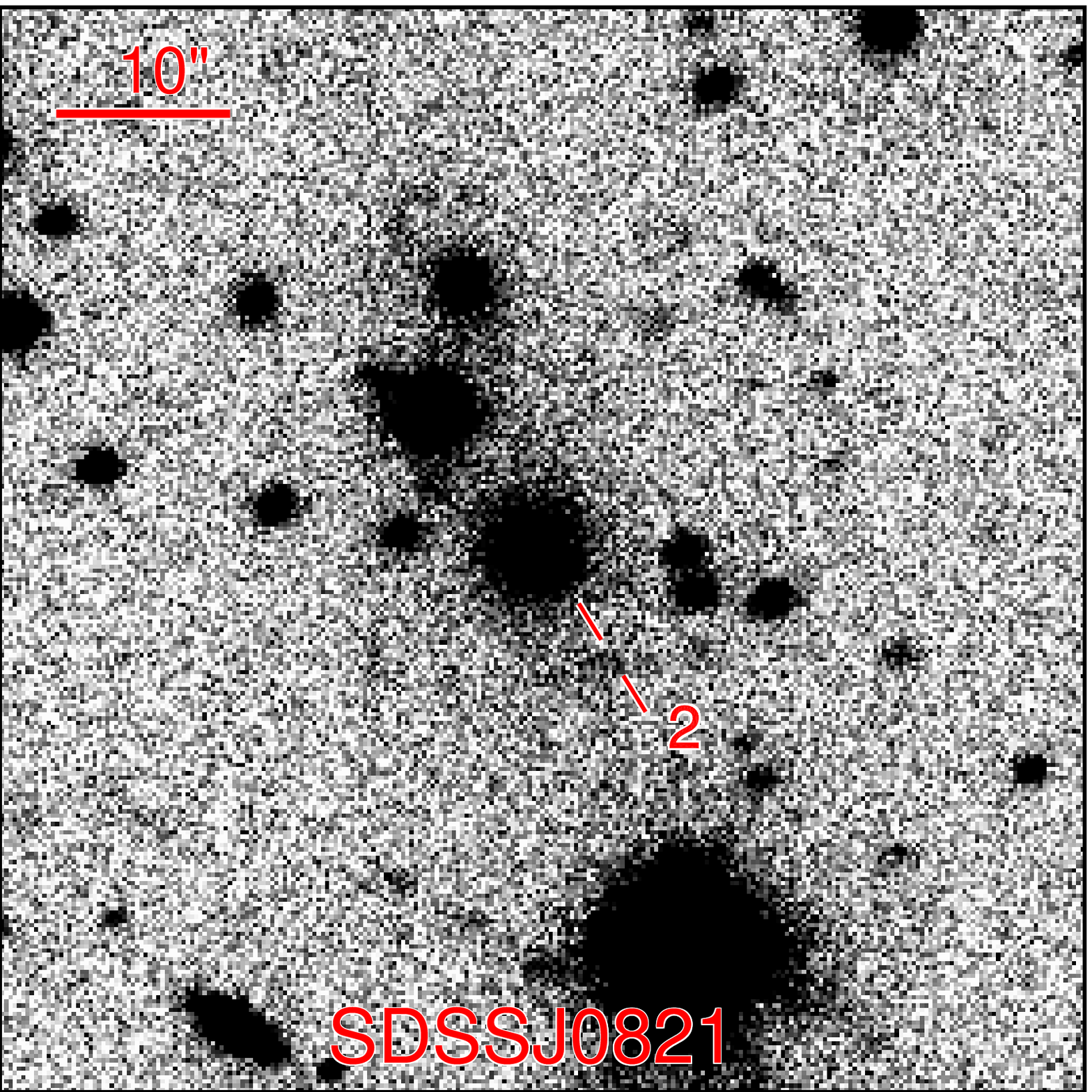,width=0.45\textwidth}
    \epsfig{file=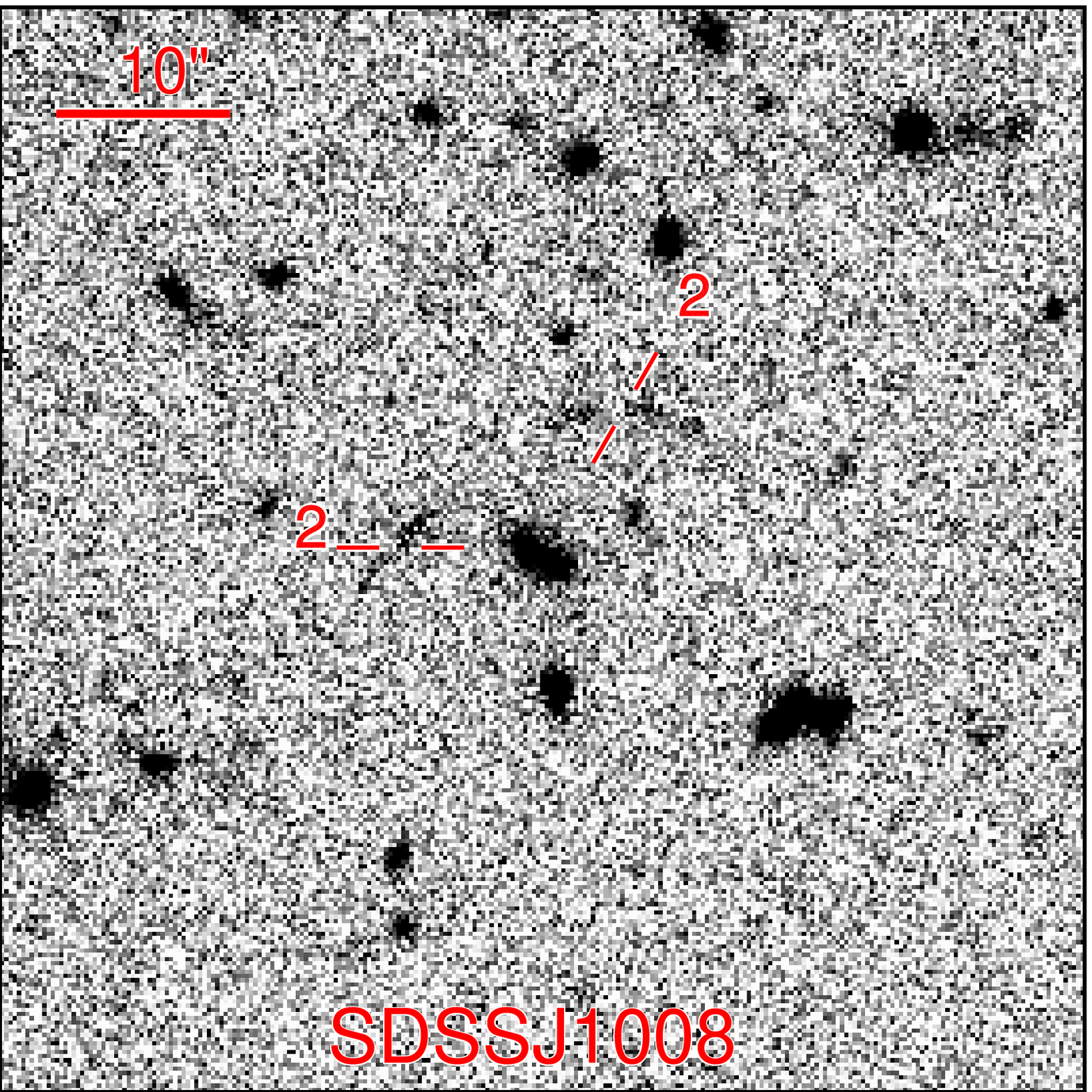,width=0.45\textwidth}}
  \vskip 0.05cm
  \centerline{\epsfig{file=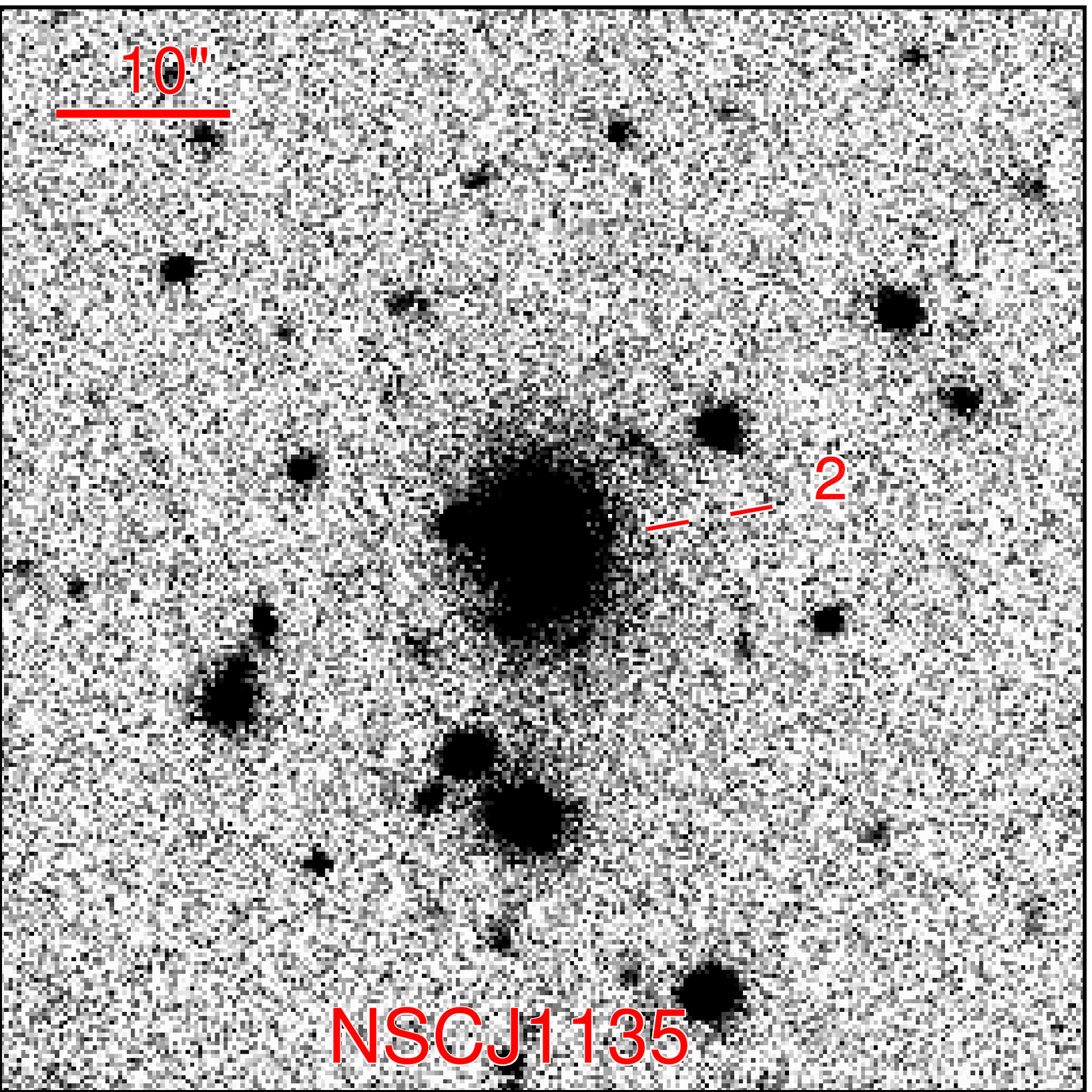,width=0.45\textwidth}
    \epsfig{file=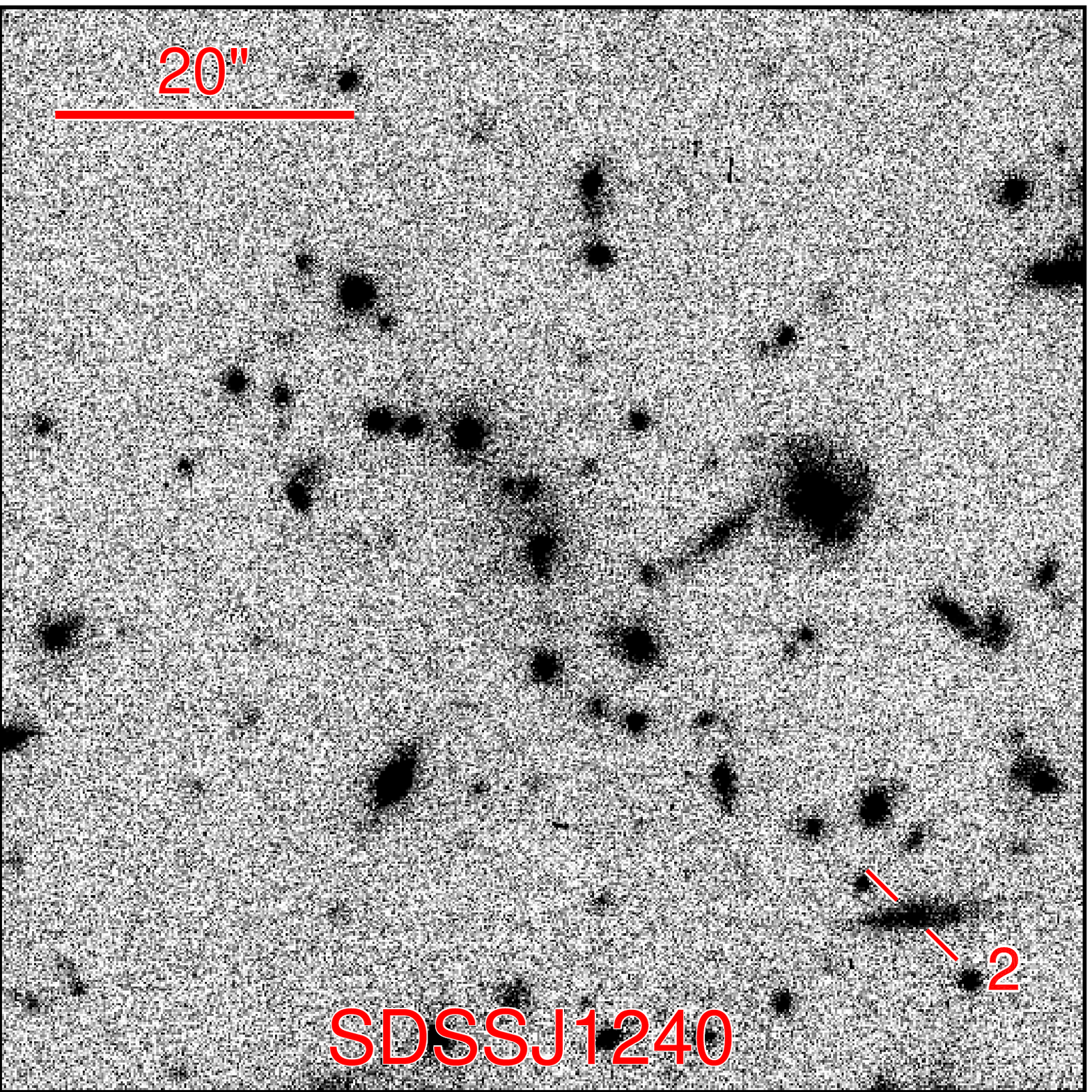,width=0.45\textwidth}}
  \vskip 0.05cm
  \centerline{\epsfig{file=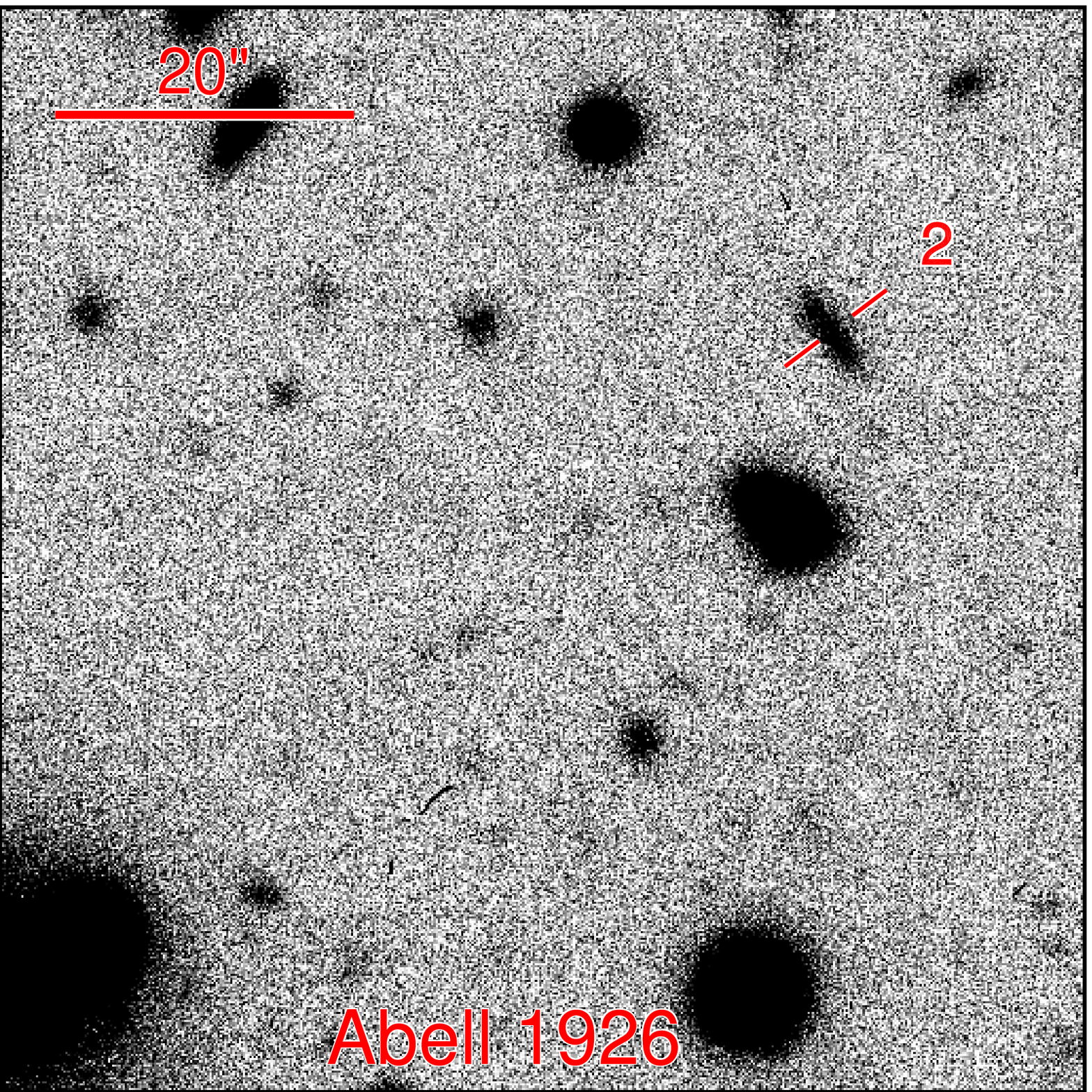,width=0.45\textwidth}
    \epsfig{file=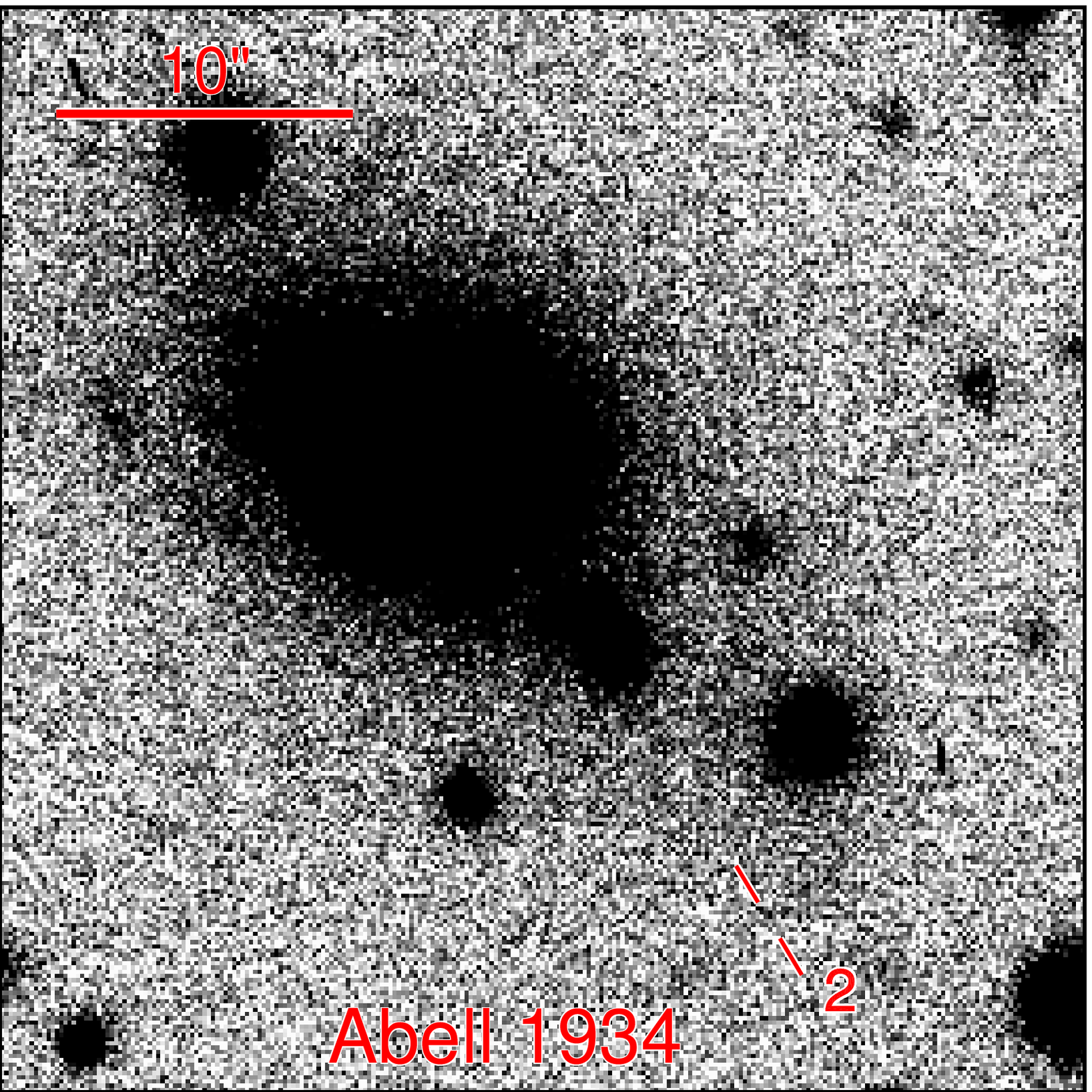,width=0.45\textwidth}}
  \caption{Same as Figure~\ref{fig:cutout1}, but for the possible lensing 
    clusters listed in Table~\ref{table:cutout3}.\label{fig:cutout3}}
\end{figure*}
\addtocounter{figure}{-1}
\begin{figure*}
  \vskip -0.1cm
  \centerline{\epsfig{file=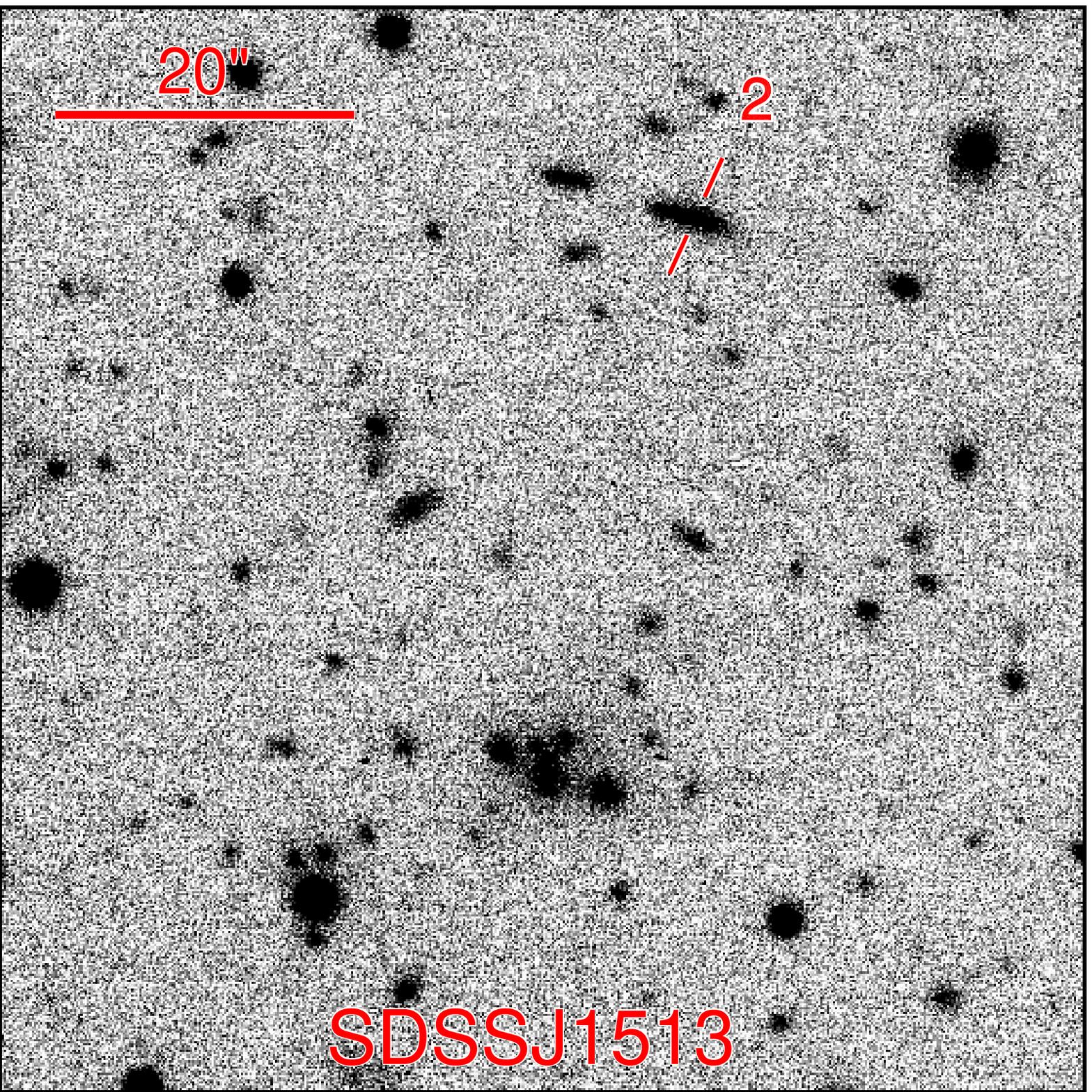,width=0.45\textwidth}
    \epsfig{file=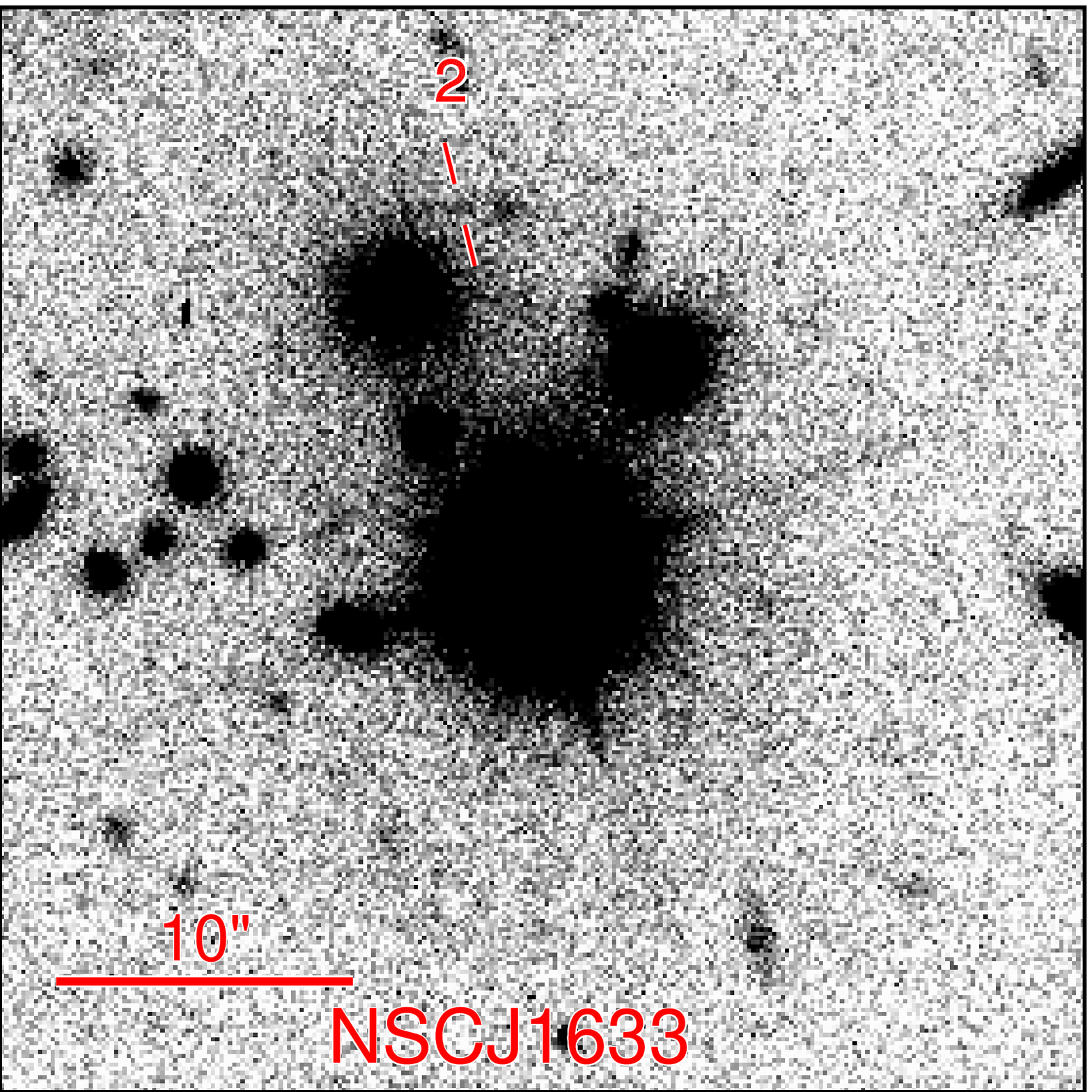,width=0.45\textwidth}}
  \hskip 0.85cm \epsfig{file=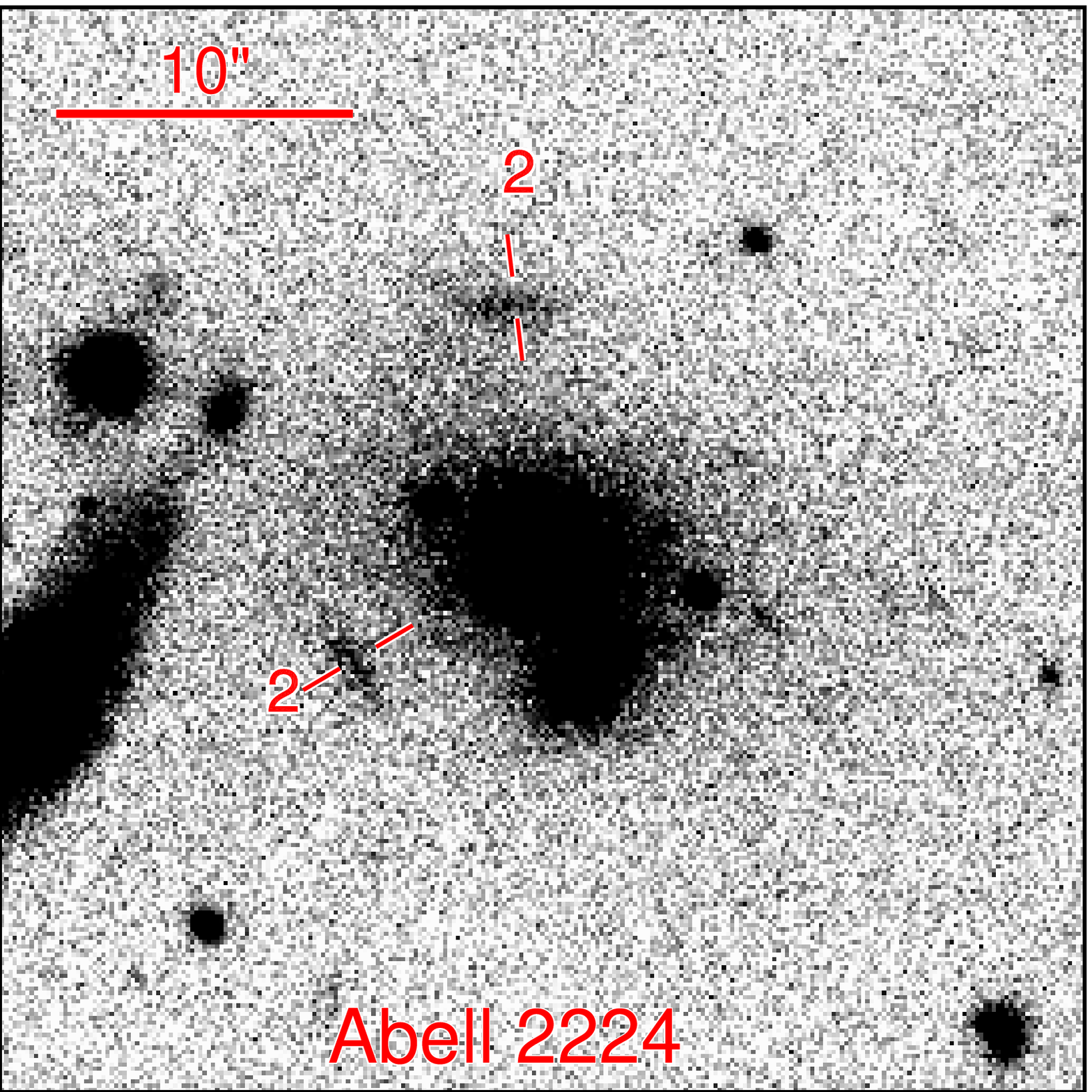,width=0.45\textwidth}
  \caption{ continued.}
\end{figure*}

\acknowledgments 

We are grateful to Nikhil Padmanabhan, Erin Sheldon, and David
Schlegel for help with photometric redshifts for SDSS galaxies.  We
would like to thank Steve Howell, George Will, Doug Williams and the 
other members of the WIYN observatory staff for assistance and superb
technical support over the course of our survey runs.  JFH and MG are
supported by NASA through Hubble Fellowship grants \# HF-01172.01-A
and HF-01184.01 respectively, awarded by the Space Telescope Science
Institute, which is operated by the Association of Universities for
Research in Astronomy, Inc., for NASA, under contract NAS
5-26555. This work was supported in part by the Department of Energy
contract DE-AC02-76SF00515. Use of the UH 2.2-m telescope for the
observations is supported by NAOJ.

Based in part on observations obtained at the 3.5 m and 0.9 m WIYN
Telescopes. The WIYN Observatory is a joint facility of the University
of Wisconsin-Madison, Indiana University, Yale University, and the
National Optical Astronomy Observatory (NOAO).

The authors wish to recognize and acknowledge the very significant
cultural role and reverence that the summit of Mauna Kea has always
had within the indigenous Hawaiian community.  We are most fortunate
to have the opportunity to conduct observations from this mountain.

Funding for the SDSS and SDSS-II has been provided by the Alfred
P. Sloan Foundation, the Participating Institutions, the National
Science Foundation, the U.S. Department of Energy, the National
Aeronautics and Space Administration, the Japanese Monbukagakusho,
the Max Planck Society, and the Higher Education Funding Council
for England. The SDSS Web Site is http://www.sdss.org/.

The SDSS is managed by the Astrophysical Research Consortium for the
Participating Institutions. The Participating Institutions are the
American Museum of Natural History, Astrophysical Institute Potsdam,
University of Basel, Cambridge University, Case Western Reserve
University, University of Chicago, Drexel University, Fermilab, the
Institute for Advanced Study, the Japan Participation Group, Johns
Hopkins University, the Joint Institute for Nuclear Astrophysics, the
Kavli Institute for Particle Astrophysics and Cosmology, the Korean
Scientist Group, the Chinese Academy of Sciences (LAMOST), Los Alamos
National Laboratory, the Max-Planck-Institute for Astronomy (MPIA),
the Max-Planck-Institute for Astrophysics (MPA), New Mexico State
University, Ohio State University, University of Pittsburgh,
University of Portsmouth, Princeton University, the United States
Naval Observatory, and the University of Washington.

\end{document}